%% file: SPRLD.tex
\documentclass[journal]{IEEEtran}
\IEEEsettopmargin{t}{0.75in}

\usepackage[cmyk,dvipsnames,svgnames]{xcolor}
\usepackage{stfloats}
\usepackage{lipsum}
\usepackage{cite}
\usepackage{graphicx}
\usepackage{subcaption}
\usepackage{pgfplots}
\usepackage{multirow}
\usepackage{amssymb}
\usepackage{amsmath, mathtools, upgreek}
\usepackage{cases}
\usepackage{pifont}
\usepackage{bm}
\usepackage{bbm}
\usepackage{amsthm}
\usepackage{tabularx}
\usepackage{booktabs}
\usepackage{array}
\usepackage{float}
\usepackage{xcolor}
\usepackage{setspace}
\usepackage[linesnumbered,ruled,vlined]{algorithm2e}
\usepackage{changes}
\usepackage[hidelinks]{hyperref}
\usepackage{bbm}
\usepackage{soul}
\usepackage{physics}
\usepackage{tikz}
\usepackage{etoolbox}
\usepackage{scalerel}
\usepackage[section]{placeins}
\usepackage{url}
\usepackage{makecell}

\usetikzlibrary{decorations,decorations.markings}
\usetikzlibrary{optics, patterns}
\usetikzlibrary{arrows,positioning,shapes.geometric,spy}

\newcolumntype{Y}{>{\centering\arraybackslash}X}

\newcommand{\fixme}[2]{\ifx&#2&{\leavevmode\color{red}#1}\else{\leavevmode\color{red}FIXME\{}#1{\leavevmode\color{red}\}}\footnote{{\leavevmode\color{red}#2}}\PackageWarning{Fixme}{#1: #2}\fi}

\SetCommentSty{mycommfont}

\DeclareMathOperator*{\argmin}{arg\,min}
\DeclareMathOperator*{\argmax}{arg\,max}

\DeclareMathOperator*{\sgn}{sgn}

\definecolor{alizarin}{rgb}{0.82, 0.1, 0.26}

\allowdisplaybreaks

\begin{document}
	
	\title{\textcolor{black}{Decoding Reed-Muller Codes with Successive Codeword Permutations}}
	
	\author{
		Nghia~Doan, %~\IEEEmembership{Student~Member,~IEEE},
		Seyyed~Ali~Hashemi, %~\IEEEmembership{Member,~IEEE},\\
		Marco~Mondelli, %~\IEEEmembership{Senior~Member,~IEEE},
		and Warren~J.~Gross %~\IEEEmembership{Senior~Member,~IEEE}
		%\thanks{N.~Doan and W.~J.~Gross are with the Department of Electrical and Computer Engineering, McGill University, Montr\'eal, QC H3A~0E9, Canada (emails: nghia.doan@mail.mcgill.ca, warren.gross@mcgill.ca).}
		%\thanks{S.~A.~Hashemi is with the Department of Electrical Engineering, Stanford University, Stanford, CA 94305, USA (email: ahashemi@stanford.edu).}
		%\thanks{M.~Mondelli is with the Institute of Science and Technology Austria, Am Campus 1, 3400 Klosterneuburg, Austria (email: marco.mondelli@ist.ac.at).}
	}
	
	\maketitle
	\begin{abstract}
		A novel recursive list decoding (RLD) algorithm for Reed-Muller (RM) codes based on successive permutations (SP) of the codeword is presented. A low-complexity SP scheme applied to a subset of the symmetry group of RM codes is first proposed to carefully select a good codeword permutation on the fly. Then, the proposed SP technique is integrated into an improved RLD algorithm that initializes different decoding paths with random codeword permutations, which are sampled from the full symmetry group of RM codes. Finally, efficient latency and complexity reduction schemes are introduced that virtually preserve the error-correction performance of the proposed decoder. Simulation results demonstrate that at the target frame error rate of $10^{-3}$ for the RM code of length $256$ with $163$ information bits, the proposed decoder reduces $6\%$ of the computational complexity and $22\%$ of the decoding latency of the state-of-the-art semi-parallel simplified successive-cancellation decoder with fast Hadamard transform (SSC-FHT) that uses $96$ permutations from the full symmetry group of RM codes, while relatively maintaining the error-correction performance and memory consumption of the semi-parallel permuted SSC-FHT decoder.
	\end{abstract}
	\begin{IEEEkeywords}
		Reed-Muller codes, polar codes, 5G, codeword permutations.
	\end{IEEEkeywords}
	
	\IEEEpeerreviewmaketitle
	\section{Introduction} \label{sec:intro}
	
	% Introductions of Reed-Muller and polar codes, and their representative decoders.
	Reed-Muller (RM) codes are a class of error-correction codes discovered by Muller \cite{Muller} and Reed \cite{Reed}, which were proven to achieve the capacity of erasure channels thanks to their large symmetry (automorphism) group \cite{Shrinivas}. RM codes are similar to polar codes under the factor-graph representation of the codes. The main difference between RM and polar codes is that RM codes are constructed to maximize the minimum distance among all the codewords \cite{Muller, Reed}, while polar codes are constructed to minimize the error probability under successive-cancellation (SC) decoding \cite{Pedarsani, PolarConst_Trifonov, PolarConst_Vardy, Marco} or SC list (SCL) decoding \cite{Huang19, Liao20}. An advantage of RM codes over polar codes is that the code construction of RM codes is channel-independent, which does not impose additional complexity when the codes are constructed for different communication mediums, even under variable channel conditions.
	
	It was shown in \cite{arikan, hussami2009performance} that with maximum likelihood (ML) decoding, RM codes achieve a better error-correction performance compared to polar codes. However, ML decoding is generally impractical due to its exponential complexity. As a consequence, RM codes are often decoded using sub-optimal decoders in practice, e.g., SC-based decoding \cite{arikan, Schnabl, niu2012stack, TrifonovScore}, SCL decoding \cite{tal_list, Niu_SCL, Alexios_LLR_SCLD}, and recursive list decoding (RLD) \cite{dumerFHT, Dumer2002, Dumer06, Ardakani_TCOM, Ali_FSSCL}. Recently, a recursive projection-aggregation (RPA) decoding algorithm has been introduced to decode RM codes, whose error-correction performance is close to that of ML decoding \cite{Ye20}. However, RPA decoding in general is of higher complexity when compared with RLD decoding \cite{fathollahi2020sparse}.
	
	The error-correction performance of RM codes under various decoding algorithms can be significantly improved by utilizing their rich symmetry group \cite{Dumer06, Dumer2002, key2010reed, Ali_SP, Kamenev19, Geiselhart}. Specifically, in \cite{Kamenev19, Geiselhart} a list of codeword permutations is selected and the decoding algorithm is performed on them. As RM and polar codes share the same factor-graph representation, it was observed in \cite{Kor09thesis, BPList, Doan_GLOBECOM_18, Doan_GLOBECOM_20, CABPList} that the error-correction performance of polar codes is also improved by running the decoding algorithms on a list of their factor-graph permutations. The decoding algorithms introduced in \cite{Kamenev19, Geiselhart, Kor09thesis, BPList, Doan_GLOBECOM_18, Doan_GLOBECOM_20, CABPList} provide flexibility to optimize the decoding latency and memory requirement trade-off as the constituent decoders on the list of codeword permutations are completely independent of each other, allowing either a fully-parallel or a semi-parallel implementation. In addition, it was discovered in \cite{Geiselhart} that utilizing the codeword permutations sampled from the full symmetry group of RM codes provides significant error-correction performance gain compared to the permutations sampled from the factor-graph permutation group of the codes. Alternatively, instead of running each constituent decoder independently on a different permutation, the improved RLD algorithm introduced in \cite{Dumer06} only performs permutation decoding until the first information bit is visited. Then, only the decoding operations in the information bit domain are carried out to select the $L$ distinct best decoding paths, while keeping the permutations of all the active paths unchanged \cite{Dumer06}.
	
	A successive permutation (SP) scheme was introduced in \cite{Ali_SP} to improve the error-correction performance of RM codes under SCL decoding. Specifically, during the course of SCL decoding, the SP technique recursively selects a cyclic factor-graph permutation of a RM code to maximize the reliability of the channel seen by its constituent codes \cite{Ali_SP}. It was shown that the error-correction performance of the SP-aided SCL (SP-SCL) decoder with a list size of $L$ ($L \geq 2$) is similar to that of a conventional SCL decoder with a list size of $2L$ \cite{Ali_SP}. However, the permutation selection scheme in \cite{Ali_SP} assumes that all the constituent codes of a RM code are of maximum order. Consequently, the SP-SCL decoder does not provide significant error-correction performance gains for low-order RM codes when a relatively small list size is used.
	
	In this paper, we present a novel RLD-based algorithm of RM codes that provides better error-correction performance and complexity trade-offs in comparison with the state-of-the-art RM decoders introduced in \cite{Dumer06, Geiselhart}. The contributions of this paper are as follows:
	\begin{enumerate}
		\item We generalize the SP scheme initially proposed in \cite{Ali_SP} as a decoding problem and perform low-complexity decoding operations on a subset of the full symmetry group of RM codes to carefully select a good codeword permutation on the fly.
		\item We propose an improved RLD algorithm that utilizes the generalized SP scheme. In particular, at the beginning of the decoding process, the proposed decoder initializes $L$ decoding paths with $L$ random codeword permutations sampled from the full symmetry group of RM codes. Then, the proposed SP scheme is independently applied to each active decoding path for all the constituent RM codes visited. Furthermore, efficient complexity and latency reduction schemes are incorporated in the proposed algorithm that relatively preserve its error-correction performance.
	\end{enumerate}
	Note that the proposed RLD algorithm utilizes existing fast and efficient decoding techniques to perform ML decoding for the first-order and single parity-check (SPC) constituent RM codes as introduced in \cite{dumerFHT} and \cite{Ali_FSSCL}, respectively. Our simulation results demonstrate that for the RM code of length $256$ with $163$ information bits, at a target frame error rate (FER) of $10^{-3}$, the sequential implementation of the proposed decoder reduces $6\%$ of the computational complexity and $22\%$ of the decoding latency of a semi-parallel successive-cancellation decoder with fast Hadamard transform (SSC-FHT) \cite{dumerFHT, gabi_fast_pcd} that uses $96$ permutations from the full symmetry group of RM codes \cite{Geiselhart}, while relatively achieving similar error-correction performance and memory consumption of the permuted SSC-FHT decoder. For the same RM code and at a similar FER performance, the proposed decoder reduces $34\%$ of the computational complexity, $76\%$ of the decoding latency, and provides a memory reduction of $49\%$ compared to the RLD algorithm with list size $64$ that also uses permutations from the full symmetry group of RM codes \cite{Dumer06, Geiselhart}.
	
	The remainder of this paper is organized as follows. Section~\ref{sec:pre} introduces the backgrounds on RM codes and their graph-based decoding algorithms. Section~\ref{sec:proposed} provides details of the proposed decoding techniques and their performance evaluation considering the error-correction performance, computational complexity, decoding latency, and memory requirement. Finally, concluding remarks are drawn in Section~\ref{sec:conclud}.
	
	\section{Preliminaries}
	\label{sec:pre}
	
	Throughout this paper, boldface letters indicate vectors and matrices. Unless otherwise specified, non-boldface letters indicate either binary, integer or real numbers. Greek letters are used to denote a RM code (node), the log-likelihood ratio (LLR) values, the hard decisions associated with a RM code, and complexity metrics. Finally, sets are denoted by blackboard bold letters, e.g., $\mathbb{R}$ is the set containing real numbers.
	
	\subsection{Reed-Muller Codes}
	A RM code is specified by a pair of integers ${0 \leq r \leq m}$ and is denoted as $\mathcal{RM}(r,m)$, where $r$ is the order of the code. $\mathcal{RM}(r,m)$ has a code length ${N=2^m}$ with ${K=\sum_{i=0}^{r} {m \choose i}}$ information bits, and a minimum distance ${d=2^{m-r}}$. A RM code is constructed by applying a linear transformation to the binary message word $\bm{u} = \{u_0,u_1,\ldots,u_{N-1}\}$ as $\bm{x} = \bm{u}\bm{G}^{\otimes m}$ where $\bm{x} = \{x_0,x_1,\ldots,x_{N-1}\}$ is the codeword and $\bm{G}^{\otimes m}$ is the $m$-th Kronecker power of the matrix $\bm{G}=\bigl[\begin{smallmatrix} 1&0\\ 1&1 \end{smallmatrix} \bigr]$ \cite{Arikan10}. The element $u_i$ of $\bm{u}$ is fixed to $0$ if the weight of the $i$-th row of $\bm{G}^{\otimes m}$, denoted as $w_i$, is smaller than $d$. Formally, $u_i = 0$ $\forall i \in \mathbb{F}$, where $\mathbb{F} = \{i|0\leq i < N, w_i < d\}$. In addition, we denote by $\mathbb{I}$ the set of information bits, i.e., $\mathbb{I} = \{i|0\leq i < N, w_i \ge d\}$, and $\mathbb{I}$ and $\mathbb{F}$ are known to both the encoder and the decoder. 
	
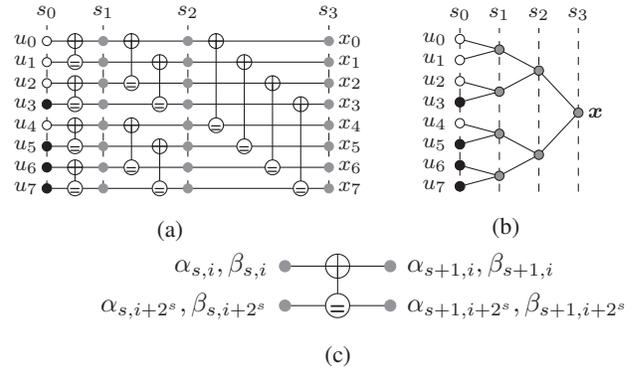
\begin{figure}[t!]
		\centering
		{
			\begin{subfigure}{0.49\columnwidth}
				\centering
				\input{PolarFactorGraph.tikz}\\
				\caption{}
				\label{fig:rm:fg}
			\end{subfigure}	
			\begin{subfigure}{0.49\columnwidth}
				\centering
				\input{PolarBinaryTree.tikz}\\
				\caption{}
				\label{fig:rm:tree}
			\end{subfigure}	
		}
		\begin{subfigure}{1\columnwidth}
			\centering
			\hspace*{20pt} \input{SCPE.tikz}
			\caption{}
			\label{fig:rm:PE}
		\end{subfigure}
		
		\caption{(a) Factor-graph representation of $\mathcal{RM}(1,3)$, (b) binary tree representation of $\mathcal{RM}(1,3)$, and (c) a PE.}
		
		%	\vspace*{-0.5\baselineskip}
	\end{figure}

	In this paper, the codeword $\bm{x}$ is modulated using binary phase-shift keying (BPSK) modulation, and an additive white Gaussian noise (AWGN) channel model is considered. Therefore, the soft vector of the transmitted codeword received by the decoder is given as ${\bm{y}=(\mathbf{1}-2\bm{x})+\bm{z}}$, where $\mathbf{1}$ is an all-one vector of size $N$, and $\bm{z} \in \mathbb{R}^N$ is a Gaussian noise vector with variance $\sigma^2$ and zero mean. In the log-likelihood ratio (LLR) domain, the LLR vector of the transmitted codeword is given as
	${\bm{\alpha}_m=\ln{\frac{Pr(\bm{x}=0|\bm{y})}{Pr(\bm{x}=1|\bm{y})}}=\frac{2\bm{y}}{\sigma^2}}$. Fig.~\ref{fig:rm:fg} illustrates the encoding process of $\mathcal{RM}(1,3)$ using the factor-graph representation of the code, where $N=8$, $K=4$, and $\mathbb{I}=\{3,5,6,7\}$. It was shown in \cite[Chapter~14]{macwilliams1977theory} that for RM codes of order $r=1$, the ML decoding algorithm can be efficiently implemented by utilizing a fast Hadamard transform (FHT). In the next sections, we summarize various decoding algorithms used to decode RM codes of order $r>1$.
	
	\subsection{Successive-Cancellation and Successive-Cancellation List Decoding} % OK
	SC decoding is executed on the factor-graph representation of the code \cite{arikan}. To obtain the message word, the soft LLR values and the hard bit estimations are propagated through all the processing elements (PEs), which are depicted in Fig.~\ref{fig:rm:PE}. Each PE performs the following computations: $\alpha_{s,i} = f(\alpha_{s+1,i},\alpha_{s+1,i+2^s})$ and $\alpha_{s,i+2^s} = g(\alpha_{s+1,i},\alpha_{s+1,i+2^s},\beta_{s,i})$, where $\alpha_{s,i}$ and $\beta_{s,i}$ are the soft LLR value and the hard-bit estimation at the $s$-th stage and the $i$-th bit, respectively. The min-sum approximation formulations of $f$ and $g$ are $f(a,b) = \min(|a|,|b|)\sgn(a)\sgn(b)$, and $g(a,b,c) = b + (1-2c)a$, where $a,b \in \mathbb{R}$ and $c\in \{0,1\}$. The soft LLR values at the $m$-th stage are initialized to $\bm{\alpha}_m$ and the hard-bit estimation of an information bit at the $0$-th stage is obtained as $\hat{u}_{i} = \beta_{0,i}=\frac{1 - \sgn(\alpha_{0,i})}{2}$, $\forall i \in \mathbb{I}$. The hard-bit values of the PE are then computed as $\beta_{s+1,i} = \beta_{s,i} \oplus \beta_{s,i+2^s}$ and $\beta_{s+1,i+2^s} = \beta_{s,i+2^s}$.
	
	SCL decoding was introduced in \cite{tal_list, Niu_SCL, Alexios_LLR_SCLD} to improve SC decoding of polar and RM codes by maintaining a list of $L$ best SC decoding paths. Under SCL decoding, the estimation of a message bit $\hat{u}_i$ $(i \in \mathbb{I})$ is considered to be both $0$ and $1$, i.e., a path split. Thus, the number of candidate codewords (decoding paths) doubles after each information bit is estimated. To prevent the exponential growth of the number of decoding paths, a path metric is utilized to select the $L$ most probable decoding paths after each information bit is decoded. In the LLR domain, the low-complexity path metric can be obtained as \cite{Alexios_LLR_SCLD}
	\begin{equation}
	\label{equ:polar:PM_LLR}
	\text{PM}_l=
	\begin{cases}
	\text{PM}_l + \abs{\alpha_{{0,i}_l}} & \text{ if } \hat{u}_i \neq \frac{1-\sgn(\alpha_{{0,i}_l})}{2},\\
	\text{PM}_l & \text{ otherwise,}
	\end{cases}
	\end{equation}
	where $\alpha_{{0,i}_l}$ denotes the soft value of the $i$-th bit at stage $0$ of the $l$-th path, and initially $\text{PM}_l=0$ $\forall l$. %After each information bit is decoded, only $L$ paths with the smallest PM values are kept to continue the decoding.
	At the end of the decoding process, only the path that has the smallest path metric is selected as the decoding output.
	
	\subsection{Successive Permutations for SCL Decoding}
	
	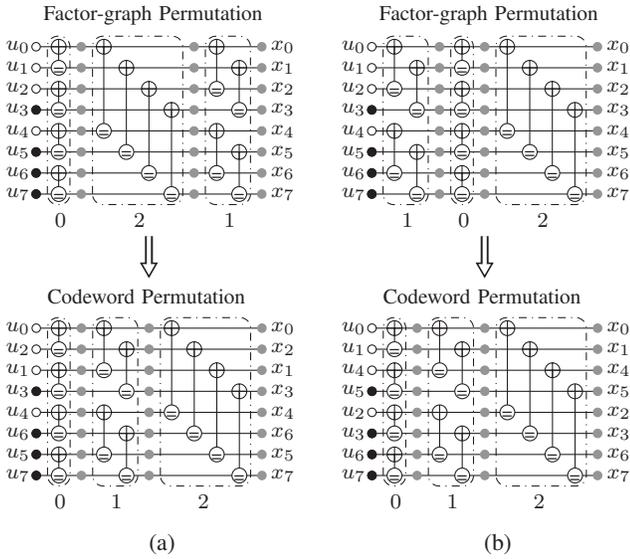
\begin{figure}
		\centering
		\begin{subfigure}{0.49\columnwidth}
			\centering
			\input{PolarFactorGraph_p1.tikz}
			\input{PolarFactorGraphPer1.tikz}
			\caption{}
		\end{subfigure}
		\begin{subfigure}{0.49\columnwidth}
			\centering
			\input{PolarFactorGraph_p2.tikz}
			\input{PolarFactorGraphPer2.tikz}
			\caption{}
		\end{subfigure}
		%{(a) \hspace*{110pt} (b) \hspace*{10pt}}
		\caption{The equivalent codeword permutations of the factor-graph permutations (a) $\{0,2,1\}$ and (b) $\{1,0,2\}$, under the original factor-graph representation with the PE layers indexed as $\{0,1,2\}$.}
		\label{fig:bit_per}
	\end{figure}
	
	SC and SCL decoding can also be illustrated on a binary tree representation of the code \cite{Schnabl, gabi_fast_pcd}. Fig.~\ref{fig:rm:tree} shows a full binary tree representation of $\mathcal{RM}(1,3)$, whose factor graph is depicted in Fig.~\ref{fig:rm:fg}. Consider a parent node $\nu$ located at the $s$-th stage $(s>0)$ of the binary tree, which is a RM code specified by a pair of parameters $(r_\nu, m_\nu)$ with $m_\nu=s$. There are $N_\nu$ LLR values and $N_\nu$ hard decisions associated with this node, where $N_\nu=2^{m_\nu}$. Let $\bm{\alpha}^{(\nu)}_{l}$ and $\bm{\beta}^{(\nu)}_{l}$ be the soft and hard values associated with the parent node $\nu$ of the $l$-th decoding path, respectively. $\bm{\alpha}^{(\nu)}_{l}$ and $\bm{\beta}^{(\nu)}_{l}$ are defined as
	\begin{equation*}
	\begin{cases}
	\bm{\alpha}^{(\nu)}_{l} = \{\alpha^{(\nu)}_{{s,i_{{\min}_{\nu_l}}}},\ldots,\alpha^{(\nu)}_{{s,i_{{\max}_{\nu_l}}}}\},\\
	\bm{\beta}^{(\nu)}_{l} = \{\beta^{(\nu)}_{{s,i_{{\min}_{\nu_l}}}},\ldots,\beta^{(\nu)}_{{s,i_{{\max}_{\nu_l}}}}\},\\
	\end{cases}
	\end{equation*}
	where $i_{{\min}_{\nu_l}}$ and $i_{{\max}_{\nu_l}}$ are the bit indices such that $0 \leq i_{{\min}_{\nu_l}} < i_{{\max}_{\nu_l}} \leq N-1$ and $i_{{\max}_{\nu_l}} - i_{{\min}_{\nu_l}} = N_\nu-1$. The hard-decision values of $\nu$ in the bipolar form are denoted as $\bm{\eta}^{(\nu)}_{l} = \{\eta^{(\nu)}_{{s,i_{{\min}_{\nu_l}}}},\ldots,\eta^{(\nu)}_{{s,i_{{\max}_{\nu_l}}}}\}$, where $\eta^{(\nu)}_{s,i}=1-2\beta^{(\nu)}_{s,i}$, $i_{{\min}_{\nu_l}} \leq i \leq i_{{\max}_{\nu_l}}$.
	
	A factor-graph permutation is constructed by permuting the PE stages of the RM code's factor graph \cite{hussami2009performance}. Fig.~\ref{fig:bit_per} illustrates examples of the factor-graph permutations of $\mathcal{RM}(1,3)$ whose original factor-graph representation is presented in Fig.~\ref{fig:rm:fg}, where the PE layers of the original factor graph in Fig.~\ref{fig:rm:fg} are indexed as $\{0,1,2\}$. Fig.~\ref{fig:bit_per} also illustrates the conversion from the factor-graph permutations to the codeword permutations used in \cite{Dumer06}, which is defined as follows. Let $\bm{b}_i=\{b_{{m-1}_i},\ldots,b_{0_i}\}$ be a binary expansion of the bit index $i$, and $\pi: \{0,\ldots,m-1\} \rightarrow \{0,\ldots,m-1\}$ be a permutation of the PE layers of $\mathcal{RM}(r,m)$. The permuted bit index of $i$ given $\pi$ in the binary expansion is $\bm{b}_{\pi(i)}=\{b_{\pi({m-1})_i},\ldots,b_{\pi({0})_i}\}$, where $0 \leq i < 2^m$ \cite{Dumer06}. 
	
	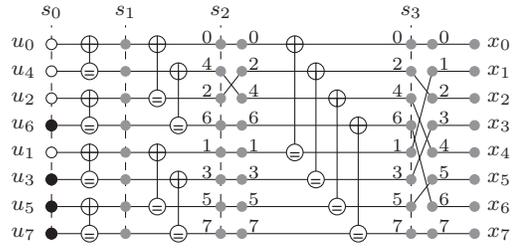
\begin{figure}[t!]
		\centering
		\input{PermutedPolarFactorGraph.tikz}
		\caption{An example of the SP technique introduced in \cite{Ali_SP} when applied to $\mathcal{RM}(1,3)$.}
		\label{fig:rmsp}
	\end{figure}
	
	The SP technique carefully selects a single codeword permutation on the fly to significantly improve the error probability of SCL decoding. Given a permutation $\pi \in \mathcal{P}_s$, the $f(\cdot)$ functions when applied to the permuted LLR values associated with the parent node $\nu$ generate an LLR vector corresponding to its left-child node $\lambda$, denoted as $\bm{\alpha}^{(\lambda)}=\{\alpha^{(\lambda)}_{s-1,i_{\min_\lambda}},\ldots,\alpha^{(\lambda)}_{s-1,i_{\max_\lambda}}\}$\footnote{In the rest of the paper, as the permutation selection scheme is applied independently for each decoding path, we drop the subscript $l$ for clarity.}. The factor-graph permutation $\pi^*$ of $\nu$ is selected to maximize the channel reliabilities corresponding to $\lambda$, which allows for a better estimation of $\lambda$ under SC and SCL decoding \cite{Ali_SP}. The selection criteria of $\pi^*$ is given as \cite{Ali_SP}
	\begin{equation}
	\pi^* = \argmax_{\pi \in \mathcal{P}_s} \sum_{i=i_{{\min}_{\lambda}}}^{i_{{\max}_{\lambda}}} \abs{\alpha^{(\lambda)}_{s-1,i}}.
	\label{equ:SP:pisel}
	\end{equation}
	The SP technique considers cyclic factor-graph permutations to be included in the set $\mathcal{P}_s$ \cite{Ali_SP}. For instance, the cyclic permutations of $\mathcal{RM}(1,3)$ are $\{0,1,2\}$, $\{2,0,1\}$, and $\{1,2,0\}$. Fig.~\ref{fig:rmsp} shows an example of the SP technique applied to $\mathcal{RM}(1,3)$, where the permuted factor graphs are transformed to the permuted bit indices as in \cite{Dumer06}. In Fig.~\ref{fig:rmsp}, let $\pi^*$ be the factor-graph permutation applied to the bit indices $I=\{0,2,4,6\}$ at stage $s_2$, corresponding to $\mathcal{RM}(0,2)$. The equivalent codeword permutation of $\pi^*$ is $\{0,1,2,3\} \rightarrow \{0,2,1,3\}$. Thus, $I=\{0,2,4,6\} \xrightarrow{\pi^*} I_{\pi^*}=\{0,4,2,6\}$, where $I_{\pi^*}$ is the resulting permuted bit indices of $I$. On the other hand, the best factor-graph permutation selected for the $\mathcal{RM}(1,2)$ code at stage $s_2$ is the original permutation.
	
	\subsection{Recursive List Decoding with Permutations} % OK
	\label{sec:polar:FSCL}
	
	\begin{figure}
		\begin{subfigure}{0.49\linewidth}
			\centering
			\input{RLD.tikz.tex}
			\caption{RLD \cite{Dumer06}}
		\end{subfigure}
		\begin{subfigure}{0.49\linewidth}
			\centering
			\input{FSCL.tikz.tex}
			\caption{FSCL \cite{Ali_FSSCL}}
		\end{subfigure}
		\caption{Binary tree representations of $\mathcal{RM}(1,3)$ under (a) RLD \cite{Dumer06} and (b) FSCL \cite{Ali_FSSCL}.}
		\label{fig:bindec}
	\end{figure}
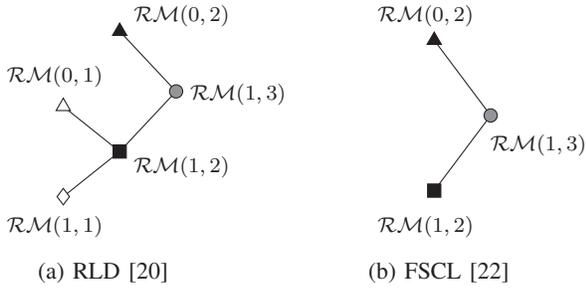
	
	The RLD algorithms introduced in \cite{dumerFHT, Dumer06} utilize ML decoding to decode the special constituent RM codes at the parent node level, instead of fully traversing the binary tree as in SCL decoding. Similarly, in \cite{Ali_FSSCL, Ardakani_TCOM}, the authors proposed fast SCL (FSCL) decoding algorithms for various special constituent polar codes, which are also RM codes. The RLD and FSCL decoding algorithms preserve the error-correction performance of SCL decoding and completely remove the need to traverse the binary tree when special nodes are encountered. Therefore, the latency of RLD and FSCL decoding is significantly smaller than SCL decoding. In particular, $\mathcal{RM}(0,m_\nu)$ (repetition (Rep) code) and $\mathcal{RM}(m_\nu-1,m_\nu)$ (SPC code) are considered under FSCL decoding \cite{Ali_FSSCL}. The RLD algorithm based on FHT (FHT-RLD) in \cite{dumerFHT} and the RLD algorithm in \cite{Dumer06} perform fast decoding for $\mathcal{RM}(1,m_\nu)$ (first-order RM code) and $\mathcal{RM}(m_\nu,m_\nu)$ (Rate-1 code). Fig.~\ref{fig:bindec} shows an example of the binary tree representations used by RLD \cite{Dumer06} and FSCL \cite{Ali_FSSCL} for decoding $\mathcal{RM}(1,3)$, while under FHT-RLD \cite{dumerFHT}, $\mathcal{RM}(1,3)$ is directly decoded using FHT without decomposing the code to smaller RM codes.
	
	An improvement scheme that runs the RLD algorithm on random factor-graph permutations was proposed in \cite{Dumer06} and is referred to as the RLDP algorithm in this paper. The RLDP algorithm on $\mathcal{RM}(r,m)$ initially runs $m \choose r$ independent decoding paths with each path performing SC decoding on a different factor-graph permutation \cite{Dumer06}. After each information bit is decoded, RLDP performs the path extension/pruning operations in the information bit domain to maintain the list of $L$ best decoding paths, while keeping the predetermined codeword permutations of the active decoding paths unchanged \cite{Dumer06}.In this paper, we run the RLD algorithm \cite{Dumer06} on the full symmetry group of the RM codes with the special nodes used in \cite{Ali_FSSCL} and we refer to this decoding algorithm as RLDA.
	
	\section{Successive Permutations for Recursive List Decoding of Reed-Muller Codes}
	\label{sec:proposed}
	
	In this section, a generalized SP scheme for the RLD-based algorithms of RM codes is first proposed. We then provide details on the integration of the proposed SP scheme into an improved RLD-based algorithm. Finally, we numerically analyze the error-correction performance, computational complexity, decoding latency, and memory requirement of the proposed decoder and compare them with those of the state-of-the-art RM decoders.
	
	\subsection{Improved Successive Permutation Scheme}
	\label{sec:proposed:SP}
	The selection criteria in accordance with (\ref{equ:SP:pisel}), which is used in \cite{Ali_SP}, is an oversimplification that does not take into account the existing parity constraints in the code. In fact, it treats all the constituent RM codes $\lambda$ as Rate-1 codes. This oversimplification becomes inaccurate, especially for low-order RM codes, as the number of information bits is significantly smaller than $N_\lambda=2^{s-1}$. Consequently, the error probability of the SP scheme in \cite{Ali_SP} for SCL decoding on low-order RM codes is not satisfactory, especially when a small to moderate list size is used.
	
	To tackle this issue, we propose an accurate SP scheme that selects the best codeword permutation $\pi^*$ by performing ML decoding on the symmetry group of RM codes. The proposed selection criteria is given as
	\begin{equation}
	\pi^* = \argmax_{\pi \in \mathcal{P}_s} M_\pi \left(\bm{\alpha}^{(\lambda)}\right),
	\label{equ:SP:sel}
	\end{equation}
	where $M_\pi \left(\bm{\alpha}^{(\lambda)}\right)$ is the permutation metric of $\pi$ when $\pi$ is applied to the parent node $\nu$ that is calculated as
	\begin{equation}
	M_\pi \left(\bm{\alpha}^{(\lambda)}\right)=\max_{\forall \bm{\eta}^{(\lambda)}} \sum_{i=i_{{\min}_{\lambda}}}^{i_{{\max}_{\lambda}}} \eta^{(\lambda)}_{s-1,i} \alpha^{(\lambda)}_{s-1,i},
	\end{equation}
	with $\bm{\eta}^{(\lambda)}$ being the hard decisions of a valid codeword corresponding to $\lambda$. It can be observed that if $\lambda$ is a Rate-1 code, (\ref{equ:SP:sel}) reverts to (\ref{equ:SP:pisel}) as $\eta^{(\lambda)}_{s-1,i}$ is set to $\sgn(\alpha^{(\lambda)}_{s-1,i})$ to maximize the likelihood of $\bm{\eta}^{(\lambda)}$ and $\bm{\alpha}^{(\lambda)}$. The elements of $\bm{\eta}^{(\lambda)}$ can be calculated by performing ML decoding on $\lambda$. However, ML decoding is generally of high complexity. Therefore, in this paper we derive $M_\pi \left(\bm{\alpha}^{(\lambda)}\right)$ for special cases of $\lambda$ for which the ML decoding operations can be realized with low complexity. Unlike \cite{Ali_SP}, the set $\mathcal{P}_s$ considered in this paper contains the general codeword permutations sampled from the full symmetry group of the codes \cite{Geiselhart}. Nevertheless, we limit the maximum number of permutations stored in $\mathcal{P}_s$ to $s$, which is equal to the number of cyclic factor-graph permutations as considered in \cite{Ali_SP}.
	
	Since first-order constituent RM codes can be decoded efficiently using ML decoding \cite[Chapter~14]{macwilliams1977theory}, the permutation metric $M_\pi \left(\bm{\alpha}^{(\lambda)}\right)$ can be efficiently calculated if $\lambda$ is a first-order RM code. When $\lambda$ is of order $2$ or higher, we propose to simplify the computation of $M_\pi \left(\bm{\alpha}^{(\lambda)}\right)$ by using (\ref{equ:SP:pisel}). The calculation of $M_\pi \left(\bm{\alpha}^{(\lambda)}\right)$ for the considered special nodes in this paper is summarized as follows.
	
	\subsubsection{$\mathcal{RM}(1,s-1)$}
	The metric $M_\pi \left(\bm{\alpha}^{(\lambda)}\right)$ is the likelihood of the best decoding path of $\lambda$ given $\bm{\alpha}^{(\lambda)}$, which can be calculated efficiently using FHT decoding \cite[Chapter~14]{macwilliams1977theory}.
	
	\subsubsection{$\mathcal{RM}(r_\lambda,s-1)$ ($r_\lambda \ge 2$)}
	We use (\ref{equ:SP:pisel}) to calculate $M_\pi \left(\bm{\alpha}^{(\lambda)}\right)$ for constituent RM codes of order $r_\lambda \geq 2$ as
	\begin{equation}
	M_\pi \left(\bm{\alpha}^{(\lambda)}\right) = \sum_{i=i_{{\min}_{\lambda}}}^{i_{{\max}_{\lambda}}} \abs{\alpha^{(\lambda)}_{s-1,i}}.
	\label{equ:SP:FPM:Rate-1}
	\end{equation}
	
	\subsection{Improved Recursive List Decoding with Successive Permutation}
		
	We now propose an improved RLD algorithm that utilizes the SP scheme introduced in Section~\ref{sec:proposed:SP}. The details of the proposed algorithm are provided in Algorithm~\ref{algo}. In the beginning of Algorithm~\ref{algo}, the proposed algorithm with list size $L$ initializes all the $L$ decoding paths with $L$ random codeword permutations sampled from the symmetry group of RM codes. Each decoding path with index $l$ $(0\leq l < L)$ associated with $\mathcal{RM}(r,m)$ is characterized by a data structure $\mathcal{Q}_{(r,m)}$ that stores the channel LLR vector $\bm{\alpha}$, the path metric $\text{PM}$, the estimated codeword $\hat{\bm{x}}$, and the initial codeword permutation $\pi_\text{init}$. The permutation $\pi_\text{tmp}$, assigned to $\pi_\text{init}$ for each decoding path, is a random codeword permutation sampled from the full symmetry group of RM codes. The recursive decoding algorithm utilizing the SP scheme, denoted as the $\texttt{SP-RLD}(\cdot)$ function, is then applied to the initialized data structures $\{\mathcal{Q}_{(r,m)}[0],\ldots,\mathcal{Q}_{(r,m)}[L-1]\}$, and returns the updated data structures of the $L$ best decoding paths. Next, the updated path metrics of all the estimated decoding paths given by the $\texttt{SP-RLD}(\cdot)$ function are used to identify the best decoding path with index $l^*$ that has the smallest path metric. The initial codeword permutation $\pi_\text{init}^*$ associated with the best decoding path $l^*$ is then obtained. Finally, an inverted permutation $\left(\pi_\text{init}^*\right)^{-1}$ is applied to the best candidate codeword $\mathcal{Q}_{(r,m)}[l^*].\bm{\hat{x}}$ to obtain the final estimated codeword $\hat{\bm{x}}$.
	
	\begin{algorithm}[t]
		\setstretch{1.0}
		\DontPrintSemicolon
		\caption{Improved RLD with SP of RM Codes}
		\label{algo}
		\SetKwInOut{Input}{Input}
		\SetKwInOut{Output}{Output}
		\SetKwInput{kwIn}{in}
		\SetKwFunction{SPRLD}{SP-RLD}
		
		\Input{$\bm{y}$}
		\Output{$\hat{\bm{x}}$}
		\tcc{Initialize $L$ random codeword permutations associated with $L$ decoding paths}
		\For{$l \leftarrow 0$ \KwTo $L-1$}{
			$\pi_\text{tmp}:\bm{y} \xrightarrow{\pi_\text{tmp}} \bm{y}_{\pi_l}$\\
			$\mathcal{Q}_{(r,m)}[l].{\pi_\text{init}}\leftarrow \pi_\text{tmp}$; $\mathcal{Q}_{(r,m)}[l].\bm{\alpha}\leftarrow \bm{y}_{\pi_l}$\\
			$\mathcal{Q}_{(r,m)}[l].\text{PM} \leftarrow 0$; $\mathcal{Q}_{(r,m)}[l].\bm{\hat{x}} \leftarrow \textbf{0}$\\
		}
		
		\vspace*{5pt}
		\tcc{Improved RLD with SP}
		$\mathcal{Q}_{(r,m)}[0],
		\ldots,\mathcal{Q}_{(r,m)}[L-1] \leftarrow \SPRLD\left(\mathcal{Q}_{(r,m)}[0],
		\ldots,\mathcal{Q}_{(r,m)}[L-1]\right)$\\
		
		\vspace*{5pt}
		\tcc{Selection of the best decoding path}
		$l^* \leftarrow \argmin_{0\leq l\leq L-1} \{\mathcal{Q}_{(r,m)}[l].\text{PM}\}$\\
		$\pi_\text{init}^* \leftarrow \mathcal{Q}_{(r,m)}[l^*].{\pi_\text{init}}$\\
		$\left(\pi_\text{init}^*\right)^{-1}: \mathcal{Q}_{(r,m)}[l^*].\bm{\hat{x}}  \xrightarrow{\left(\pi_\text{init}^*\right)^{-1}} \hat{\bm{x}}$\\
		\Return $\hat{\bm{x}}$		
	\end{algorithm}
	
	In Algorithm~\ref{algo:SPRLD}, we provide the details of the $\texttt{SP-RLD}(\cdot)$ function. If the constituent RM codes are of order $1$ or $m-1$, the ML decoders of the first-order RM codes ($\texttt{FHT-List}(\cdot)$ \cite{dumerFHT}) or that of the SPC codes ($\texttt{SPC-List}(\cdot)$ \cite{Ali_FSSCL}) is queried to obtain the estimated codewords of the best $L$ decoding paths and their path metrics, respectively. Note that as the $\texttt{FHT-List}(\cdot)$ \cite{dumerFHT} and $\texttt{SPC-List}(\cdot)$ functions perform ML decoding at the parent node level, no codeword permutation is required to obtain the optimal decoding outputs of the $L$ best decoding paths. On the other hand, if the constituent code $\mathcal{RM}(r,m)$ satisfies $1<r<m-1$, the $\texttt{SP-RLD}(\cdot)$ function is recursively queried to decode the left and right child nodes $\mathcal{RM}(r-1,m-1)$ and $\mathcal{RM}(r,m-1)$ of $\mathcal{RM}(r,m)$, respectively, whose decoding results are used to construct the decoding output of $\mathcal{RM}(r,m)$. Specifically, from line~6 to line~20 of Algorithm~\ref{algo:SPRLD}, the best permutations of $\mathcal{RM}(r,m)$ are obtained independently for each decoding path with index $l$, using the proposed SP scheme. By $\pi_\text{tmp}$, we indicate a random permutation sampled from the full symmetry group of $\mathcal{RM}(r,m)$, which is used to obtain the LLR values associated with the right child $\mathcal{RM}(r-1,m-1)$, i.e., $\bm{\alpha}^{(\lambda)}$, followed by the permutation metric computation specified in Section~\ref{sec:proposed:SP}. If a better permutation $\pi_\text{tmp}$ is found for the $l$-th decoding path, the selected permutation $\pi_\text{SP}$ is updated in the data structure $\mathcal{Q}_{(r,m)}[l]$, which is required to perform the inverted permutation after the right-child node $\mathcal{RM}(r,m-1)$ is decoded. In addition, the data structures of the left-child nodes $\mathcal{Q}_{(r-1,m-1)}$ are also initialized during the permutation selection of $\mathcal{RM}(r,m)$. Given the initialized data structures of the left-child node $\mathcal{Q}_{(r-1,m-1)}$, the $\texttt{SP-RLD}(\cdot)$ function is then queried to obtain the updated data structures $\mathcal{Q}_{(r-1,m-1)}$ corresponding to the $L$ best decoding paths of the left-child node.
	
	\begin{algorithm}
		\DontPrintSemicolon
		\setstretch{1.0}
		\caption{$\texttt{SP-RLD}(\cdot)$}
		\label{algo:SPRLD}
		\SetKwInOut{Input}{Input}
		\SetKwInOut{Output}{Output}
		\SetKwInput{kwIn}{in}
		\SetKwFunction{SPRLD}{SP-RLD}
		\SetKwFunction{FHTL}{FHT-List}
		\SetKwFunction{SPCL}{SPC-List}
		
		\Input{$\mathcal{Q}_{(r,m)}[0],
			\ldots,\mathcal{Q}_{(r,m)}[L-1]$}
		\Output{$\mathcal{Q}_{(r,m)}[0],
			\ldots,\mathcal{Q}_{(r,m)}[L-1]$}
		\If{$r=1$}{
			$\mathcal{Q}_{(r,m)}[0],
			\ldots,\mathcal{Q}_{(r,m)}[L-1]\leftarrow \FHTL\left(\mathcal{Q}_{(r,m)}[0],
			\ldots,\mathcal{Q}_{(r,m)}[L-1]\right)$\\
		}
		\ElseIf{$r=m-1$}{
			$\mathcal{Q}_{(r,m)}[0],
			\ldots,\mathcal{Q}_{(r,m)}[L-1]\leftarrow \SPCL\left(\mathcal{Q}_{(r,m)}[0],
			\ldots,\mathcal{Q}_{(r,m)}[L-1]\right)$\\
		}
		\Else{
			\tcc{Decode the left-child node with SP}
			\For{$l \leftarrow 0$ \KwTo $L-1$}{
				%				\tcp{Initialize the datastructure for the left-child node $\mathcal{RM}(r-1,m-1)$}
				$\mathcal{Q}_{(r-1,m-1)}[l].{\pi_\text{init}}\leftarrow \mathcal{Q}_{(r,m)}[l].{\pi_\text{init}}$\\
				$\mathcal{Q}_{(r-1,m-1)}[l].\text{PM} \leftarrow \mathcal{Q}_{(r,m)}[l].\text{PM}$\\ 
				%$\mathcal{Q}_{(r-1,m-1)}[l].\bm{\hat{x}} \leftarrow \textbf{0}$;
				%$\mathcal{Q}_{(r-1,m-1)}[l].{\pi_\text{SP}}\leftarrow \emptyset$\\
				%\tcp{Select the codeword permutation of the parent node for each decoding path}
				$M^* \leftarrow -\infty$\\
				\For{$p \leftarrow 0$ \KwTo $m-1$}{
					$\pi_\text{tmp}:$ 			 	$\mathcal{Q}_{(r,m)}[l].\bm{\alpha}\xrightarrow{\pi_\text{tmp}} \bm{\alpha}_\text{tmp}$\\
					$\bm{\alpha}^{(\lambda)} \leftarrow f(\bm{\alpha}_\text{tmp})$\\
					\If{$r=2$}{
						Compute $M_{\pi_\text{tmp}} \left(\bm{\alpha}^{(\lambda)}\right)$ using FHT\\
					}
					\ElseIf{$r>2$}{
						Compute $M_{\pi_\text{tmp}} \left(\bm{\alpha}^{(\lambda)}\right)$ using (\ref{equ:SP:FPM:Rate-1})\\
					}
					\If{$M_{\pi_\text{tmp}} \left(\bm{\alpha}^{(\lambda)}\right)>M^*$}{
						$\mathcal{Q}_{(r-1,m-1)}[l].\bm{\alpha}\leftarrow \bm{\alpha}^{(\lambda)}$\\
						$\mathcal{Q}_{(r,m)}[l].\pi_\text{SP} \leftarrow \pi_\text{tmp}$\\ $M^* \leftarrow M_{\pi_\text{tmp}} \left(\bm{\alpha}^{(\lambda)}\right)$\\
					}
				}
			}
			$\mathcal{Q}_{(r-1,m-1)}[0],
			\ldots,\mathcal{Q}_{(r-1,m-1)}[L-1] \leftarrow \SPRLD\left(\mathcal{Q}_{(r-1,m-1)}[0],
			\ldots,\mathcal{Q}_{(r-1,m-1)}[L-1]\right)$\\
			\tcc{Decode the right-child node}
			\For{$l \leftarrow 0$ \KwTo $L-1$}{
				$\mathcal{Q}_{(r,m-1)}[l].{\pi_\text{init}}\leftarrow \mathcal{Q}_{(r-1,m-1)}[l].{\pi_\text{init}}$\\
				$\mathcal{Q}_{(r,m-1)}[l].\text{PM} \leftarrow \mathcal{Q}_{(r-1,m-1)}[l].\text{PM}$\\ 
				%$\mathcal{Q}_{(r,m-1)}[l].\bm{\hat{x}} \leftarrow \textbf{0}$;
				%$\mathcal{Q}_{(r,m-1)}[l].{\pi_\text{SP}}\leftarrow\emptyset$\\
				$\mathcal{Q}_{(r,m-1)}[l].{\bm{\alpha}}\leftarrow g\left(\mathcal{Q}_{(r-1,m-1)}[l].\hat{\bm{x}}, \mathcal{Q}_{(r,m)}[l^{(r,m)}_\text{org}].\bm{\alpha}\right)$\\
			}
			$\mathcal{Q}_{(r,m-1)}[0],
			\ldots,\mathcal{Q}_{(r,m-1)}[L-1] \leftarrow \SPRLD\left(\mathcal{Q}_{(r,m-1)}[0],
			\ldots,\mathcal{Q}_{(r,m-1)}[L-1]\right)$\\
			\tcc{Repermute the decoded codewords}
			\For{$l \leftarrow 0$ \KwTo $L-1$}{
				$\mathcal{Q}^{\text{tmp}}_{(r,m)}[l].{\pi_\text{init}}\leftarrow \mathcal{Q}_{(r,m-1)}[l].{\pi_\text{init}}$\\
				$\mathcal{Q}^{\text{tmp}}_{(r,m)}[l].\text{PM} \leftarrow \mathcal{Q}_{(r,m-1)}[l].\text{PM}$\\ 
				$\mathcal{Q}^{\text{tmp}}_{(r,m)}[l].\hat{\bm{x}} \leftarrow \texttt{Concat}\large(\mathcal{Q}_{(r-1,m-1)}[l^{(r-1,m-1)}_\text{org}].{\hat{\bm{x}}},$\\
				\hspace*{20pt}$\mathcal{Q}_{(r-1,m-1)}[l^{(r-1,m-1)}_\text{org}].{\hat{\bm{x}}}\oplus\mathcal{Q}_{(r,m-1)}[l].{\hat{\bm{x}}}\large)$\\
				$\pi_\text{SP} \leftarrow \mathcal{Q}_{(r,m)}[l^{(r,m)}_\text{org}].{\pi_\text{SP}}$\\
				$\left(\pi_\text{SP}\right)^{-1}: \mathcal{Q}_{(r,m)}^{\text{tmp}}[l].\bm{\hat{x}}  \xrightarrow{\left(\pi_\text{SP}\right)^{-1}} \mathcal{Q}_{(r,m)}^{\text{tmp}}[l].\bm{\hat{x}}$
			}
			\For{$l \leftarrow 0$ \KwTo $L-1$}{
				$\mathcal{Q}_{(r,m)}[l]\leftarrow\mathcal{Q}_{(r,m)}^{\text{tmp}}[l]$
			}
		}
		\Return $\mathcal{Q}_{(r,m)}[0],
		\ldots,\mathcal{Q}_{(r,m)}[L-1]$
	\end{algorithm}
	
	The decoding of the right-child nodes $\mathcal{RM}(r,m-1)$ is specified in line~22 to line~26 of Algorithm~\ref{algo:SPRLD}. The $g(\cdot)$ functions are used to obtain the LLR values of the right-child nodes, given the hard estimations of the left-child node, i.e., $\mathcal{Q}_{(r-1,m-1)}[l].\hat{\bm{x}}$, and the corresponding LLR values of the parent node where the left-child node is originated from, i.e., $\mathcal{Q}_{(r,m)}[l^{(r,m)}_\text{org}].\bm{\alpha}$. Here, by $l^{(r,m)}_\text{org}$ we indicate the path index of the parent node from which the surviving left-child node is derived. After the decoding of the right-child nodes is finished, the estimated codewords of the $L$ best paths associated with the parent node $\mathcal{RM}(r,m)$ are obtained based on the estimated hard values of the left-child and the right-child nodes (see lines 30 and 31), where $\texttt{Concat}(\bm{a},\bm{b})$ indicates the concatenation of the binary vectors $\bm{a}$ and $\bm{b}$. In addition, $\mathcal{Q}_{(r-1,m-1)}[l^{(r-1,m-1)}_\text{org}].{\hat{\bm{x}}}$ indicates the hard values of the left-child node that corresponds to the $l$-th active decoding path $\mathcal{Q}_{(r,m-1)}[l].{\hat{\bm{x}}}$ of the right-child node. Next, $\mathcal{Q}_{(r,m)}[l^{(r,m)}_\text{org}].{\pi_\text{SP}}$, the codeword permutation previously selected for the parent node from which the $l$-th active decoding path of the right-child node is originated from, is used to re-permute the estimated codeword of the parent node. Finally, the data structures $\{\mathcal{Q}_{(r,m)}[0],\ldots,\mathcal{Q}_{(r,m)}[L-1]\}$ of the $L$ best decoding paths for the parent node $\mathcal{RM}(r,m)$ are returned as the outputs of the recursive $\texttt{SP-RLD}(\cdot)$ function. Note that we keep track of the initial permutation $\pi_\text{init}$ applied to the parent code for each active decoding path during the course of decoding.

	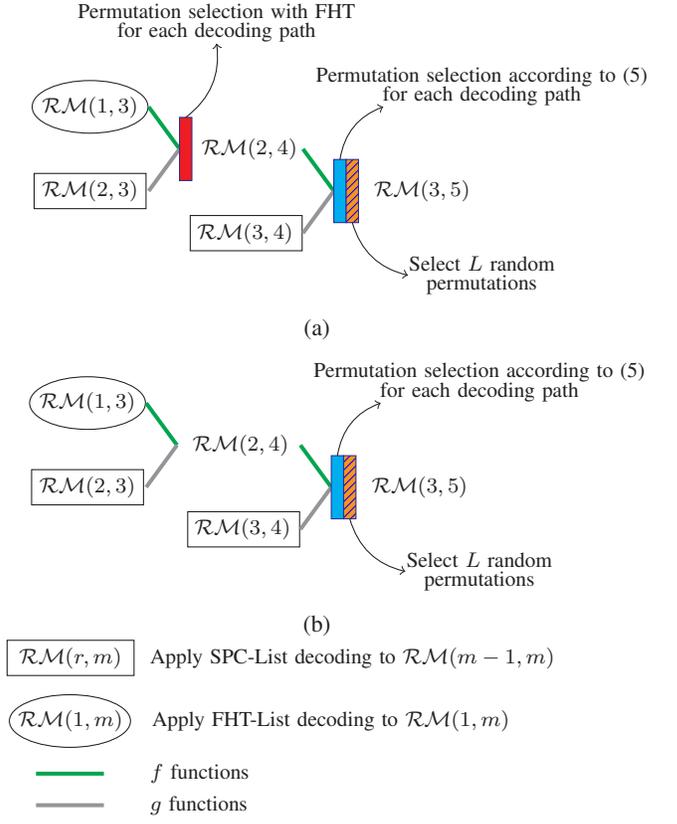
\begin{figure}
		\centering
		\begin{subfigure}{\linewidth}
			\centering
			\input{example_FSC.tikz.tex}\\
			\caption{}
		\end{subfigure}
		\begin{subfigure}{\linewidth}
			\centering
			\vspace*{-1\baselineskip}
			\input{example_FSC_simple.tikz.tex}\\
			\caption{}
		\end{subfigure}
		\begin{subfigure}{\linewidth}
			\centering
			\input{example_FSC_legend.tikz.tex}\\
		\end{subfigure}		
		\caption{Examples of the proposed decoder when applied to $\mathcal{RM}(3,5)$ with (a) $S=2$ and (b) $S=1$.}
		\label{fig:example}
	\end{figure}
	
	As the constituent RM codes are decoded successively under the proposed decoder, we propose a complexity and decoding latency reduction scheme that only applies the SP operations for the first $S$ $(S>0)$ left-child nodes. We refer to this simplified decoding algorithm as SSP-RLD. Fig.~\ref{fig:example} illustrates an example of SSP-RLD decoding when applied to $\mathcal{RM}(3,5)$ using $S \in \{2,1\}$. In Fig.~\ref{fig:example}(a), since the left-child node of $\mathcal{RM}(3,5)$ is a RM code of order $2$, the codeword permutation of $\mathcal{RM}(3,5)$ is selected in accordance with $(\ref{equ:SP:FPM:Rate-1})$. Then, the LLR values associated with $\mathcal{RM}(2,4)$ are obtained with the $f$ functions. Since the left-child node of $\mathcal{RM}(2,4)$ is a first-order RM code, FHT decoding is used to select the best permutation for $\mathcal{RM}(2,4)$, followed by the FHT-List decoder applied on the list of $L$ decoding paths with the selected codeword permutations for $\mathcal{RM}(2,4)$. Finally, as all the right-child RM codes are SPC codes, SPC-List decoding is used to decode them. The similar decoding operations are carried out for $\mathcal{RM}(2,5)$ in Fig.~\ref{fig:example}(b) except that the SP operations are only applied to the parent node of the first left-child node, while the original permutations are used for the parent node $\mathcal{RM}(2,4)$ of the second left-child node.
	
	The computational complexity and the decoding latency of SP-RLD and SSP-RLD decoders significantly increase as the list size $L$ increases. This is mainly caused by path metric and LLR sorting operations in the constituent $\texttt{SPC-List}(\cdot)$ and $\texttt{FHT-List}(\cdot)$ functions. Since random subsets of the full symmetry group of RM codes are utilized for the SSP-RLD decoder, we propose to further reduce the computational complexity and the decoding latency of the SSP-RLD decoder by using a variation of the ensemble decoding technique in \cite{Geiselhart}. In particular, we run $T$ $(T\ge1)$ independent SSP-RLD decoders with a small list size $L$ in parallel and select the output codeword that has the smallest path metric among the $T$ resulting codewords from the $T$ constituent SSP-RLD decoders. This variation of the proposed decoder is referred to as the ensemble SSP-RLD (Ens-SSP-RLD) decoding. Note that Ens-SSP-RLD decoding enables flexible design choices where the error-correction performance and complexity trade-offs can be explored with different choices of $S$, $L$, and $T$.
	
	\subsection{Performance Evaluation}
	\label{sec:results}
	
	\subsubsection{Quantitative Complexity Analysis}
	
	In this paper, we consider sequential and parallel implementations of the permutation selection scheme in Section~\ref{sec:proposed:SP}. Under the sequential implementation of the proposed SP scheme, a similar memory consumption as SC-based decoders is required to store the internal LLR values \cite{Ali_SP}. On the other hand, under the parallel implementation, a memory of $mLNQ$ bits is required to store the internal LLR values, where $m$ is the maximum number of the candidate permutations for the SP scheme, and $Q$ is the number of quantization bits. In addition, the SP scheme in both the sequential and parallel implementations requires $mQ$ memory bits to store the permutation metric $M_\pi \left(\bm{\alpha}^{(\lambda)}\right)$. Throughout this paper, we use $Q=32$ for all the considered decoders. Note that the FHT operations compute and store the new LLR values directly to $\bm{\alpha}^{(\lambda)}$, thus no extra memory is needed under FHT decoding. Table~\ref{tab:mem:general} summarizes the memory requirements of the SP-RLD-$L$ $(L\ge1)$ and SSP-RLD-$S$-$L$ $(S>0)$ decoders, while Table~\ref{tab:mem:general:ens} provides the memory consumption of the Ens-SSP-RLD-$S$-$L$-$T$ $(T \ge 1)$ decoder. Note that with $T=1$, Ens-SSP-RLD-$S$-$L$-$T$ reverts to SSP-RLD-$S$-$L$. In addition, with $S$ being the number of left-child nodes visited following the course of decoding and with $T=1$, Ens-SSP-RLD-$S$-$L$-$T$ reverts to SP-RLD-$L$.
	
	{\begin{table}[t!]
		\caption{Memory requirement in terms of the number of bits required by the SP-RLD and SSP-RLD decoders.}
		\footnotesize
		\setlength{\tabcolsep}{3pt}
		\def\arraystretch{1.25}
		\centering
		\begin{tabular}{c c | c}	
			\toprule
			\multicolumn{2}{c|}{Decoding Algorithm} & Memory Requirement\\
			\midrule
			\makecell{SP-RLD-$1$\\SSP-RLD-$S$-$1$} &($L=1$, sequential SP)& $2NQ + mQ + N$\\
			\midrule
			\makecell{SP-RLD-$1$\\SSP-RLD-$S$-$1$} &($L=1$, parallel SP)& $(m+1)NQ + mQ + N$\\
			\midrule
			\makecell{SP-RLD-$L$\\SSP-RLD-$S$-$L$} &($L>1$, sequential SP)& $N(L+1)Q+ mQ + 2NL$\\
			\midrule
			\makecell{SP-RLD-$L$\\SSP-RLD-$S$-$L$} &($L>1$, parallel SP)& $N(mL+1)Q+ mQ + 2NL$\\
			\midrule
		\end{tabular}
		\label{tab:mem:general}
	\end{table}
	\begin{table}[t!]
		\caption{Memory requirement in terms of the number of bits required by the Ens-SSP-RLD decoder.}
		\footnotesize
		\setlength{\tabcolsep}{3pt}
		\def\arraystretch{1.25}
		\centering
		\begin{tabular}{ c | c}	
			\toprule
			{Decoding Algorithm} & Memory Requirement\\
			\midrule
			\makecell{Ens-SSP-RLD-$S$-$1$-$T$\\($L=1$, sequential SP)}& $(NQ+mQ+N)T + NQ$\\
			\midrule
			\makecell{Ens-SSP-RLD-$S$-$1$-$T$\\($L=1$, parallel SP)}& $(mNQ + mQ + N)T + NQ$\\
			\midrule
			\makecell{Ens-SSP-RLD-$S$-$L$-$T$\\($L>1$, sequential SP)}& $(NLQ+ mQ + 2NL)T+ NQ$\\
			\midrule
			\makecell{Ens-SSP-RLD-$S$-$L$-$T$\\($L>1$, parallel SP)}&$(mNLQ+ mQ + 2NL)T+ NQ$\\
			\midrule
		\end{tabular}
		\label{tab:mem:general:ens}
	\end{table}
	}
	The decoding latency of the proposed decoders is computed by counting the number of time steps required by all the floating point operations. We assume that there is no resource constraint. Thus, all concurrent operations in $f(\cdot)$ and $g(\cdot)$ functions, the path metric computations, and the computations of $M_\pi \left(\bm{\alpha}^{(\lambda)}\right)$ in Section~\ref{sec:proposed:SP} require one time step \cite{Ali_FSSCL, Ardakani_TCOM}. The permutation metric $M_\pi \left(\bm{\alpha}^{(\lambda)}\right)$ obtained from FHT requires $s$ time steps for a first-order RM code located at the $s$-th stage \cite[Chapter~14]{macwilliams1977theory}. Furthermore, for the sequential implementation, the proposed SP scheme requires $s$ time steps if $r_\lambda \ge 2$ and $s^2$ time steps if $r_\lambda = 1$, since each permutation is evaluated sequentially. In the parallel implementation the proposed SP scheme, a single time step is required if $r_\lambda \ge 2$ and $s$ time steps are required if $r_\lambda=1$. In addition, the hard decisions obtained from the LLR values and binary operations are computed instantaneously \cite{Ali_FSSCL,Ardakani_TCOM,Alexios_LLR_SCLD}. Finally, we assume that the number of time steps required by a merge sort algorithm to sort an array of $N$ elements is $\log_2(N)$ \cite[Chapter 2]{cormen2009introduction}. We also use similar assumptions to compute the decoding latency of all the other decoders considered in this paper.
	
	To calculate the computational complexity of the decoders considered in this paper, we count the number of floating point operations, namely, the number of additions, subtractions, and comparisons, required during the course of decoding. Note that the merge sort algorithm requires $N\log_2 N$ comparisons to sort an array of length $N$ \cite[Chapter~2]{cormen2009introduction}.
	
	\subsubsection{Comparison with FSCL, SC-Stack and SP-SCL Decoding Algorithms}
	
	\begin{figure*}
		\centering
		\input{ferRM512_46.tikz}
		\input{ferRM512_130.tikz}
		\input{ferRM512_256.tikz}\\
		\ref{perf-legend-SP-RLD-3-9}
		\caption{Error-correction performance of the proposed decoders and that of the SCS, SP-SCL, and FSCL decoders.}
		\label{fig:fer:SPSCL}
	\end{figure*}
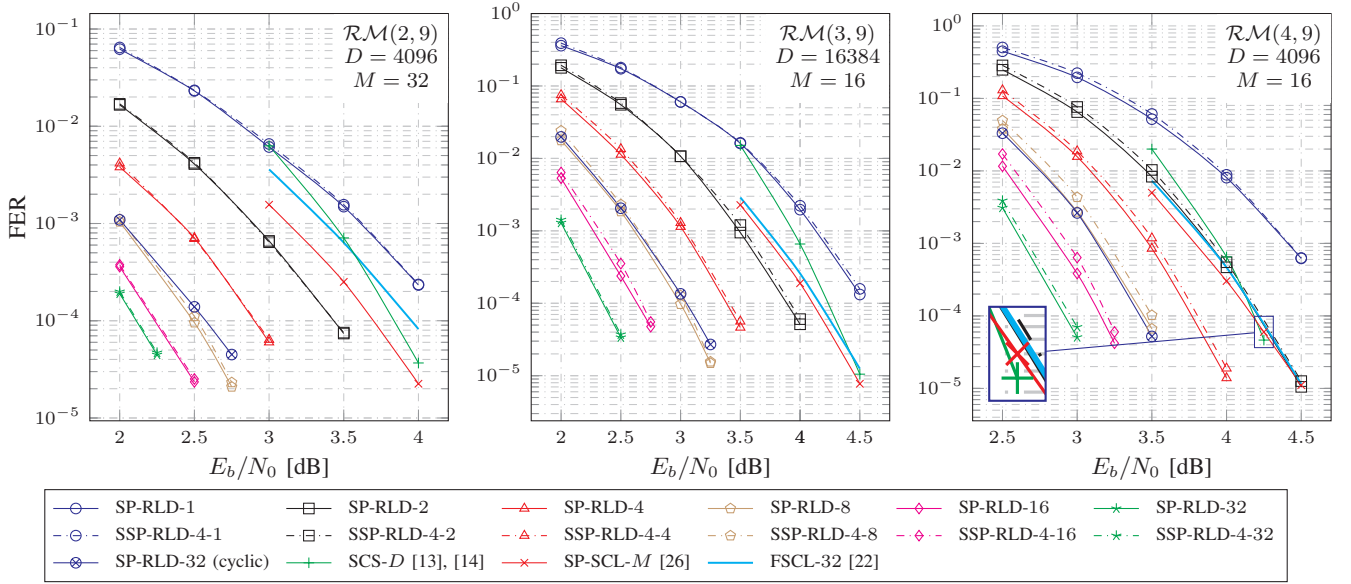
		
	Fig.~\ref{fig:fer:SPSCL} provides the error-correction performance in terms of FER of the proposed SP-RLD-$L$ and SSP-RLD-$S$-$L$ decoders, and that of the FSCL, SC-Stack (SCS), and SP-SCL decoders for $\mathcal{RM}(r, 9)$, $r\in\{2,3,4\}$. The SCS decoder considered in this paper utilizes the enhanced score function introduced in \cite{TrifonovScore} to reduce the stack size when compared with the conventional SCS decoder introduced in \cite{niu2012stack}. In Fig.~\ref{fig:fer:SPSCL}, the SCS decoder with stack size $D$ is denoted as SCS-$D$, while the SP-SCL decoder with list size $M$ is denoted as SP-SCL-$M$. The values of $D$ and $M$ are selected to allow an FER performance comparable to that of the FSCL decoder with list size $32$ (FSCL-$32$). With $S=4$, the SSP-RLD decoder has a negligible error-correction performance degradation when compared to the SP-RLD decoder with the same list size. Furthermore, we also provide the FER performance of the proposed SP-RLD-$32$ decoder where only cyclic factor-graph permutations are considered.
		
	It can be observed from Fig.~\ref{fig:fer:SPSCL} that the FER of the SP-RLD-$32$ decoder with cyclic factor-graph permutations is relatively similar to that of the SP-RLD-$8$ decoder with the codeword permutations sampled from the full symmetry group. In addition, at no additional cost, the SP-RLD-$32$ decoder that utilizes the general codeword permutations obtains a maximum gain of $0.7$ dB at the target FER of $10^{-4}$, when compared to the SP-RLD-$32$ decoder that only uses cyclic factor-graph permutations.	

	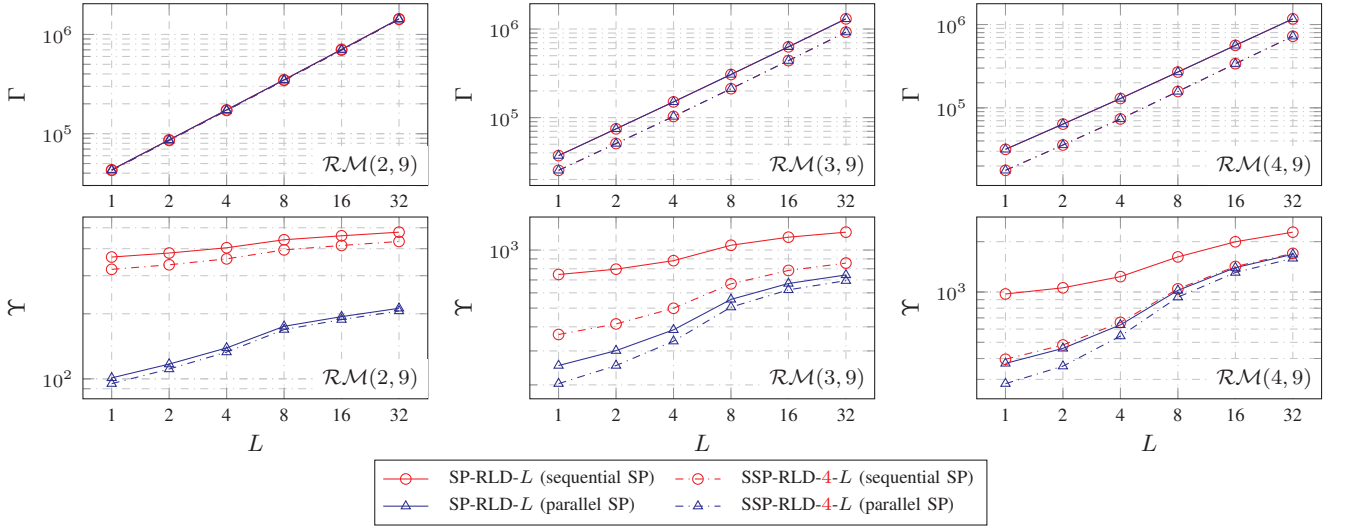
\begin{figure*}
		\centering
		\input{comp_RM512_46.tikz}
		\input{comp_RM512_130.tikz}
		\input{comp_RM512_256.tikz}\\
		\input{lat_RM512_46.tikz}
		\input{lat_RM512_130.tikz}
		\input{lat_RM512_256.tikz}\\		
		\hspace{15pt}\ref{perf-legend-comp-3-9}
		\caption{Computational complexity and decoding latency of the proposed decoders under the sequential and parallel implementations of the SP scheme.}
		\label{fig:comp}
%		\vspace*{-1\baselineskip}
	\end{figure*}

	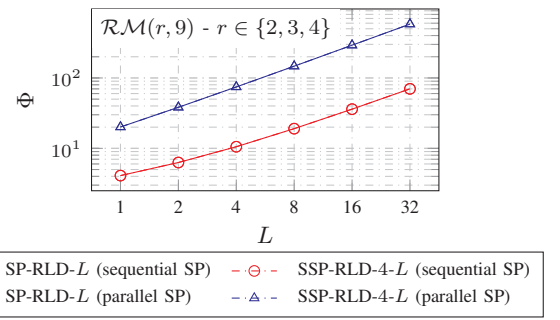
\begin{figure}
		\centering
		\input{mem_RM512_46.tikz}\\
		{\hspace*{10pt}\ref{perf-legend-mem}}\\
		\caption{Memory consumption in kB $(\Phi)$ of the proposed decoders whose FER curves are provided in Fig.~\ref{fig:fer:SPSCL}.}
		\label{fig:mem}
		\vspace*{-1.25\baselineskip}
	\end{figure}

	\begin{table*}
		\centering
		\setlength{\tabcolsep}{2.1pt}
		\renewcommand{\arraystretch}{1.25}
		\caption{Comparison of computational complexity $(\Gamma)$, decoding latency in time steps $(\Upsilon)$, and memory requirement in kB $(\Phi)$ of SCS, SP-SCL, FSCL, and proposed SSP-RLD decoders considered in Fig.~\ref{fig:fer:SPSCL}.}
		\footnotesize
		\begin{tabular}{c|cccc|cccc|ccc|ccccc}
			\toprule
			&\multicolumn{4}{c|}{SCS-$D$ \cite{TrifonovScore,niu2012stack}}&\multicolumn{4}{c|}{SP-SCL-$M$ \cite{Ali_SP}}&\multicolumn{3}{c|}{FSCL-$32$ \cite{Ali_FSSCL}}&\multicolumn{5}{c}{SSP-RLD-$4$-$2$}\\
			
			&\multicolumn{1}{c}{${D}$}&\multicolumn{1}{c}{$\Gamma$}&\multicolumn{1}{c}{$\Upsilon$}&\multicolumn{1}{c|}{$\Phi$}&\multicolumn{1}{c}{$M$}&\multicolumn{1}{c}{$\Gamma$}&\multicolumn{1}{c}{$\Upsilon$}&\multicolumn{1}{c|}{$\Phi$}&\multicolumn{1}{c}{$\Gamma$}&\multicolumn{1}{c}{$\Upsilon$}&\multicolumn{1}{c|}{$\Phi$}&\multicolumn{1}{c}{$\Gamma$}&\multicolumn{1}{c}{$\Upsilon_{\text{s}}$}&\multicolumn{1}{c}{$\Phi_{\text{s}}$}&\multicolumn{1}{c}{$\Upsilon_{\text{p}}$}&\multicolumn{1}{c}{$\Phi_{\text{p}}$}\\
			
			\midrule
			$\mathcal{RM}(2,9)$&4096&4.21$\times10^7$&3.0$\times10^5$&8208&32&8.52$\times10^5$&1.8$\times10^3$&70&9.62$\times10^4$&373&70&8.33$\times10^4$&337&6.3&111&38.3\\
			$\mathcal{RM}(3,9)$&16384&4.83$\times10^6$&1.0$\times10^4$&32832&16&4.37$\times10^5$&2.3$\times10^3$&36&1.64$\times10^5$&1039&70&4.86$\times10^4$&414&6.3&252&38.3\\
			$\mathcal{RM}(4,9)$&8192&1.13$\times10^7$&4.2$\times10^4$&16416&16&4.58$\times10^5$&3.1$\times10^3$&36&2.25$\times10^5$&1991&70&3.41$\times10^4$&482&6.3&369&38.3\\
		\end{tabular}
		\label{tab:SCS_comp}
	\end{table*}
		
	Fig.~\ref{fig:comp} illustrates the computational complexity $(\Gamma)$ and the decoding latency $(\Upsilon)$ of SP-RLD-$L$ and SSP-RLD-$4$-$L$ under the sequential and parallel implementations of the proposed SP scheme. In addition, the memory requirement $(\Phi)$ in kilobytes (kB) of SP-RLD-$L$ and SSP-RLD-$4$-$L$ is provided in Fig.~\ref{fig:mem}. It can be observed from Fig.~\ref{fig:comp} that the SSP-RLD-$4$-$L$ decoder relatively maintains the computational complexity when compared with SP-RLD-$L$. However, SSP-RLD-$4$-$L$ significantly reduces the decoding latency of SP-RLD-$L$ while only incurring negligible error-correction performance degradation as seen from Fig.~\ref{fig:fer:SPSCL}. Furthermore, Fig.~\ref{fig:comp} and Fig.~\ref{fig:mem} reveal the trade-offs between the decoding latency and memory requirement of the proposed decoders under the sequential and parallel implementations of the SP scheme. In particular, the improvements in the decoding latency of the parallel implementation over the sequential implementation come at the cost of memory consumption overheads.	
		
	Table~\ref{tab:SCS_comp} summarizes the computational complexity $(\Gamma)$, the decoding latency in time steps $(\Upsilon)$, and the memory requirement in kB $(\Phi)$ of the FSCL, SCS, and SP-SCL decoders, and those of the SSP-RLD decoder with $L=2$ and $S=4$, whose FER values are plotted in Fig.~\ref{fig:fer:SPSCL}. For the SSP-RLD-$4$-$2$ decoder, $\Upsilon_{\text{s}}$ and $\Phi_{\text{s}}$ indicate the decoding latency and the memory requirement of the sequential SP implementation, while $\Upsilon_{\text{p}}$ and $\Phi_{\text{p}}$ indicate the decoding latency and the memory requirement of the parallel implementation of the SP scheme, respectively. It can be seen in Fig.~\ref{fig:fer:SPSCL} that the FER performance of SSP-RLD-$4$-$2$ is similar to or better than that of FSCL, SCS, and SP-SCL decoders at the target FER of $10^{-4}$ for all the considered RM codes. In addition, under both sequential and parallel implementations of the SP scheme, SSP-RLD-$4$-$2$ significantly outperforms the FSCL, SCS, and SP-SCL decoders in various complexity metrics as shown in Table~\ref{tab:SCS_comp}.
	
	\subsubsection{Comparison with State-of-the-Art RM Decoders}
			
	Fig.~\ref{fig:fer_rpa} compares the FER performance of SSP-RLD-$S$-$L$, Ens-SSP-RLD-$S$-$L'$-$T$, and that of the state-of-the-art decoders for various RM codes. Note that the list size $L'$ and the number of decoding attempts $T$ used by the Ens-SSP-RLD decoder satisfy the constraint $L=L'T$, where $L$ is the list size used by the SSP-RLD decoder. $L'$ is selected as the smallest list size that allows the Ens-SSP-RLD decoder to have an error-correction performance that is within $0.1$ dB of that of the SSP-RLD decoder at the target FER of $10^{-3}$. We consider the RLDP \cite{Dumer06} and the RLDA \cite{Dumer06, Geiselhart} algorithms with list size $M$, the SSC-FHT decoder \cite{dumerFHT, gabi_fast_pcd} when applied to $P$ factor-graph permutations (Per-SSC-FHT-$P$), and $P$ general permutations (Aut-SSC-FHT-$P$) sampled from the full symmetry group of RM codes\footnote{The C++ implementations of RLDA, Aut-SSC-FHT, and all variations of the proposed decoders are available at \href{https://github.com/nghiadt05/SPRLD}{https://github.com/nghiadt05/SPRLD}.}. The empirical lower bounds of the error-correction performance of ML decoding \cite{Dumer06} are also provided for all the RM configurations in Fig.~\ref{fig:fer_rpa}. In addition, the FER performance curves of the sparse-RPA (SRPA) decoder introduced in \cite{fathollahi2020sparse} are shown for $\mathcal{RM}(2,8)$, $\mathcal{RM}(3,8)$, $\mathcal{RM}(4,8)$, and $\mathcal{RM}(2,9)$.
	
	{
	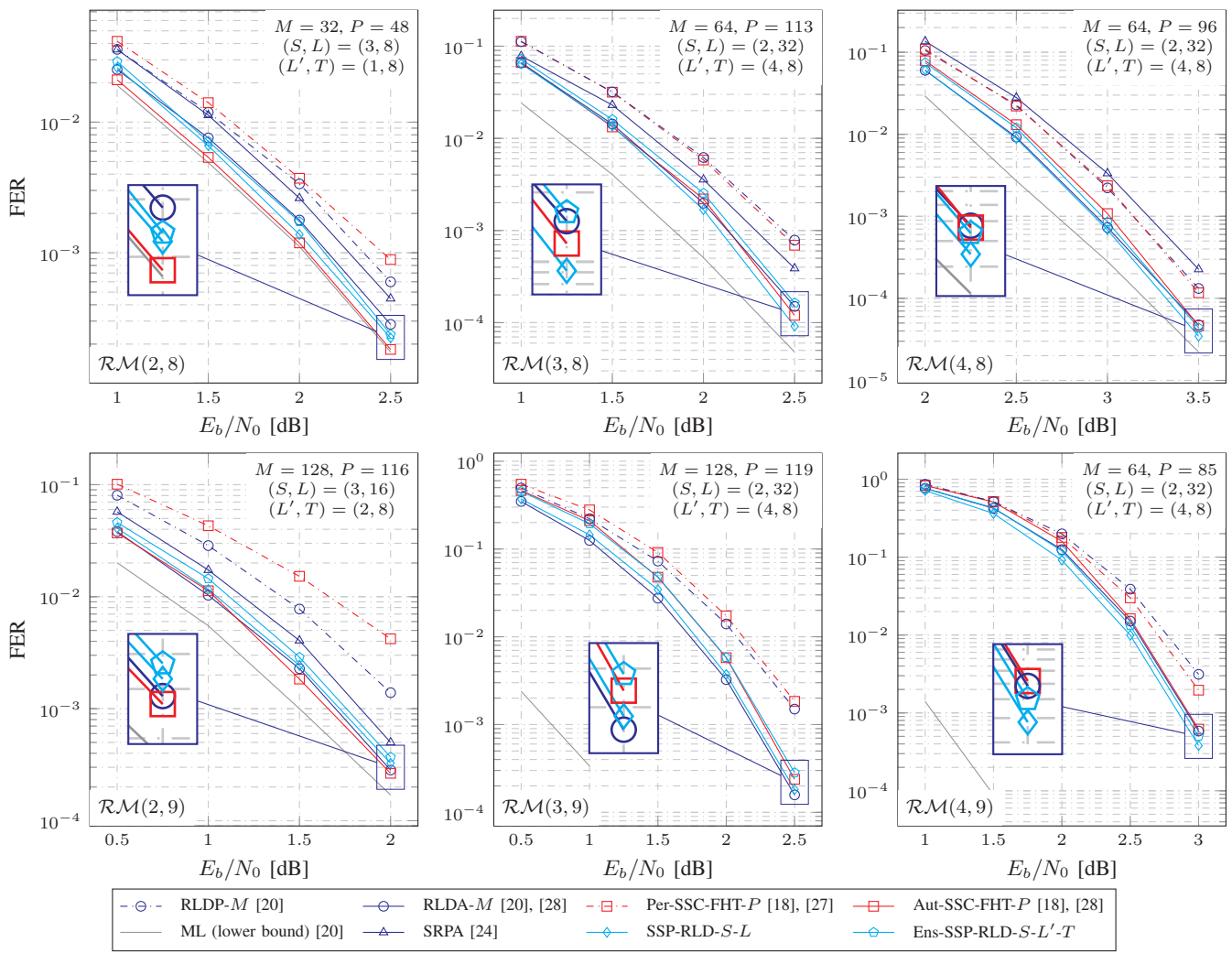
\begin{figure*}
		\centering
		{\input{ferRM256_37_FSC_per_rpa.tikz.tex}
			\input{ferRM256_93_FSC_per_rpa.tikz.tex}
			\input{ferRM256_163_FSC_per_rpa.tikz.tex}}\\
		{\input{ferRM512_46_FSC_per_rpa.tikz.tex}
			\input{ferRM512_130_FSC_per_rpa.tikz.tex}
			\input{ferRM512_256_FSC_per_rpa.tikz.tex}}\\
		\ref{perf-legend-RPA-fer-2-8}
		\caption{Error-correction performance of various permutation decoding algorithms of RM codes. The FER of the SRPA decoder and the lower bound of ML decoding are also plotted for comparison.}
		\label{fig:fer_rpa}
	\end{figure*}	
	
	}
		
	Table~\ref{tab:decoder_comp:a} summarizes the computational complexity, the decoding latency, and the memory requirement of SSP-RLD and Ens-SSP-RLD decoding, while Table~\ref{tab:decoder_comp:b} provides the complexity metrics of SRPA, Aut-SSC-FHT, and RLDA decoding. Note that the decoders provided in Table~\ref{tab:decoder_comp:a} and Table~\ref{tab:decoder_comp:b} have similar error-correction performance at the target FER of $10^{-3}$ as shown in Fig.~\ref{fig:fer_rpa}. In this paper, a fully-parallel implementation of the SRPA decoder, in which all the operations that can be carried out concurrently are executed at the same time, is considered. The SRPA decoding algorithm runs two fully-parallel RPA decoders with each decoder using a quarter of the code projections at each recursion step \cite{fathollahi2020sparse}. Thus, the SRPA decoder effectively reduces $50\%$ of the total number of projections used by the conventional RPA algorithm \cite{Ye20}. This configuration incurs negligible error-correction performance loss with respect to the conventional RPA decoder in \cite{Ye20} for the second and third order RM codes of size $256$.

	In Table~\ref{tab:decoder_comp:b}, we consider fully-parallel and semi-parallel implementations of the Aut-SSC-FHT decoder. Under the semi-parallel implementation, the number of parallel SSC-FHT decoders is set to the list size $L$ used by SSP-RLD decoding for the same RM code. This configuration enables the Aut-SSC-FHT-$P$ decoder to have a relatively similar memory consumption in comparison with SSP-RLD-$S$-$L$ and Ens-SSP-RLD-$S$-$L'$-$T$ decoding when the sequential SP scheme is used. On the other hand, in the fully-parallel implementation of Aut-SSC-FHT decoding, $P$ concurrent SSC-FHT decoders are used. For Aut-SSC-FHT decoding, $\Upsilon_\text{sp}$ and $\Phi_\text{sp}$ indicate the latency and memory requirement of the semi-parallel implementation, while $\Upsilon_{\text{p}}$ and $\Phi_{\text{p}}$ indicate the decoding latency and memory requirement of the fully-parallel implementation, respectively.

	With the fully-parallel implementation, Aut-SSC-FHT provides the best decoding latency in comparison with RLDA, SP-RLD, and Ens-SSP-RLD decoders. However, the fully-parallel implementation of the Aut-SSC-FHT decoder and the parallel SP scheme of the proposed decoders require large memory consumption, rendering them less attractive in practical applications. In addition, RLDA decoding suffers from high computational complexity and high decoding latency that are mainly caused by sorting operations, especially with large values of $M$ and $r$. In particular, for $\mathcal{RM}(4,9)$ and with relatively similar FER performance, the Ens-SSP-RLD-$2$-$4$-$8$ decoder with the sequential SP scheme reduces $19\%$ of the computational complexity, $75\%$ of the number of time steps, and $49\%$ of the memory consumption of RLDA-$64$. It can be seen from Table~\ref{tab:decoder_comp:a} that Ens-SSP-RLD significantly reduces the computational complexity and decoding latency of SSP-RLD, especially for $r\in\{3,4\}$, while relatively preserving the error-correction performance and memory requirement of SSP-RLD.
		
	As observed in Fig.~\ref{fig:fer_rpa}, Table~\ref{tab:decoder_comp:a}, and Table~\ref{tab:decoder_comp:b}, under similar computational complexity and memory consumption for $\mathcal{RM}(2,8)$, $\mathcal{RM}(3,8)$, $\mathcal{RM}(2,9)$, and $\mathcal{RM}(3,9)$, the sequential Ens-SSP-RLD decoder obtains significant latency reductions, ranging from $22\%$ to $48\%$, while incurring negligible error-correction performance degradation, ranging from $0.02$~dB to $0.1$~dB, compared to the semi-parallel Aut-SSC-FHT decoder at the target FER of $10^{-3}$. Furthermore, for $\mathcal{RM}(4,8)$ at the same target FER, a small gain of $0.05$~dB in error-correction performance is obtained for the proposed decoder, while it provides a $22\%$ reduction in the decoding latency and a $6\%$ reduction in the computational complexity in comparison with the semi-parallel Aut-SSC-FHT decoder. However, for $\mathcal{RM}(4,9)$, the semi-parallel Aut-SSC-FHT decoder is slightly better than the proposed sequential Ens-SSP-RLD decoder in all the complexity metrics.
	
	Note that with the same list size $M$ or the same number of permutations $P$, using the permutations randomly sampled from the full symmetry group of the codes provides significant error-correction performance improvement for the RLDA-$M$ and Aut-SSC-FHT-$P$ decoders at no additional cost, when compared to the RLDP-$M$ and Per-SSC-FHT-$P$ decoders, respectively. In addition, all permutation decoding algorithms, RLDA, Aut-SSC-FHT, SSP-RLD, and Ens-SSP-RLD, provide significantly better error-correction performance with significantly lower computational complexity and decoding latency compared to the SRPA decoder for various RM code configurations.
	
	\section{Conclusion}
	\label{sec:conclud}
	
	In this paper, a novel successive permutation (SP) scheme is proposed to significantly improve the error-correction performance of Reed-Muller (RM) codes under an improved recursive list decoding (RLD) algorithm. We performed low-complexity decoding operations on the rich symmetry group of RM codes to select a good codeword permutation of the code on the fly. Efficient decoding latency and complexity reduction schemes were introduced that relatively maintain the error-correction performance. We performed a numerical analysis of the proposed decoders in terms of error-correction performance, computational complexity, decoding latency, and memory requirement and compared them with those of the state-of-the-art RM decoders. The simulation results confirmed the effectiveness of the proposed decoder under various configurations of RM codes. Specifically, for the RM codes of lengths $256$ and $512$ and with code orders $2$ and $3$, with relatively similar computational complexity and memory requirement, the proposed decoder significantly reduces the decoding latency of the state-of-the-art permuted successive-cancellation decoder with fast Hadamard transform (Aut-SSC-FHT) under a semi-parallel implementation, while incurring negligible error-correction performance degradation at a target frame error rate of $10^{-3}$.
	
	\begin{table*}
		\centering
		\vspace*{-1\baselineskip}
		\setlength{\tabcolsep}{3pt}
		\renewcommand{\arraystretch}{1.25}
		\caption{Computational complexity $(\Gamma)$, decoding latency in time steps $(\Upsilon)$, and memory requirement in kB $(\Phi)$ of the SSP-RLD and Ens-SSP-RLD decoders considered in Fig.~\ref{fig:fer_rpa}.}
		\footnotesize
		\begin{tabular}{c|ccccccc|cccccccc}
			\toprule
			&\multicolumn{7}{c|}{SSP-RLD-$S$-$L$}&\multicolumn{8}{c}{Ens-SSP-RLD-$S$-$L'$-$T$}\\
			&\multicolumn{1}{c}{$S$}&\multicolumn{1}{c}{$L$}&\multicolumn{1}{c}{$\Gamma$}&\multicolumn{1}{c}{$\Upsilon_\text{s}$}&\multicolumn{1}{c}{$\Phi_\text{s}$}&\multicolumn{1}{c}{$\Upsilon_\text{p}$}&\multicolumn{1}{c|}{$\Phi_\text{p}$}&\multicolumn{1}{c}{$S$}&\multicolumn{1}{c}{$L'$}&\multicolumn{1}{c}{$T$}&\multicolumn{1}{c}{$\Gamma$}&\multicolumn{1}{c}{$\Upsilon_\text{s}$}&\multicolumn{1}{c}{$\Phi_\text{s}$}&\multicolumn{1}{c}{$\Upsilon_\text{p}$}&\multicolumn{1}{c}{$\Phi_\text{p}$}\\
			\midrule
			$\mathcal{RM}(2,8)$&3&8&1.32$\times10^5$&287&9.5&141&65.5&3&1&8&1.21$\times10^5$&222&9.5&76&65.5\\
			$\mathcal{RM}(3,8)$&2&32&3.39$\times10^5$&588&35.0&526&259.0&2&4&8&2.72$\times10^5$&298&35.3&236&259.3\\
			$\mathcal{RM}(4,8)$&2&32&2.63$\times10^5$&1092&35.0&1066&259.0&2&4&8&1.86$\times10^5$&341&35.3&315&259.3\\
			\midrule
			$\mathcal{RM}(2,9)$&3&16&6.60$\times10^5$&376&36.0&185&292.0&3&2&8&6.48$\times10^5$&302&36.3&111&292.3\\
			$\mathcal{RM}(3,9)$&2&32&7.55$\times10^5$&767&70.0&688&582.0&2&4&8&6.48$\times10^5$&413&70.3&334&582.3\\
			$\mathcal{RM}(4,9)$&2&32&5.66$\times10^5$&1613&70.0&1583&582.0&2&4&8&4.23$\times10^5$&572&70.3&542&582.3\\
		\end{tabular}
		\label{tab:decoder_comp:a}
	\end{table*}
	
	\begin{table*}
		\centering
		\setlength{\tabcolsep}{3pt}
		\renewcommand{\arraystretch}{1.25}
		\caption{Computational complexity $(\Gamma)$, decoding latency in time steps $(\Upsilon)$, and memory requirement in kB $(\Phi)$ of the SRPA, RLDA, and Aut-SSC-FHT decoders considered in Fig.~\ref{fig:fer_rpa}.}
		\footnotesize
		\begin{tabular}{c|ccc|cccc|cccccc}
			\toprule
			&\multicolumn{3}{c|}{SRPA \cite{fathollahi2020sparse}}&\multicolumn{4}{c|}{RLDA-$M$ \cite{Geiselhart, Dumer06}}&\multicolumn{6}{c}{Aut-SSC-FHT-$P$ \cite{dumerFHT, Geiselhart}}\\
			%			\cmidrule{2-15}
			&\multicolumn{1}{c}{$\Gamma$}&\multicolumn{1}{c}{$\Upsilon$}&\multicolumn{1}{c|}{$\Phi$}&\multicolumn{1}{c}{$M$}&\multicolumn{1}{c}{$\Gamma$}&\multicolumn{1}{c}{$\Upsilon$}&\multicolumn{1}{c|}{$\Phi$}
			
			&\multicolumn{1}{c}{$P$}&\multicolumn{1}{c}{$\Gamma$}&\multicolumn{1}{c}{$\Upsilon_\text{sp}$}&\multicolumn{1}{c}{$\Phi_\text{sp}$}&\multicolumn{1}{c}{$\Upsilon_\text{p}$}&\multicolumn{1}{c}{$\Phi_\text{p}$}\\
			
			\midrule
			$\mathcal{RM}(2,8)$&6.55$\times10^5$&3592&69.2&32&6.20$\times10^4$&317&35.0&48&1.22$\times10^5$&390&9.4&70&50.7\\
			$\mathcal{RM}(3,8)$&7.92$\times10^7$&6184&281.5&64&2.04$\times10^5$&854&69.0&113&2.71$\times10^5$&382&34.4&132&118.0\\
			$\mathcal{RM}(4,8)$&3.63$\times10^9$&7816&465.2&64&2.83$\times10^5$&1433&69.0&96&1.98$\times10^5$&439&34.4&151&100.4\\
			\midrule
			$\mathcal{RM}(2,9)$&3.44$\times10^6$&10250&271.6&128&4.82$\times10^5$&490&274.0&116&6.50$\times10^5$&581&35.5&89&241.7\\
			$\mathcal{RM}(3,9)$&-&-&-&128&7.64$\times10^5$&1336&274.0&119&6.48$\times10^5$&577&68.5&197&247.9\\
			$\mathcal{RM}(4,9)$&-&-&-&64&5.20$\times10^5$&2283&138.0&85&4.23$\times10^5$&547&68.3&277&177.6\\
		\end{tabular}
		\label{tab:decoder_comp:b}
	\end{table*}
	
	\section*{Acknowledgment}
	The authors would like to thank the anonymous reviewers for their helpful and constructive comments that significantly improved the quality of the manuscript.

%	\bibliographystyle{IEEEtran}	
%	\bibliography{IEEEabrv,myrefs}


	\begin{IEEEbiography}[{\includegraphics[width=1in,height=1.25in,clip,keepaspectratio]{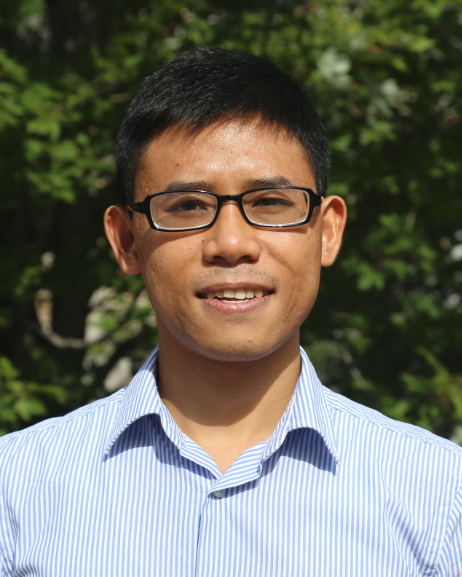}}]{Nghia~Doan} (Student Member, IEEE) received the B.Sc. degree from Posts and Telecommunications Institute of Technology, Hanoi, Vietnam, in 2014 and the M.Sc. degree from Seoul National University, Seoul, South Korea, in 2017, both in electrical and computer engineering. He is currently working toward the Ph.D. degree in electrical and computer engineering at McGill University, Montreal, QC, Canada. His research interests include channel coding, machine learning for communications, and hardware-aware algorithm optimization of digital signal processing applications.
	\end{IEEEbiography}

	\begin{IEEEbiography}[{\includegraphics[width=1in,height=1.25in,clip,keepaspectratio]{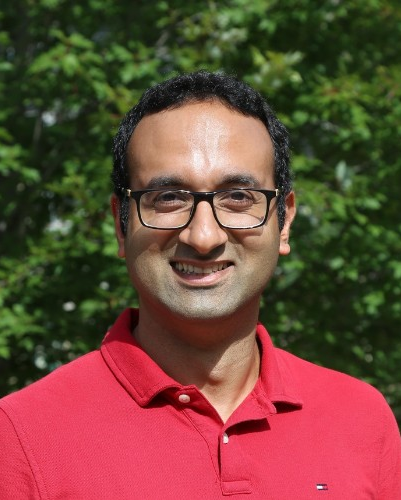}}]{Seyyed~Ali~Hashemi} (Member, IEEE) was born in Qaemshahr, Iran. He received the B.S. degree from Sharif University of Technology, Iran, the M.S. degree from the University of Alberta, Canada, and the Ph.D. degree from McGill University, Canada, all in Electrical Engineering. He is currently a Senior Engineer at Qualcomm Technologies, Inc., USA. Prior to that, he was a Postdoctoral Fellow with the Department of Electrical Engineering, Stanford University, USA, and a Visiting Lecturer with the Department of Electrical and Computer Engineering, Princeton University, USA. His research interests include machine learning for communications, error-correcting codes, and VLSI implementation of digital signal processing systems. He was a recipient of the Best Student Paper Award at ISCAS 2016 and the Postdoctoral Fellowship from Natural Sciences and Engineering Research Council of Canada (NSERC) in 2018.
	\end{IEEEbiography}

	\begin{IEEEbiography}[{\includegraphics[width=1in,height=1.25in,clip,keepaspectratio]{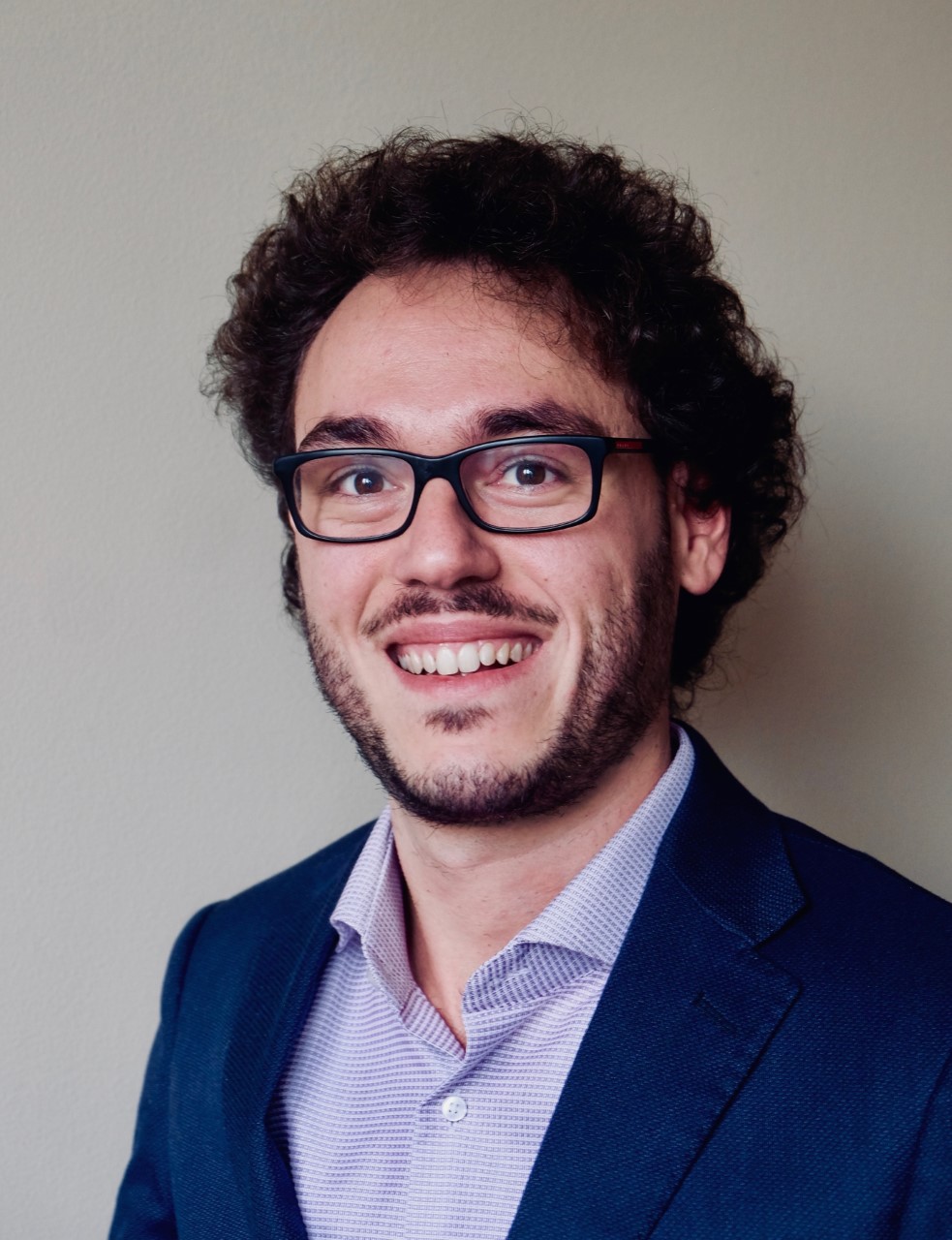}}]{Marco Mondelli} (Senior~Member) received the B.S. and M.S. degree in Telecommunications Engineering from the University of Pisa, Italy, in 2010 and 2012, respectively. In 2016, he obtained his Ph.D. degree in Computer and Communication Sciences at the École Polytechnique Fédérale de Lausanne (EPFL), Switzerland. He is currently an Assistant Professor at the Institute of Science and Technology Austria (IST Austria). Prior to that, he was a Postdoctoral Scholar in the Department of Electrical Engineering at Stanford University, CA, USA, from February 2017 to August 2019. He was also a Research Fellow with the Simons Institute for the Theory of Computing, UC Berkeley, CA, USA, for the program on Foundations of Data Science from August to December 2018. His research interests include data science, machine learning, information theory, wireless communication systems, and modern coding theory. He was the recipient of a number of fellowships and awards, including the Jack K. Wolf ISIT Student Paper Award in 2015, the STOC Best Paper Award in 2016, the EPFL Doctorate Award in 2018, the Simons-Berkeley Research Fellowship in 2018, the Lopez-Loreta Prize in 2019, and the Information Theory Society Paper Award in 2021.
	\end{IEEEbiography}
	
	\begin{IEEEbiography}[{\includegraphics[width=1in,height=1.25in,clip,keepaspectratio]{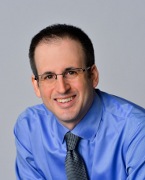}}]{Warren~J.~Gross} (Senior Member, IEEE) received the B.A.Sc. degree in electrical engineering from the University of Waterloo, Waterloo, ON, Canada, in 1996, and the M.A.Sc. and Ph.D. degrees from the University of Toronto, Toronto, ON, Canada, in 1999 and 2003, respectively. He is currently a James McGill Professor and the Chair of the Department of Electrical and Computer Engineering, McGill University, Montreal, QC, Canada. His research interests are in the design and implementation of signal processing systems and custom computer architectures. Dr. Gross served as the Chair for the IEEE Signal Processing Society Technical Committee on Design and Implementation of Signal Processing Systems. He served as the General Co-Chair for the IEEE GlobalSIP 2017 and the IEEE SiPS 2017 and the Technical Program Co-Chair for SiPS 2012. He also served as an Organizer for the Workshop on Polar Coding in Wireless Communications at WCNC 2017, the Symposium on Data Flow Algorithms and Architecture for Signal Processing Systems (GlobalSIP 2014), and the IEEE ICC 2012 Workshop on Emerging Data Storage Technologies. He served as an Associate Editor for the IEEE TRANSACTIONS ON SIGNAL PROCESSING and as a Senior Area Editor. He is a Licensed Professional Engineer in the Province of Ontario.
	\end{IEEEbiography}
\end{document}

%% file: PolarFactorGraph.tikz
\begin{tikzpicture}[scale=0.8]
	\footnotesize	
	\def\N{16}
	\def\xM{7.5}
	\def\xss{\xM/\N}
	\def\xs{\xss/1}
	\def\ys{0.35}
	\def\gain{1.0}
	\def\markSize{2.125}
	\def\PEmarkSize{3.5}
	
	%% Stage lines
	\draw[dashed] (0*\xs,0*\ys) -- (0*\xs,7.75*\ys*\gain) node[above]{\footnotesize{$s_0$}};
	\draw[dashed] (2*\xs,0*\ys) -- (2*\xs,7.75*\ys*\gain) node[above]{\footnotesize{$s_1$}};
	\draw[dashed] (5*\xs,0*\ys) -- (5*\xs,7.75*\ys*\gain) node[above]{\footnotesize{$s_2$}};
	\draw[dashed] (10*\xs,0*\ys) -- (10*\xs,7.75*\ys*\gain) node[above]{\footnotesize{$s_3$}};
	
	% Node and bit indcies
	\foreach \i in{0,...,7}
	{
		\pgfmathsetmacro\bIndex{int(7-\i)};
		\draw[] (0,\i*\ys) -- (10*\xs,\i*\ys);
		
		\draw[] plot[mark=*, mark size = \markSize, mark options={fill=white}] coordinates {(0*\xs,\i*\ys)};
		
		\ifthenelse{\i = 4}{\draw[] plot[mark=*, mark size = \markSize, mark options={fill=black}] coordinates {(0*\xs,\i*\ys)};}
		
		\ifthenelse{\i = 8}{\draw[] plot[mark=*, mark size = \markSize, mark options={fill=black}] coordinates {(0*\xs,\i*\ys)};}
		
		\ifthenelse{\i < 3}{\draw[] plot[mark=*, mark size = \markSize, mark options={fill=black}] coordinates {(0*\xs,\i*\ys)};}
		
		\draw[] plot[mark=*, mark size = \markSize, mark options={color=gray}] coordinates {(2*\xs,\i*\ys)};
		\draw[] plot[mark=*, mark size = \markSize, mark options={color=gray}] coordinates {(5*\xs,\i*\ys)};
		\draw[] plot[mark=*, mark size = \markSize, mark options={color=gray}] coordinates {(10*\xs,\i*\ys)};
		%	\draw[] plot[mark=*, mark size = \markSize, mark options={color=black}] coordinates {(10*\xs,\i*\ys)};
		
		\node[text width=0.5cm] at (-0.5*\xs,\i*\ys) {\footnotesize{$u_{\bIndex}$}};
		\node[text width=0.5cm] at (11*\xs,\i*\ys) {\footnotesize{$x_{\bIndex}$}};
	}
	
	%% PEs
	% Stage 0
	\draw[] plot[mark=oplus, mark size = \PEmarkSize, mark options={fill=white}] coordinates {(\xs,\ys)} --
	plot[mark=*, mark size = \PEmarkSize, mark options={fill=white}] coordinates {(\xs,0)} node[above=-0.19]{\scriptsize{=}};
	
	\draw[] plot[mark=oplus, mark size = \PEmarkSize, mark options={fill=white}] coordinates {(\xs,3*\ys)} --
	plot[mark=*, mark size = \PEmarkSize, mark options={fill=white}] coordinates {(\xs,2*\ys)} node[above=-0.19]{\scriptsize{=}};
	
	\draw[] plot[mark=oplus, mark size = \PEmarkSize, mark options={fill=white}] coordinates {(\xs,5*\ys)} --
	plot[mark=*, mark size = \PEmarkSize, mark options={fill=white}] coordinates {(\xs,4*\ys)} node[above=-0.19]{\scriptsize{=}};
	
	\draw[] plot[mark=oplus, mark size = \PEmarkSize, mark options={fill=white}] coordinates {(\xs,7*\ys)} --
	plot[mark=*, mark size = \PEmarkSize, mark options={fill=white}] coordinates {(\xs,6*\ys)} node[above=-0.19]{\scriptsize{=}};

	% Stage 1
	\draw[] plot[mark=oplus, mark size = \PEmarkSize, mark options={fill=white}] coordinates {(3*\xs,3*\ys)} --
	plot[mark=*, mark size = \PEmarkSize, mark options={fill=white}] coordinates {(3*\xs,\ys)} node[above=-0.19]{\scriptsize{=}};
	
	\draw[] plot[mark=oplus, mark size = \PEmarkSize, mark options={fill=white}] coordinates {(3*\xs,7*\ys)} --
	plot[mark=*, mark size = \PEmarkSize, mark options={fill=white}] coordinates {(3*\xs,5*\ys)} node[above=-0.19]{\scriptsize{=}};
	
	\draw[] plot[mark=oplus, mark size = \PEmarkSize, mark options={fill=white}] coordinates {(4*\xs,2*\ys)} --
	plot[mark=*, mark size = \PEmarkSize, mark options={fill=white}] coordinates {(4*\xs,0)} node[above=-0.19]{\scriptsize{=}};
	
	\draw[] plot[mark=oplus, mark size = \PEmarkSize, mark options={fill=white}] coordinates {(4*\xs,6*\ys)} --
	plot[mark=*, mark size = \PEmarkSize, mark options={fill=white}] coordinates {(4*\xs,4*\ys)} node[above=-0.19]{\scriptsize{=}};
	
	% Stage 2
	\draw[] plot[mark=oplus, mark size = \PEmarkSize, mark options={fill=white}] coordinates {(9*\xs,4*\ys)} --
	plot[mark=*, mark size = \PEmarkSize, mark options={fill=white}] coordinates {(9*\xs,0*\ys)} node[above=-0.19]{\scriptsize{=}};
	
	\draw[] plot[mark=oplus, mark size = \PEmarkSize, mark options={fill=white}] coordinates {(8*\xs,5*\ys)} --
	plot[mark=*, mark size = \PEmarkSize, mark options={fill=white}] coordinates {(8*\xs,1*\ys)} node[above=-0.19]{\scriptsize{=}};
	
	\draw[] plot[mark=oplus, mark size = \PEmarkSize, mark options={fill=white}] coordinates {(7*\xs,6*\ys)} --
	plot[mark=*, mark size = \PEmarkSize, mark options={fill=white}] coordinates {(7*\xs,2*\ys)} node[above=-0.19]{\scriptsize{=}};
	
	\draw[] plot[mark=oplus, mark size = \PEmarkSize, mark options={fill=white}] coordinates {(6*\xs,7*\ys)} --
	plot[mark=*, mark size = \PEmarkSize, mark options={fill=white}] coordinates {(6*\xs,3*\ys)} node[above=-0.19]{\scriptsize{=}};
	
	%\draw[rounded corners, dashdotted] (0.5*\xs, -0.5*\ys) rectangle (1.5*\xs, 7.5*\ys) {};
	%\draw[rounded corners, dashdotted] (2.5*\xs, -0.5*\ys) rectangle (4.5*\xs, 7.5*\ys) {};
	%\draw[rounded corners, dashdotted] (5.5*\xs, -0.5*\ys) rectangle (9.5*\xs, 7.5*\ys) {};
	
	%\node[text width=1.25cm, fill=none] at (2.25*\xs, -1*\ys) {\footnotesize{$0$}};
	%\node[text width=1.25cm, fill=none] at (4.75*\xs, -1*\ys) {\tiny{$1$}};
	%\node[text width=1.25cm, fill=none] at (8.5*\xs, -1*\ys) {\tiny{$2$}};
	
\end{tikzpicture}

%% file: PolarBinaryTree.tikz
\usetikzlibrary{arrows, decorations}

% for double arrows a la chef
% adapt line thickness and line width, if needed
\tikzstyle{vecArrow} = [thick, decoration={markings,mark=at position
	1 with {\arrow[semithick]{open triangle 60}}},
double distance=1.4pt, shorten >= 5.5pt,
preaction = {decorate},
postaction = {draw,line width=1.4pt, white,shorten >= 4.5pt}]
\tikzstyle{innerWhite} = [semithick, white,line width=1.4pt, shorten >= 4.5pt]

\begin{tikzpicture}[scale=0.7]
\footnotesize
\def\N{16}
\def\xM{7.5}
\def\xss{\xM/\N}
\def\xs{\xss/1}
\def\ys{0.4}
\def\Ygain{1.075}
\def\Xgain{0.8}
\def\markSize{2.5}
\def\PEmarkSize{3.8}

% Indices
%\node[text width=2cm] at (2.6*\xs,9.5*\ys) {\small{Stage Indices}};
%\node[text width=2cm, rotate=90] at (-2.5*\xs,4.5*\ys) {\small{Bit Indices}};

%% Stage lines
\draw[dashed] (0*\xs*\Xgain,-0.25*\ys) -- (0*\xs*\Xgain,7*\ys*\Ygain) node[above]{\footnotesize{$s_0$}};
\draw[dashed] (2*\xs*\Xgain,-0.25*\ys) -- (2*\xs*\Xgain,7*\ys*\Ygain) node[above]{\footnotesize{$s_1$}};
\draw[dashed] (4*\xs*\Xgain,-0.25*\ys) -- (4*\xs*\Xgain,7*\ys*\Ygain) node[above]{\footnotesize{$s_2$}};
\draw[dashed] (6*\xs*\Xgain,-0.25*\ys) -- (6*\xs*\Xgain,7*\ys*\Ygain) node[above]{\footnotesize{$s_3$}};
%\draw[dashed] (8*\xs*\Xgain,-0.25*\ys) -- (8*\xs*\Xgain,14.5*\ys*\Ygain) node[above]{\footnotesize{$s_4$}};

% Node and bit indcies
%\draw[] (8*\xs*\Xgain,7.5*\ys) -- (6*\xs*\Xgain,3.5*\ys);
%\draw[] (8*\xs*\Xgain,7.5*\ys) -- (6*\xs*\Xgain,11.5*\ys);
%\draw[] plot[mark=*, mark size = \markSize, mark options={fill=gray}] coordinates {(8*\xs*\Xgain,7.5*\ys)};

\foreach \i in{0,...,0}
{
	\draw[] (6*\xs*\Xgain,8*\i*\ys+3.5*\ys) -- (4*\xs*\Xgain,8*\i*\ys+1.5*\ys);
	\draw[] (6*\xs*\Xgain,8*\i*\ys+3.5*\ys) -- (4*\xs*\Xgain,8*\i*\ys+5.5*\ys);
	\draw[] plot[mark=*, mark size = \markSize, mark options={fill=gray}] coordinates {(6*\xs*\Xgain,8*\i*\ys+3.5*\ys)};
}

\foreach \i in{0,...,1}
{
	\draw[] (4*\xs*\Xgain,4*\i*\ys+1.5*\ys) -- (2*\xs*\Xgain,4*\i*\ys+0.5*\ys);
	\draw[] (4*\xs*\Xgain,4*\i*\ys+1.5*\ys) -- (2*\xs*\Xgain,4*\i*\ys+2.5*\ys);
	\draw[] plot[mark=*, mark size = \markSize, mark options={fill=gray}] coordinates {(4*\xs*\Xgain,4*\i*\ys+1.5*\ys)};
}

\foreach \i in{0,...,3}
{
	\draw[] (2*\xs*\Xgain,2*\i*\ys+0.5*\ys) -- (0*\xs*\Xgain,2*\i*\ys);
	\draw[] (2*\xs*\Xgain,2*\i*\ys+0.5*\ys) -- (0*\xs*\Xgain,2*\i*\ys+\ys);
	\draw[] plot[mark=*, mark size = \markSize, mark options={fill=gray}] coordinates {(2*\xs*\Xgain,2*\i*\ys+0.5*\ys)};	
}

\foreach \i in{0,...,7}
{
	\pgfmathsetmacro\bIndex{int(7-\i)};
%	\ifthenelse{\i > 10}{\draw[] plot[mark=*, mark size = \markSize, mark options={fill=white}] coordinates {(0*\xs,\i*\ys)};}
%	
%	\ifthenelse{\i < 11 \AND \i > 7}{\draw[] plot[mark=*, mark size = \markSize, mark options={fill=black}] coordinates {(0*\xs,\i*\ys)};}
%	
%	\ifthenelse{\i < 8 \AND \i > 4}{\draw[] plot[mark=*, mark size = \markSize, mark options={fill=white}] coordinates {(0*\xs,\i*\ys)};}
%	
%	\ifthenelse{\i < 5}{\draw[] plot[mark=*, mark size = \markSize, mark options={fill=black}] coordinates {(0*\xs,\i*\ys)};}

	\draw[] plot[mark=*, mark size = \markSize, mark options={fill=white}] coordinates {(0*\xs,\i*\ys)};
	
	\ifthenelse{\i = 4}{\draw[] plot[mark=*, mark size = \markSize, mark options={fill=black}] coordinates {(0*\xs,\i*\ys)};}
	
	\ifthenelse{\i = 8}{\draw[] plot[mark=*, mark size = \markSize, mark options={fill=black}] coordinates {(0*\xs,\i*\ys)};}
	
	\ifthenelse{\i < 3}{\draw[] plot[mark=*, mark size = \markSize, mark options={fill=black}] coordinates {(0*\xs,\i*\ys)};}
	
	\node[text width=0.5cm] at (-0.75*\xs,\i*\ys) {\footnotesize{$u_{\bIndex}$}};
}

\node[text width=0cm] at (6.5*\xs*\Xgain,3.5*\ys) {\footnotesize{$\bm{x}$}};

\end{tikzpicture}

%% file: SCPE.tikz
\begin{tikzpicture}[scale=0.7]
%\footnotesize
\def\markSize{3}
\draw[] plot[mark=*, mark size = \markSize, mark options={color=gray}] coordinates {(0,0.75)} node[left=0.1] {$\alpha_{s,i},\beta_{s,i}$} -- plot[mark=*, mark size = \markSize, mark options={color=gray}] coordinates {(2,0.75)} node[right=0.1] {$\alpha_{s+1,i},\beta_{s+1,i}$};

\draw[] plot[mark=*, mark size = \markSize, mark options={color=gray}] coordinates {(0,0)} node[left=0.1] {$\alpha_{s,i+2^s},\beta_{s,i+2^s}$} -- plot[mark=*, mark size = \markSize, mark options={color=gray}] coordinates {(2,0)} node[right=0.1] {$\alpha_{s+1,i+2^s},\beta_{s+1,i+2^s}$};

\draw[] plot[mark=oplus, mark size = 6.5, mark options={fill=white}] coordinates {(1,0.75)} --
plot[mark=*, mark size = 6.5, mark options={fill=white}] coordinates {(1,0)} node[above=-0.25]{=};
\end{tikzpicture}

%% file: PolarFactorGraph_p1.tikz.tex
\begin{tikzpicture}[scale=0.8]	
	% for double arrows a la chef
	% adapt line thickness and line width, if needed
	\footnotesize
	\tikzstyle{vecArrow} = [thick, decoration={markings,mark=at position
		1 with {\arrow[semithick]{open triangle 60}}},
	double distance=1.4pt, shorten >= 5.5pt,
	preaction = {decorate},
	postaction = {draw,line width=1.4pt, white,shorten >= 4.5pt}]
	\tikzstyle{innerWhite} = [semithick, white,line width=1.4pt, shorten >= 4.5pt]
	
	\def\N{16}
	\def\xM{7.5}
	\def\xss{\xM/\N}
	\def\xs{\xss/1.25}
	\def\ys{0.35}
	\def\gain{1.0}
	\def\markSize{2}
	\def\PEmarkSize{3.5}
	
	%% Stage lines
	%	\draw[dashed] (0*\xs,0*\ys) -- (0*\xs,7.75*\ys*\gain) node[above]{\footnotesize{$s_0$}};
	%	\draw[dashed] (2*\xs,0*\ys) -- (2*\xs,7.75*\ys*\gain) node[above]{\footnotesize{$s_1$}};
	%	\draw[dashed] (7*\xs,0*\ys) -- (7*\xs,7.75*\ys*\gain) node[above]{\footnotesize{$s_2$}};
	%	\draw[dashed] (10*\xs,0*\ys) -- (10*\xs,7.75*\ys*\gain) node[above]{\footnotesize{$s_3$}};
	
	% Node and bit indcies
	\foreach \i in{0,...,7}
	{
		\pgfmathsetmacro\bIndex{int(7-\i)};
		\draw[] (0,\i*\ys) -- (10*\xs,\i*\ys);
		
		\draw[] plot[mark=*, mark size = \markSize, mark options={fill=white}] coordinates {(0*\xs,\i*\ys)};
		
		\ifthenelse{\i = 4}{\draw[] plot[mark=*, mark size = \markSize, mark options={fill=black}] coordinates {(0*\xs,\i*\ys)};}
		
		\ifthenelse{\i = 8}{\draw[] plot[mark=*, mark size = \markSize, mark options={fill=black}] coordinates {(0*\xs,\i*\ys)};}
		
		\ifthenelse{\i < 3}{\draw[] plot[mark=*, mark size = \markSize, mark options={fill=black}] coordinates {(0*\xs,\i*\ys)};}
		
		\draw[] plot[mark=*, mark size = \markSize, mark options={color=gray}] coordinates {(2*\xs,\i*\ys)};
		\draw[] plot[mark=*, mark size = \markSize, mark options={color=gray}] coordinates {(7*\xs,\i*\ys)};
		\draw[] plot[mark=*, mark size = \markSize, mark options={color=gray}] coordinates {(10*\xs,\i*\ys)};
		%	\draw[] plot[mark=*, mark size = \markSize, mark options={color=black}] coordinates {(10*\xs,\i*\ys)};
		
		\node[text width=0.5cm] at (-0.5*\xs,\i*\ys) {\footnotesize{$u_{\bIndex}$}};
		\node[text width=0.5cm] at (11.25*\xs,\i*\ys) {\footnotesize{$x_{\bIndex}$}};
	}
	
	%% PEs
	% Stage 0
	\draw[] plot[mark=oplus, mark size = \PEmarkSize, mark options={fill=white}] coordinates {(\xs,\ys)} --
	plot[mark=*, mark size = \PEmarkSize, mark options={fill=white}] coordinates {(\xs,0)} node[above=-0.19]{\scriptsize{=}};
	
	\draw[] plot[mark=oplus, mark size = \PEmarkSize, mark options={fill=white}] coordinates {(\xs,3*\ys)} --
	plot[mark=*, mark size = \PEmarkSize, mark options={fill=white}] coordinates {(\xs,2*\ys)} node[above=-0.19]{\scriptsize{=}};
	
	\draw[] plot[mark=oplus, mark size = \PEmarkSize, mark options={fill=white}] coordinates {(\xs,5*\ys)} --
	plot[mark=*, mark size = \PEmarkSize, mark options={fill=white}] coordinates {(\xs,4*\ys)} node[above=-0.19]{\scriptsize{=}};
	
	\draw[] plot[mark=oplus, mark size = \PEmarkSize, mark options={fill=white}] coordinates {(\xs,7*\ys)} --
	plot[mark=*, mark size = \PEmarkSize, mark options={fill=white}] coordinates {(\xs,6*\ys)} node[above=-0.19]{\scriptsize{=}};

	%	 Stage 1
	\draw[] plot[mark=oplus, mark size = \PEmarkSize, mark options={fill=white}] coordinates {(6*\xs,4*\ys)} --
	plot[mark=*, mark size = \PEmarkSize, mark options={fill=white}] coordinates {(6*\xs,0*\ys)} node[above=-0.19]{\scriptsize{=}};
	
	\draw[] plot[mark=oplus, mark size = \PEmarkSize, mark options={fill=white}] coordinates {(5*\xs,5*\ys)} --
	plot[mark=*, mark size = \PEmarkSize, mark options={fill=white}] coordinates {(5*\xs,1*\ys)} node[above=-0.19]{\scriptsize{=}};
	
	\draw[] plot[mark=oplus, mark size = \PEmarkSize, mark options={fill=white}] coordinates {(4*\xs,6*\ys)} --
	plot[mark=*, mark size = \PEmarkSize, mark options={fill=white}] coordinates {(4*\xs,2*\ys)} node[above=-0.19]{\scriptsize{=}};
	
	\draw[] plot[mark=oplus, mark size = \PEmarkSize, mark options={fill=white}] coordinates {(3*\xs,7*\ys)} --
	plot[mark=*, mark size = \PEmarkSize, mark options={fill=white}] coordinates {(3*\xs,3*\ys)} node[above=-0.19]{\scriptsize{=}};
	
	% Stage 2
	
	\draw[] plot[mark=oplus, mark size = \PEmarkSize, mark options={fill=white}] coordinates {(8*\xs,3*\ys)} --
	plot[mark=*, mark size = \PEmarkSize, mark options={fill=white}] coordinates {(8*\xs,\ys)} node[above=-0.19]{\scriptsize{=}};
	
	\draw[] plot[mark=oplus, mark size = \PEmarkSize, mark options={fill=white}] coordinates {(8*\xs,7*\ys)} --
	plot[mark=*, mark size = \PEmarkSize, mark options={fill=white}] coordinates {(8*\xs,5*\ys)} node[above=-0.19]{\scriptsize{=}};
	
	\draw[] plot[mark=oplus, mark size = \PEmarkSize, mark options={fill=white}] coordinates {(9*\xs,2*\ys)} --
	plot[mark=*, mark size = \PEmarkSize, mark options={fill=white}] coordinates {(9*\xs,0)} node[above=-0.19]{\scriptsize{=}};
	
	\draw[] plot[mark=oplus, mark size = \PEmarkSize, mark options={fill=white}] coordinates {(9*\xs,6*\ys)} --
	plot[mark=*, mark size = \PEmarkSize, mark options={fill=white}] coordinates {(9*\xs,4*\ys)} node[above=-0.19]{\scriptsize{=}};
	
	\draw[rounded corners, dashdotted] (0.5*\xs, -0.5*\ys) rectangle (1.5*\xs, 7.5*\ys) {};
	\draw[rounded corners, dashdotted] (2.5*\xs, -0.5*\ys) rectangle (6.5*\xs, 7.5*\ys) {};	
	\draw[rounded corners, dashdotted] (7.5*\xs, -0.5*\ys) rectangle (9.5*\xs, 7.5*\ys) {};
	
	\node[text width=1cm, fill=none] at (2.5*\xs, -1.25*\ys) {\footnotesize{$0$}};
	\node[text width=1cm, fill=none] at (6*\xs, -1.25*\ys) {\footnotesize{$2$}};
	\node[text width=1cm, fill=none] at (10*\xs, -1.25*\ys) {\footnotesize{$1$}};
	
	\node[text width=4cm, fill=none] at (7*\xs, 8.5*\ys) {\footnotesize{Factor-graph Permutation}};
	
	\draw[vecArrow] (5*\xs,-2*\ys) -- (5*\xs,-4*\ys);
	
\end{tikzpicture}

%% file: PolarFactorGraphPer1.tikz.tex
\begin{tikzpicture}[scale=0.8]
	\footnotesize

	\def\N{16}
	\def\xM{7.5}
	\def\xss{\xM/\N}
	\def\xs{\xss/1.25}
	\def\ys{0.35}
	\def\gain{1.0}
	\def\markSize{2}
	\def\PEmarkSize{3.5}
	
	%% Stage lines
	%	\draw[dashed] (0*\xs,0*\ys) -- (0*\xs,7.75*\ys*\gain) node[above]{\footnotesize{$s_0$}};
	%	\draw[dashed] (2*\xs,0*\ys) -- (2*\xs,7.75*\ys*\gain) node[above]{\footnotesize{$s_1$}};
	%	\draw[dashed] (5*\xs,0*\ys) -- (5*\xs,7.75*\ys*\gain) node[above]{\footnotesize{$s_2$}};
	%	\draw[dashed] (10*\xs,0*\ys) -- (10*\xs,7.75*\ys*\gain) node[above]{\footnotesize{$s_3$}};
	
	% Node and bit indcies
	\foreach \i in{0,...,7}
	{
		\pgfmathsetmacro\bIndex{int(7-\i)};
		\draw[] (0,\i*\ys) -- (10*\xs,\i*\ys);
		
		\draw[] plot[mark=*, mark size = \markSize, mark options={fill=white}] coordinates {(0*\xs,\i*\ys)};
		
		\ifthenelse{\i = 4}{\draw[] plot[mark=*, mark size = \markSize, mark options={fill=black}] coordinates {(0*\xs,\i*\ys)};}
		
		\ifthenelse{\i = 8}{\draw[] plot[mark=*, mark size = \markSize, mark options={fill=black}] coordinates {(0*\xs,\i*\ys)};}
		
		\ifthenelse{\i < 3}{\draw[] plot[mark=*, mark size = \markSize, mark options={fill=black}] coordinates {(0*\xs,\i*\ys)};}
		
		\draw[] plot[mark=*, mark size = \markSize, mark options={color=gray}] coordinates {(2*\xs,\i*\ys)};
		\draw[] plot[mark=*, mark size = \markSize, mark options={color=gray}] coordinates {(5*\xs,\i*\ys)};
		\draw[] plot[mark=*, mark size = \markSize, mark options={color=gray}] coordinates {(10*\xs,\i*\ys)};
		%	\draw[] plot[mark=*, mark size = \markSize, mark options={color=black}] coordinates {(10*\xs,\i*\ys)};
		
		%		\node[text width=0.5cm] at (-0.5*\xs,\i*\ys) {\footnotesize{$u_{\bIndex}$}};
		%		\node[text width=0.5cm] at (11.25*\xs,\i*\ys) {\footnotesize{$x_{\bIndex}$}};
	}
	
	\node[text width=0.5cm] at (-0.5*\xs,0*\ys) {\footnotesize{$u_{7}$}};
	\node[text width=0.5cm] at (-0.5*\xs,1*\ys) {\footnotesize{$u_{5}$}};
	\node[text width=0.5cm] at (-0.5*\xs,2*\ys) {\footnotesize{$u_{6}$}};
	\node[text width=0.5cm] at (-0.5*\xs,3*\ys) {\footnotesize{$u_{4}$}};
	\node[text width=0.5cm] at (-0.5*\xs,4*\ys) {\footnotesize{$u_{3}$}};
	\node[text width=0.5cm] at (-0.5*\xs,5*\ys) {\footnotesize{$u_{1}$}};
	\node[text width=0.5cm] at (-0.5*\xs,6*\ys) {\footnotesize{$u_{2}$}};
	\node[text width=0.5cm] at (-0.5*\xs,7*\ys) {\footnotesize{$u_{0}$}};
	
	\node[text width=0.5cm] at (11.25*\xs,0*\ys) {\footnotesize{$x_{7}$}};
	\node[text width=0.5cm] at (11.25*\xs,1*\ys) {\footnotesize{$x_{5}$}};
	\node[text width=0.5cm] at (11.25*\xs,2*\ys) {\footnotesize{$x_{6}$}};
	\node[text width=0.5cm] at (11.25*\xs,3*\ys) {\footnotesize{$x_{4}$}};
	\node[text width=0.5cm] at (11.25*\xs,4*\ys) {\footnotesize{$x_{3}$}};
	\node[text width=0.5cm] at (11.25*\xs,5*\ys) {\footnotesize{$x_{1}$}};
	\node[text width=0.5cm] at (11.25*\xs,6*\ys) {\footnotesize{$x_{2}$}};
	\node[text width=0.5cm] at (11.25*\xs,7*\ys) {\footnotesize{$x_{0}$}};
	
	%% PEs
	% Stage 0
	\draw[] plot[mark=oplus, mark size = \PEmarkSize, mark options={fill=white}] coordinates {(\xs,\ys)} --
	plot[mark=*, mark size = \PEmarkSize, mark options={fill=white}] coordinates {(\xs,0)} node[above=-0.19]{\scriptsize{=}};
	
	\draw[] plot[mark=oplus, mark size = \PEmarkSize, mark options={fill=white}] coordinates {(\xs,3*\ys)} --
	plot[mark=*, mark size = \PEmarkSize, mark options={fill=white}] coordinates {(\xs,2*\ys)} node[above=-0.19]{\scriptsize{=}};
	
	\draw[] plot[mark=oplus, mark size = \PEmarkSize, mark options={fill=white}] coordinates {(\xs,5*\ys)} --
	plot[mark=*, mark size = \PEmarkSize, mark options={fill=white}] coordinates {(\xs,4*\ys)} node[above=-0.19]{\scriptsize{=}};
	
	\draw[] plot[mark=oplus, mark size = \PEmarkSize, mark options={fill=white}] coordinates {(\xs,7*\ys)} --
	plot[mark=*, mark size = \PEmarkSize, mark options={fill=white}] coordinates {(\xs,6*\ys)} node[above=-0.19]{\scriptsize{=}};

	% Stage 1
	\draw[] plot[mark=oplus, mark size = \PEmarkSize, mark options={fill=white}] coordinates {(3*\xs,3*\ys)} --
	plot[mark=*, mark size = \PEmarkSize, mark options={fill=white}] coordinates {(3*\xs,\ys)} node[above=-0.19]{\scriptsize{=}};
	
	\draw[] plot[mark=oplus, mark size = \PEmarkSize, mark options={fill=white}] coordinates {(3*\xs,7*\ys)} --
	plot[mark=*, mark size = \PEmarkSize, mark options={fill=white}] coordinates {(3*\xs,5*\ys)} node[above=-0.19]{\scriptsize{=}};
	
	\draw[] plot[mark=oplus, mark size = \PEmarkSize, mark options={fill=white}] coordinates {(4*\xs,2*\ys)} --
	plot[mark=*, mark size = \PEmarkSize, mark options={fill=white}] coordinates {(4*\xs,0)} node[above=-0.19]{\scriptsize{=}};
	
	\draw[] plot[mark=oplus, mark size = \PEmarkSize, mark options={fill=white}] coordinates {(4*\xs,6*\ys)} --
	plot[mark=*, mark size = \PEmarkSize, mark options={fill=white}] coordinates {(4*\xs,4*\ys)} node[above=-0.19]{\scriptsize{=}};
	
	% Stage 2
	\draw[] plot[mark=oplus, mark size = \PEmarkSize, mark options={fill=white}] coordinates {(9*\xs,4*\ys)} --
	plot[mark=*, mark size = \PEmarkSize, mark options={fill=white}] coordinates {(9*\xs,0*\ys)} node[above=-0.19]{\scriptsize{=}};
	
	\draw[] plot[mark=oplus, mark size = \PEmarkSize, mark options={fill=white}] coordinates {(8*\xs,5*\ys)} --
	plot[mark=*, mark size = \PEmarkSize, mark options={fill=white}] coordinates {(8*\xs,1*\ys)} node[above=-0.19]{\scriptsize{=}};
	
	\draw[] plot[mark=oplus, mark size = \PEmarkSize, mark options={fill=white}] coordinates {(7*\xs,6*\ys)} --
	plot[mark=*, mark size = \PEmarkSize, mark options={fill=white}] coordinates {(7*\xs,2*\ys)} node[above=-0.19]{\scriptsize{=}};
	
	\draw[] plot[mark=oplus, mark size = \PEmarkSize, mark options={fill=white}] coordinates {(6*\xs,7*\ys)} --
	plot[mark=*, mark size = \PEmarkSize, mark options={fill=white}] coordinates {(6*\xs,3*\ys)} node[above=-0.19]{\scriptsize{=}};
	
	\draw[rounded corners, dashdotted] (0.5*\xs, -0.5*\ys) rectangle (1.5*\xs, 7.5*\ys) {};
	\draw[rounded corners, dashdotted] (2.5*\xs, -0.5*\ys) rectangle (4.5*\xs, 7.5*\ys) {};
	\draw[rounded corners, dashdotted] (5.5*\xs, -0.5*\ys) rectangle (9.5*\xs, 7.5*\ys) {};
	
	\node[text width=1.25cm, fill=none] at (2.9*\xs, -1.25*\ys) {\footnotesize{$0$}};
	\node[text width=1.25cm, fill=none] at (5.4*\xs, -1.25*\ys) {\footnotesize{$1$}};
	\node[text width=1.25cm, fill=none] at (9.25*\xs, -1.25*\ys) {\footnotesize{$2$}};
	
	\node[text width=3cm, fill=none] at (5.5*\xs, 8.5*\ys) {\footnotesize{Codeword Permutation}};
	
\end{tikzpicture}

%% file: PolarFactorGraph_p2.tikz.tex
\begin{tikzpicture}[scale=0.8]
	\footnotesize
	\tikzstyle{vecArrow} = [thick, decoration={markings,mark=at position
		1 with {\arrow[semithick]{open triangle 60}}},
	double distance=1.4pt, shorten >= 5.5pt,
	preaction = {decorate},
	postaction = {draw,line width=1.4pt, white,shorten >= 4.5pt}]
	\tikzstyle{innerWhite} = [semithick, white,line width=1.4pt, shorten >= 4.5pt]
	
	\def\N{16}
	\def\xM{7.5}
	\def\xss{\xM/\N}
	\def\xs{\xss/1.25}
	\def\ys{0.35}
	\def\gain{1.0}
	\def\markSize{2}
	\def\PEmarkSize{3.5}
	
	%% Stage lines
	%	\draw[dashed] (0*\xs,0*\ys) -- (0*\xs,7.75*\ys*\gain) node[above]{\footnotesize{$s_0$}};
	%	\draw[dashed] (3*\xs,0*\ys) -- (3*\xs,7.75*\ys*\gain) node[above]{\footnotesize{$s_1$}};
	%	\draw[dashed] (5*\xs,0*\ys) -- (5*\xs,7.75*\ys*\gain) node[above]{\footnotesize{$s_2$}};
	%	\draw[dashed] (10*\xs,0*\ys) -- (10*\xs,7.75*\ys*\gain) node[above]{\footnotesize{$s_3$}};
	
	% Node and bit indcies
	\foreach \i in{0,...,7}
	{
		\pgfmathsetmacro\bIndex{int(7-\i)};
		\draw[] (0,\i*\ys) -- (10*\xs,\i*\ys);
		
		\draw[] plot[mark=*, mark size = \markSize, mark options={fill=white}] coordinates {(0*\xs,\i*\ys)};
		
		\ifthenelse{\i = 4}{\draw[] plot[mark=*, mark size = \markSize, mark options={fill=black}] coordinates {(0*\xs,\i*\ys)};}
		
		\ifthenelse{\i = 8}{\draw[] plot[mark=*, mark size = \markSize, mark options={fill=black}] coordinates {(0*\xs,\i*\ys)};}
		
		\ifthenelse{\i < 3}{\draw[] plot[mark=*, mark size = \markSize, mark options={fill=black}] coordinates {(0*\xs,\i*\ys)};}
		
		\draw[] plot[mark=*, mark size = \markSize, mark options={color=gray}] coordinates {(3*\xs,\i*\ys)};
		\draw[] plot[mark=*, mark size = \markSize, mark options={color=gray}] coordinates {(5*\xs,\i*\ys)};
		
		\draw[] plot[mark=*, mark size = \markSize, mark options={color=gray}] coordinates {(10*\xs,\i*\ys)};
		%	\draw[] plot[mark=*, mark size = \markSize, mark options={color=black}] coordinates {(10*\xs,\i*\ys)};
		
		\node[text width=0.5cm] at (-0.5*\xs,\i*\ys) {\footnotesize{$u_{\bIndex}$}};
		\node[text width=0.5cm] at (11.25*\xs,\i*\ys) {\footnotesize{$x_{\bIndex}$}};
	}
	
	%% PEs
	% Stage 0
	\draw[] plot[mark=oplus, mark size = \PEmarkSize, mark options={fill=white}] coordinates {(4*\xs,\ys)} --
	plot[mark=*, mark size = \PEmarkSize, mark options={fill=white}] coordinates {(4*\xs,0)} node[above=-0.19]{\scriptsize{=}};
	
	\draw[] plot[mark=oplus, mark size = \PEmarkSize, mark options={fill=white}] coordinates {(4*\xs,3*\ys)} --
	plot[mark=*, mark size = \PEmarkSize, mark options={fill=white}] coordinates {(4*\xs,2*\ys)} node[above=-0.19]{\scriptsize{=}};
	
	\draw[] plot[mark=oplus, mark size = \PEmarkSize, mark options={fill=white}] coordinates {(4*\xs,5*\ys)} --
	plot[mark=*, mark size = \PEmarkSize, mark options={fill=white}] coordinates {(4*\xs,4*\ys)} node[above=-0.19]{\scriptsize{=}};
	
	\draw[] plot[mark=oplus, mark size = \PEmarkSize, mark options={fill=white}] coordinates {(4*\xs,7*\ys)} --
	plot[mark=*, mark size = \PEmarkSize, mark options={fill=white}] coordinates {(4*\xs,6*\ys)} node[above=-0.19]{\scriptsize{=}};

	% Stage 1
	\draw[] plot[mark=oplus, mark size = \PEmarkSize, mark options={fill=white}] coordinates {(1*\xs,3*\ys)} --
	plot[mark=*, mark size = \PEmarkSize, mark options={fill=white}] coordinates {(1*\xs,\ys)} node[above=-0.19]{\scriptsize{=}};
	
	\draw[] plot[mark=oplus, mark size = \PEmarkSize, mark options={fill=white}] coordinates {(1*\xs,7*\ys)} --
	plot[mark=*, mark size = \PEmarkSize, mark options={fill=white}] coordinates {(1*\xs,5*\ys)} node[above=-0.19]{\scriptsize{=}};
	
	\draw[] plot[mark=oplus, mark size = \PEmarkSize, mark options={fill=white}] coordinates {(2*\xs,2*\ys)} --
	plot[mark=*, mark size = \PEmarkSize, mark options={fill=white}] coordinates {(2*\xs,0)} node[above=-0.19]{\scriptsize{=}};
	
	\draw[] plot[mark=oplus, mark size = \PEmarkSize, mark options={fill=white}] coordinates {(2*\xs,6*\ys)} --
	plot[mark=*, mark size = \PEmarkSize, mark options={fill=white}] coordinates {(2*\xs,4*\ys)} node[above=-0.19]{\scriptsize{=}};
	
	% Stage 2
	\draw[] plot[mark=oplus, mark size = \PEmarkSize, mark options={fill=white}] coordinates {(9*\xs,4*\ys)} --
	plot[mark=*, mark size = \PEmarkSize, mark options={fill=white}] coordinates {(9*\xs,0*\ys)} node[above=-0.19]{\scriptsize{=}};
	
	\draw[] plot[mark=oplus, mark size = \PEmarkSize, mark options={fill=white}] coordinates {(8*\xs,5*\ys)} --
	plot[mark=*, mark size = \PEmarkSize, mark options={fill=white}] coordinates {(8*\xs,1*\ys)} node[above=-0.19]{\scriptsize{=}};
	
	\draw[] plot[mark=oplus, mark size = \PEmarkSize, mark options={fill=white}] coordinates {(7*\xs,6*\ys)} --
	plot[mark=*, mark size = \PEmarkSize, mark options={fill=white}] coordinates {(7*\xs,2*\ys)} node[above=-0.19]{\scriptsize{=}};
	
	\draw[] plot[mark=oplus, mark size = \PEmarkSize, mark options={fill=white}] coordinates {(6*\xs,7*\ys)} --
	plot[mark=*, mark size = \PEmarkSize, mark options={fill=white}] coordinates {(6*\xs,3*\ys)} node[above=-0.19]{\scriptsize{=}};
	
	\draw[rounded corners, dashdotted] (0.5*\xs, -0.5*\ys) rectangle (2.5*\xs, 7.5*\ys) {};
	\draw[rounded corners, dashdotted] (3.5*\xs, -0.5*\ys) rectangle (4.5*\xs, 7.5*\ys) {};	
	\draw[rounded corners, dashdotted] (5.5*\xs, -0.5*\ys) rectangle (9.5*\xs, 7.5*\ys) {};
	
	\node[text width=1cm, fill=none] at (3*\xs, -1.25*\ys) {\footnotesize{$1$}};
	\node[text width=1cm, fill=none] at (5.5*\xs, -1.25*\ys) {\footnotesize{$0$}};
	\node[text width=1cm, fill=none] at (9*\xs, -1.25*\ys) {\footnotesize{$2$}};
	
	\node[text width=4cm, fill=none] at (7*\xs, 8.5*\ys) {\footnotesize{Factor-graph Permutation}};
	
	\draw[vecArrow] (5*\xs,-2*\ys) -- (5*\xs,-4*\ys);
	
\end{tikzpicture}

%% file: PolarFactorGraphPer2.tikz.tex
\begin{tikzpicture}[scale=0.8]
	\footnotesize	

	\def\N{16}
	\def\xM{7.5}
	\def\xss{\xM/\N}
	\def\xs{\xss/1.25}
	\def\ys{0.35}
	\def\gain{1.0}
	\def\markSize{2}
	\def\PEmarkSize{3.5}
	
	%% Stage lines
	%	\draw[dashed] (0*\xs,0*\ys) -- (0*\xs,7.75*\ys*\gain) node[above]{\footnotesize{$s_0$}};
	%	\draw[dashed] (2*\xs,0*\ys) -- (2*\xs,7.75*\ys*\gain) node[above]{\footnotesize{$s_1$}};
	%	\draw[dashed] (5*\xs,0*\ys) -- (5*\xs,7.75*\ys*\gain) node[above]{\footnotesize{$s_2$}};
	%	\draw[dashed] (10*\xs,0*\ys) -- (10*\xs,7.75*\ys*\gain) node[above]{\footnotesize{$s_3$}};
	
	% Node and bit indcies
	\foreach \i in{0,...,7}
	{
		\pgfmathsetmacro\bIndex{int(7-\i)};
		\draw[] (0,\i*\ys) -- (10*\xs,\i*\ys);
		
		\draw[] plot[mark=*, mark size = \markSize, mark options={fill=white}] coordinates {(0*\xs,\i*\ys)};
		
		\ifthenelse{\i = 4}{\draw[] plot[mark=*, mark size = \markSize, mark options={fill=black}] coordinates {(0*\xs,\i*\ys)};}
		
		\ifthenelse{\i = 8}{\draw[] plot[mark=*, mark size = \markSize, mark options={fill=black}] coordinates {(0*\xs,\i*\ys)};}
		
		\ifthenelse{\i < 3}{\draw[] plot[mark=*, mark size = \markSize, mark options={fill=black}] coordinates {(0*\xs,\i*\ys)};}
		
		\draw[] plot[mark=*, mark size = \markSize, mark options={color=gray}] coordinates {(2*\xs,\i*\ys)};
		\draw[] plot[mark=*, mark size = \markSize, mark options={color=gray}] coordinates {(5*\xs,\i*\ys)};
		\draw[] plot[mark=*, mark size = \markSize, mark options={color=gray}] coordinates {(10*\xs,\i*\ys)};
		%	\draw[] plot[mark=*, mark size = \markSize, mark options={color=black}] coordinates {(10*\xs,\i*\ys)};
		
		%		\node[text width=0.5cm] at (-0.5*\xs,\i*\ys) {\footnotesize{$u_{\bIndex}$}};
		%		\node[text width=0.5cm] at (11.25*\xs,\i*\ys) {\footnotesize{$x_{\bIndex}$}};
	}
	
	\node[text width=0.5cm] at (-0.5*\xs,0*\ys) {\footnotesize{$u_{7}$}};
	\node[text width=0.5cm] at (-0.5*\xs,1*\ys) {\footnotesize{$u_{6}$}};
	\node[text width=0.5cm] at (-0.5*\xs,2*\ys) {\footnotesize{$u_{3}$}};
	\node[text width=0.5cm] at (-0.5*\xs,3*\ys) {\footnotesize{$u_{2}$}};
	\node[text width=0.5cm] at (-0.5*\xs,4*\ys) {\footnotesize{$u_{5}$}};
	\node[text width=0.5cm] at (-0.5*\xs,5*\ys) {\footnotesize{$u_{4}$}};
	\node[text width=0.5cm] at (-0.5*\xs,6*\ys) {\footnotesize{$u_{1}$}};
	\node[text width=0.5cm] at (-0.5*\xs,7*\ys) {\footnotesize{$u_{0}$}};
	
	\node[text width=0.5cm] at (11.25*\xs,0*\ys) {\footnotesize{$x_{7}$}};
	\node[text width=0.5cm] at (11.25*\xs,1*\ys) {\footnotesize{$x_{6}$}};
	\node[text width=0.5cm] at (11.25*\xs,2*\ys) {\footnotesize{$x_{3}$}};
	\node[text width=0.5cm] at (11.25*\xs,3*\ys) {\footnotesize{$x_{2}$}};
	\node[text width=0.5cm] at (11.25*\xs,4*\ys) {\footnotesize{$x_{5}$}};
	\node[text width=0.5cm] at (11.25*\xs,5*\ys) {\footnotesize{$x_{4}$}};
	\node[text width=0.5cm] at (11.25*\xs,6*\ys) {\footnotesize{$x_{1}$}};
	\node[text width=0.5cm] at (11.25*\xs,7*\ys) {\footnotesize{$x_{0}$}};
	
	%% PEs
	% Stage 0
	\draw[] plot[mark=oplus, mark size = \PEmarkSize, mark options={fill=white}] coordinates {(\xs,\ys)} --
	plot[mark=*, mark size = \PEmarkSize, mark options={fill=white}] coordinates {(\xs,0)} node[above=-0.19]{\scriptsize{=}};
	
	\draw[] plot[mark=oplus, mark size = \PEmarkSize, mark options={fill=white}] coordinates {(\xs,3*\ys)} --
	plot[mark=*, mark size = \PEmarkSize, mark options={fill=white}] coordinates {(\xs,2*\ys)} node[above=-0.19]{\scriptsize{=}};
	
	\draw[] plot[mark=oplus, mark size = \PEmarkSize, mark options={fill=white}] coordinates {(\xs,5*\ys)} --
	plot[mark=*, mark size = \PEmarkSize, mark options={fill=white}] coordinates {(\xs,4*\ys)} node[above=-0.19]{\scriptsize{=}};
	
	\draw[] plot[mark=oplus, mark size = \PEmarkSize, mark options={fill=white}] coordinates {(\xs,7*\ys)} --
	plot[mark=*, mark size = \PEmarkSize, mark options={fill=white}] coordinates {(\xs,6*\ys)} node[above=-0.19]{\scriptsize{=}};

	% Stage 1
	\draw[] plot[mark=oplus, mark size = \PEmarkSize, mark options={fill=white}] coordinates {(3*\xs,3*\ys)} --
	plot[mark=*, mark size = \PEmarkSize, mark options={fill=white}] coordinates {(3*\xs,\ys)} node[above=-0.19]{\scriptsize{=}};
	
	\draw[] plot[mark=oplus, mark size = \PEmarkSize, mark options={fill=white}] coordinates {(3*\xs,7*\ys)} --
	plot[mark=*, mark size = \PEmarkSize, mark options={fill=white}] coordinates {(3*\xs,5*\ys)} node[above=-0.19]{\scriptsize{=}};
	
	\draw[] plot[mark=oplus, mark size = \PEmarkSize, mark options={fill=white}] coordinates {(4*\xs,2*\ys)} --
	plot[mark=*, mark size = \PEmarkSize, mark options={fill=white}] coordinates {(4*\xs,0)} node[above=-0.19]{\scriptsize{=}};
	
	\draw[] plot[mark=oplus, mark size = \PEmarkSize, mark options={fill=white}] coordinates {(4*\xs,6*\ys)} --
	plot[mark=*, mark size = \PEmarkSize, mark options={fill=white}] coordinates {(4*\xs,4*\ys)} node[above=-0.19]{\scriptsize{=}};
	
	% Stage 2
	\draw[] plot[mark=oplus, mark size = \PEmarkSize, mark options={fill=white}] coordinates {(9*\xs,4*\ys)} --
	plot[mark=*, mark size = \PEmarkSize, mark options={fill=white}] coordinates {(9*\xs,0*\ys)} node[above=-0.19]{\scriptsize{=}};
	
	\draw[] plot[mark=oplus, mark size = \PEmarkSize, mark options={fill=white}] coordinates {(8*\xs,5*\ys)} --
	plot[mark=*, mark size = \PEmarkSize, mark options={fill=white}] coordinates {(8*\xs,1*\ys)} node[above=-0.19]{\scriptsize{=}};
	
	\draw[] plot[mark=oplus, mark size = \PEmarkSize, mark options={fill=white}] coordinates {(7*\xs,6*\ys)} --
	plot[mark=*, mark size = \PEmarkSize, mark options={fill=white}] coordinates {(7*\xs,2*\ys)} node[above=-0.19]{\scriptsize{=}};
	
	\draw[] plot[mark=oplus, mark size = \PEmarkSize, mark options={fill=white}] coordinates {(6*\xs,7*\ys)} --
	plot[mark=*, mark size = \PEmarkSize, mark options={fill=white}] coordinates {(6*\xs,3*\ys)} node[above=-0.19]{\scriptsize{=}};
	
	\draw[rounded corners, dashdotted] (0.5*\xs, -0.5*\ys) rectangle (1.5*\xs, 7.5*\ys) {};
	\draw[rounded corners, dashdotted] (2.5*\xs, -0.5*\ys) rectangle (4.5*\xs, 7.5*\ys) {};
	\draw[rounded corners, dashdotted] (5.5*\xs, -0.5*\ys) rectangle (9.5*\xs, 7.5*\ys) {};
	
	\node[text width=1.25cm, fill=none] at (2.9*\xs, -1.25*\ys) {\footnotesize{$0$}};
	\node[text width=1.25cm, fill=none] at (5.4*\xs, -1.25*\ys) {\footnotesize{$1$}};
	\node[text width=1.25cm, fill=none] at (9.25*\xs, -1.25*\ys) {\footnotesize{$2$}};
	
	\node[text width=3cm, fill=none] at (5.5*\xs, 8.5*\ys) {\footnotesize{Codeword Permutation}};
	
\end{tikzpicture}

%% file: PermutedPolarFactorGraph.tikz.tex
\begin{tikzpicture}[scale=0.9]
	\footnotesize	
	\def\N{8}
	\def\xM{5}
	\def\xss{\xM/\N}
	\def\xs{\xss/1}
	\def\ys{0.4}
	\def\gain{1.0}
	\def\markSize{2}
	\def\markSizeGain{1.125}
	\def\PEmarkSize{3.5}
	
	%% Stage lines
	\draw[dashed] (0*\xs,0*\ys) -- (0*\xs,7.7*\ys*\gain) node[above]{\footnotesize{$s_0$}};
	\draw[dashed] (1.75*\xs,0*\ys) -- (1.75*\xs,7.7*\ys*\gain) node[above]{\footnotesize{$s_1$}};
	\draw[dashed] (4*\xs,0*\ys) -- (4*\xs,7.7*\ys*\gain) node[above]{\footnotesize{$s_2$}};
	\draw[dashed] (8.5*\xs,0*\ys) -- (8.5*\xs,7.7*\ys*\gain) node[above]{\footnotesize{$s_3$}};

	%% Permutation blocks
	
	\node[text width=0.5cm] at (-0.5*\xs,0*\ys) {\footnotesize{$u_{7}$}};
	\node[text width=0.5cm] at (-0.5*\xs,1*\ys) {\footnotesize{$u_{5}$}};
	\node[text width=0.5cm] at (-0.5*\xs,2*\ys) {\footnotesize{$u_{3}$}};
	\node[text width=0.5cm] at (-0.5*\xs,3*\ys) {\footnotesize{$u_{1}$}};
	\node[text width=0.5cm] at (-0.5*\xs,4*\ys) {\footnotesize{$u_{6}$}};
	\node[text width=0.5cm] at (-0.5*\xs,5*\ys) {\footnotesize{$u_{2}$}};
	\node[text width=0.5cm] at (-0.5*\xs,6*\ys) {\footnotesize{$u_{4}$}};
	\node[text width=0.5cm] at (-0.5*\xs,7*\ys) {\footnotesize{$u_{0}$}};

	% Block 1 stage 2
	\draw[] (4*\xs,4*\ys) -- (4.5*\xs,4*\ys);
	\draw[] (4*\xs,5*\ys) -- (4.5*\xs,6*\ys);
	\draw[] (4*\xs,6*\ys) -- (4.5*\xs,5*\ys);
	\draw[] (4*\xs,7*\ys) -- (4.5*\xs,7*\ys);
	
	% Block 2 stage 2
	\draw[] (4*\xs,0*\ys) -- (4.5*\xs,0*\ys);
	\draw[] (4*\xs,1*\ys) -- (4.5*\xs,1*\ys);
	\draw[] (4*\xs,2*\ys) -- (4.5*\xs,2*\ys);
	\draw[] (4*\xs,3*\ys) -- (4.5*\xs,3*\ys);
	
	\node[text width=0.5cm] at (4*\xs,0.28*\ys) {\scriptsize{$7$}};
	\node[text width=0.5cm] at (4*\xs,1.28*\ys) {\scriptsize{$5$}};
	\node[text width=0.5cm] at (4*\xs,2.28*\ys) {\scriptsize{$3$}};
	\node[text width=0.5cm] at (4*\xs,3.28*\ys) {\scriptsize{$1$}};
	\node[text width=0.5cm] at (4*\xs,4.28*\ys) {\scriptsize{$6$}};
	\node[text width=0.5cm] at (4*\xs,5.28*\ys) {\scriptsize{$2$}};
	\node[text width=0.5cm] at (4*\xs,6.28*\ys) {\scriptsize{$4$}};
	\node[text width=0.5cm] at (4*\xs,7.28*\ys) {\scriptsize{$0$}};
	
	\node[text width=0.5cm] at (5.1*\xs,0.28*\ys) {\scriptsize{$7$}};
	\node[text width=0.5cm] at (5.1*\xs,1.28*\ys) {\scriptsize{$5$}};
	\node[text width=0.5cm] at (5.1*\xs,2.28*\ys) {\scriptsize{$3$}};
	\node[text width=0.5cm] at (5.1*\xs,3.28*\ys) {\scriptsize{$1$}};
	\node[text width=0.5cm] at (5.1*\xs,4.28*\ys) {\scriptsize{$6$}};
	\node[text width=0.5cm] at (5.1*\xs,5.28*\ys) {\scriptsize{$4$}};
	\node[text width=0.5cm] at (5.1*\xs,6.28*\ys) {\scriptsize{$2$}};
	\node[text width=0.5cm] at (5.1*\xs,7.28*\ys) {\scriptsize{$0$}};
	
	% Block 1 stage 3
	\draw[] (8.5*\xs,0*\ys) -- (9*\xs,0*\ys);
	\draw[] (8.5*\xs,1*\ys) -- (9*\xs,2*\ys);
	\draw[] (8.5*\xs,2*\ys) -- (9*\xs,4*\ys);
	\draw[] (8.5*\xs,3*\ys) -- (9*\xs,6*\ys);
	\draw[] (8.5*\xs,4*\ys) -- (9*\xs,1*\ys);
	\draw[] (8.5*\xs,5*\ys) -- (9*\xs,3*\ys);
	\draw[] (8.5*\xs,6*\ys) -- (9*\xs,5*\ys);
	\draw[] (8.5*\xs,7*\ys) -- (9*\xs,7*\ys);
	
	\node[text width=0.5cm] at (8.5*\xs,0.28*\ys) {\scriptsize{$7$}};
	\node[text width=0.5cm] at (8.5*\xs,1.28*\ys) {\scriptsize{$5$}};
	\node[text width=0.5cm] at (8.5*\xs,2.28*\ys) {\scriptsize{$3$}};
	\node[text width=0.5cm] at (8.5*\xs,3.28*\ys) {\scriptsize{$1$}};
	\node[text width=0.5cm] at (8.5*\xs,4.28*\ys) {\scriptsize{$6$}};
	\node[text width=0.5cm] at (8.5*\xs,5.28*\ys) {\scriptsize{$4$}};
	\node[text width=0.5cm] at (8.5*\xs,6.28*\ys) {\scriptsize{$2$}};
	\node[text width=0.5cm] at (8.5*\xs,7.28*\ys) {\scriptsize{$0$}};
	
	\node[text width=0.5cm] at (9.6*\xs,0.28*\ys) {\scriptsize{$7$}};
	\node[text width=0.5cm] at (9.6*\xs,1.28*\ys) {\scriptsize{$6$}};
	\node[text width=0.5cm] at (9.6*\xs,2.28*\ys) {\scriptsize{$5$}};
	\node[text width=0.5cm] at (9.6*\xs,3.28*\ys) {\scriptsize{$4$}};
	\node[text width=0.5cm] at (9.6*\xs,4.28*\ys) {\scriptsize{$3$}};
	\node[text width=0.5cm] at (9.6*\xs,5.28*\ys) {\scriptsize{$2$}};
	\node[text width=0.5cm] at (9.6*\xs,6.28*\ys) {\scriptsize{$1$}};
	\node[text width=0.5cm] at (9.6*\xs,7.28*\ys) {\scriptsize{$0$}};	
	
	% Node and bit indcies
	\foreach \i in{0,...,7}
	{
		\pgfmathsetmacro\bIndex{int(7-\i)};
		\draw[] (0,\i*\ys) -- (4*\xs,\i*\ys);
		\draw[] (4.5*\xs,\i*\ys) -- (8.5*\xs,\i*\ys);
		\draw[] (9*\xs,\i*\ys) -- (10*\xs,\i*\ys);
			
		\draw[] plot[mark=*, mark size = \markSizeGain*\markSize, mark options={fill=white}] coordinates {(0*\xs,\i*\ys)};
		
		\ifthenelse{\i = 4}{\draw[] plot[mark=*, mark size = \markSizeGain*\markSize, mark options={fill=black}] coordinates {(0*\xs,\i*\ys)};}
		
		\ifthenelse{\i = 8}{\draw[] plot[mark=*, mark size = \markSizeGain*\markSize, mark options={fill=black}] coordinates {(0*\xs,\i*\ys)};}
		
		\ifthenelse{\i < 3}{\draw[] plot[mark=*, mark size = \markSizeGain*\markSize, mark options={fill=black}] coordinates {(0*\xs,\i*\ys)};}
		
		\draw[] plot[mark=*, mark size = \markSize, mark options={color=gray}] coordinates {(1.75*\xs,\i*\ys)};
		\draw[] plot[mark=*, mark size = \markSize, mark options={color=gray}] coordinates {(4*\xs,\i*\ys)};
		\draw[] plot[mark=*, mark size = \markSize, mark options={color=gray}] coordinates {(4.5*\xs,\i*\ys)};
		\draw[] plot[mark=*, mark size = \markSize, mark options={color=gray}] coordinates {(8.5*\xs,\i*\ys)};
		\draw[] plot[mark=*, mark size = \markSize, mark options={color=gray}] coordinates {(9*\xs,\i*\ys)};
		\draw[] plot[mark=*, mark size = \markSize, mark options={color=gray}] coordinates {(10*\xs,\i*\ys)};
	
%		\node[text width=0.5cm] at (-1*\xs,\i*\ys) {\footnotesize{$u_{\bIndex}$}};
		\node[text width=0.5cm] at (10.75*\xs,\i*\ys) {\footnotesize{$x_{\bIndex}$}};
	}
	
	%% PEs
	% Stage 0
	\draw[] plot[mark=oplus, mark size = \PEmarkSize, mark options={fill=white}] coordinates {(0.9*\xs,\ys)} --
	plot[mark=*, mark size = \PEmarkSize, mark options={fill=white}] coordinates {(0.9*\xs,0)} node[above=-0.19]{\scriptsize{=}};
	
	\draw[] plot[mark=oplus, mark size = \PEmarkSize, mark options={fill=white}] coordinates {(0.9*\xs,3*\ys)} --
	plot[mark=*, mark size = \PEmarkSize, mark options={fill=white}] coordinates {(0.9*\xs,2*\ys)} node[above=-0.19]{\scriptsize{=}};
	
	\draw[] plot[mark=oplus, mark size = \PEmarkSize, mark options={fill=white}] coordinates {(0.9*\xs,5*\ys)} --
	plot[mark=*, mark size = \PEmarkSize, mark options={fill=white}] coordinates {(0.9*\xs,4*\ys)} node[above=-0.19]{\scriptsize{=}};
	
	\draw[] plot[mark=oplus, mark size = \PEmarkSize, mark options={fill=white}] coordinates {(0.9*\xs,7*\ys)} --
	plot[mark=*, mark size = \PEmarkSize, mark options={fill=white}] coordinates {(0.9*\xs,6*\ys)} node[above=-0.19]{\scriptsize{=}};
	
	% Stage 1
	\draw[] plot[mark=oplus, mark size = \PEmarkSize, mark options={fill=white}] coordinates {(2.5*\xs,3*\ys)} --
	plot[mark=*, mark size = \PEmarkSize, mark options={fill=white}] coordinates {(2.5*\xs,\ys)} node[above=-0.19]{\scriptsize{=}};
	
	\draw[] plot[mark=oplus, mark size = \PEmarkSize, mark options={fill=white}] coordinates {(2.5*\xs,7*\ys)} --
	plot[mark=*, mark size = \PEmarkSize, mark options={fill=white}] coordinates {(2.5*\xs,5*\ys)} node[above=-0.19]{\scriptsize{=}};
	
	\draw[] plot[mark=oplus, mark size = \PEmarkSize, mark options={fill=white}] coordinates {(3*\xs,2*\ys)} --
	plot[mark=*, mark size = \PEmarkSize, mark options={fill=white}] coordinates {(3*\xs,0)} node[above=-0.19]{\scriptsize{=}};
	
	\draw[] plot[mark=oplus, mark size = \PEmarkSize, mark options={fill=white}] coordinates {(3*\xs,6*\ys)} --
	plot[mark=*, mark size = \PEmarkSize, mark options={fill=white}] coordinates {(3*\xs,4*\ys)} node[above=-0.19]{\scriptsize{=}};
		
	% Stage 2
	\draw[] plot[mark=oplus, mark size = \PEmarkSize, mark options={fill=white}] coordinates {(7.25*\xs,4*\ys)} --
	plot[mark=*, mark size = \PEmarkSize, mark options={fill=white}] coordinates {(7.25*\xs,0*\ys)} node[above=-0.19]{\scriptsize{=}};
	
	\draw[] plot[mark=oplus, mark size = \PEmarkSize, mark options={fill=white}] coordinates {(6.75*\xs,5*\ys)} --
	plot[mark=*, mark size = \PEmarkSize, mark options={fill=white}] coordinates {(6.75*\xs,1*\ys)} node[above=-0.19]{\scriptsize{=}};
	
	\draw[] plot[mark=oplus, mark size = \PEmarkSize, mark options={fill=white}] coordinates {(6.25*\xs,6*\ys)} --
	plot[mark=*, mark size = \PEmarkSize, mark options={fill=white}] coordinates {(6.25*\xs,2*\ys)} node[above=-0.19]{\scriptsize{=}};
	
	\draw[] plot[mark=oplus, mark size = \PEmarkSize, mark options={fill=white}] coordinates {(5.75*\xs,7*\ys)} --
	plot[mark=*, mark size = \PEmarkSize, mark options={fill=white}] coordinates {(5.75*\xs,3*\ys)} node[above=-0.19]{\scriptsize{=}};
	
	%% Final text
%	\node[text width=2cm] at (5*\xs,9*\ys) {\small{Stage Indices}};
%	\node[text width=2cm, rotate=90] at (-1.5*\xs,4*\ys) {\small{Bit Indices}};
	
\end{tikzpicture}

%% file: RLD.tikz.tex
\usetikzlibrary{arrows, decorations}

% for double arrows a la chef
% adapt line thickness and line width, if needed
\tikzstyle{vecArrow} = [thick, decoration={markings,mark=at position
	1 with {\arrow[semithick]{open triangle 60}}},
double distance=1.4pt, shorten >= 5.5pt,
preaction = {decorate},
postaction = {draw,line width=1.4pt, white,shorten >= 4.5pt}]
\tikzstyle{innerWhite} = [semithick, white,line width=1.4pt, shorten >= 4.5pt]

\begin{tikzpicture}[scale=1.0]
\footnotesize
\def\N{16}
\def\xM{7.5}
\def\xss{\xM/\N}
\def\xs{\xss/1}
\def\ys{0.4}
\def\Ygain{1.075}
\def\Xgain{0.8}
\def\markSize{2.5}
\def\PEmarkSize{3.8}

%\draw[dashed, color=gray] (15*\xs*\Xgain,-2*\ys) -- (15*\xs*\Xgain,7*\ys*\Ygain) node[above]{\footnotesize{$s_1$}};
%\draw[dashed, color=gray] (17*\xs*\Xgain,-2*\ys) -- (17*\xs*\Xgain,7*\ys*\Ygain) node[above]{\footnotesize{$s_2$}};
%\draw[dashed, color=gray] (19*\xs*\Xgain,-2*\ys) -- (19*\xs*\Xgain,7*\ys*\Ygain) node[above]{\footnotesize{$s_3$}};

\draw[] (19*\xs*\Xgain,3.5*\ys) -- (17*\xs*\Xgain,1.5*\ys);
\draw[] (19*\xs*\Xgain,3.5*\ys) -- (17*\xs*\Xgain,5.5*\ys);
\draw[] plot[mark=*, mark size = \markSize, mark options={fill=gray}] coordinates {(19*\xs*\Xgain,3.5*\ys)};

\draw[] plot[mark=square*, mark size = \markSize, mark options={fill=black}] coordinates {(17*\xs*\Xgain,1.5*\ys)};

\draw[] plot[mark=triangle*, mark size = \markSize*1.25, mark options={fill=black}] coordinates {(17*\xs*\Xgain,4*\ys+1.5*\ys)};

\draw[] (15*\xs*\Xgain,3*\ys) -- (17*\xs*\Xgain,1.5*\ys);
\draw[] plot[mark=triangle*, mark size = 1.25*\markSize, mark options={fill=white}] coordinates {(15*\xs*\Xgain,3*\ys)};

\draw[] (15*\xs*\Xgain,0*\ys) -- (17*\xs*\Xgain,1.5*\ys);
\draw[] plot[mark=diamond*, mark size = 1.25*\markSize, mark options={fill=white}] coordinates {(15*\xs*\Xgain,0*\ys)};

\node[text width=0cm] at (13*\xs*\Xgain,4*\ys) {\footnotesize{$\mathcal{RM}(0,1)$}};
\node[text width=0cm] at (13*\xs*\Xgain,-1*\ys) {\footnotesize{$\mathcal{RM}(1,1)$}};

\node[text width=0cm] at (17.5*\xs*\Xgain,6*\ys) {\footnotesize{$\mathcal{RM}(0,2)$}};
\node[text width=0cm] at (17.5*\xs*\Xgain,1*\ys) {\footnotesize{$\mathcal{RM}(1,2)$}};
\node[text width=0cm] at (19.5*\xs*\Xgain,3.4*\ys) {\footnotesize{$\mathcal{RM}(1,3)$}};
\end{tikzpicture}

%% file: FSCL.tikz.tex
\usetikzlibrary{arrows, decorations}

% for double arrows a la chef
% adapt line thickness and line width, if needed
\tikzstyle{vecArrow} = [thick, decoration={markings,mark=at position
	1 with {\arrow[semithick]{open triangle 60}}},
double distance=1.4pt, shorten >= 5.5pt,
preaction = {decorate},
postaction = {draw,line width=1.4pt, white,shorten >= 4.5pt}]
\tikzstyle{innerWhite} = [semithick, white,line width=1.4pt, shorten >= 4.5pt]

\begin{tikzpicture}[scale=1.0]
\footnotesize
\def\N{16}
\def\xM{7.5}
\def\xss{\xM/\N}
\def\xs{\xss/1}
\def\ys{0.4}
\def\Ygain{1.075}
\def\Xgain{0.8}
\def\markSize{2.5}
\def\PEmarkSize{3.8}

%\draw[dashed, color=gray] (17*\xs*\Xgain,-2*\ys) -- (17*\xs*\Xgain,7*\ys*\Ygain) node[above]{{\footnotesize{$s_2$}}};
%\draw[dashed, color=gray] (19*\xs*\Xgain,-2*\ys) -- (19*\xs*\Xgain,7*\ys*\Ygain) node[above]{{\footnotesize{$s_3$}}};

\draw[] (19*\xs*\Xgain,3.5*\ys) -- (17*\xs*\Xgain,1*\ys);
\draw[] (19*\xs*\Xgain,3.5*\ys) -- (17*\xs*\Xgain,6*\ys);
\draw[] plot[mark=*, mark size = \markSize, mark options={fill=gray}] coordinates {(19*\xs*\Xgain,3.5*\ys)};

\draw[] plot[mark=square*, mark size = \markSize, mark options={fill=black}] coordinates {(17*\xs*\Xgain,1*\ys)};

\draw[] plot[mark=triangle*, mark size = \markSize*1.25, mark options={fill=black}] coordinates {(17*\xs*\Xgain,6*\ys)};

\node[text width=0cm] at (15*\xs*\Xgain,6.8*\ys) {\footnotesize{$\mathcal{RM}(0,2)$}};
\node[text width=0cm] at (15*\xs*\Xgain,-0.2*\ys) {\footnotesize{$\mathcal{RM}(1,2)$}};
\node[text width=0cm] at (19*\xs*\Xgain,2.5*\ys) {\footnotesize{$\mathcal{RM}(1,3)$}};

\end{tikzpicture}

%% file: example_FSC.tikz.tex
\usetikzlibrary{patterns}
\usetikzlibrary{arrows, decorations, patterns}

% for double arrows a la chef
% adapt line thickness and line width, if needed
\tikzstyle{vecArrow} = [thick, decoration={markings,mark=at position
	1 with {\arrow[semithick]{open triangle 60}}},
double distance=1.4pt, shorten >= 5.5pt,
preaction = {decorate},
postaction = {draw,line width=1.4pt, white,shorten >= 4.5pt}]
\tikzstyle{innerWhite} = [semithick, white,line width=1.4pt, shorten >= 4.5pt]

\begin{tikzpicture}[scale=0.7]
	\footnotesize	
	\def\N{16}
	\def\xM{7.5}
	\def\xss{\xM/\N}
	\def\xs{\xss/1}
	\def\ys{0.4}
	\def\Ygain{1.075}
	\def\Xgain{0.8}
	\def\markSize{3}
	\def\PEmarkSize{3.8}

	% Stage 1
	\draw (6.125*\xs,2*\ys) circle [x radius=1.1, y radius=0.5, rotate=0];	
	\node[text width=1.3cm, align=center] at (6.125*\xs,2*\ys) {\footnotesize{$\mathcal{RM}(1,3)$}};
	\node[text width=1.3cm, draw=black, align=center] at (6.125*\xs,-2*\ys) {\footnotesize{$\mathcal{RM}(2,3)$}};
	
	% Stage 2
	\draw[line width=1.5pt, color=green] (8.5*\xs,2*\ys)--(9.75*\xs,0*\ys);
	\draw[line width=1.5pt, color=gray] (8.5*\xs,-2*\ys)--(9.75*\xs,0*\ys);
	\draw[preaction={fill=red}, draw=blue] (9.75*\xs,-1.5*\ys) rectangle (10.25*\xs,1.5*\ys);
	\node[text width=1.22cm] at (12.5*\xs,0*\ys) {\footnotesize{$\mathcal{RM}(2, 4)$}};
	\node[text width=1.3cm, draw=black] at (12.45*\xs,-4*\ys) {\footnotesize{$\mathcal{RM}(3, 4)$}};
	
	% Stage 3
	\draw[line width=1.5pt, color=green] (14.75*\xs,0*\ys)--(16*\xs,-2*\ys);
	\draw[line width=1.5pt, color=gray] (14.75*\xs,-4*\ys)--(16*\xs,-2*\ys);

	\draw[preaction={fill=cyan}, draw=blue] (16*\xs,-3.5*\ys) rectangle (16.5*\xs,-0.5*\ys);
	\node[text width=1.22cm] at (19.5*\xs,-2*\ys) {\footnotesize{$\mathcal{RM}(3, 5)$}};
	
	%Stage 4
	\draw[preaction={fill=orange}, pattern=north east lines, pattern color=blue, draw=blue] (16.5*\xs,-3.5*\ys) rectangle (17*\xs,-0.5*\ys);
		
	\path[->] (10*\xs,1.5*\ys) edge [bend right] (11.25*\xs,5*\ys);
	\node[text width=6cm, font=\footnotesize\linespread{0.8}\selectfont, align=center] at (11.25*\xs,6*\ys) {Permutation selection with FHT\\for each decoding path};
	
	\path[->] (16.25*\xs,-0.5*\ys) edge [bend left] (18*\xs,2*\ys);
	\node[text width=6cm, font=\footnotesize\linespread{0.8}\selectfont, align=center] at (22*\xs,3*\ys) {Permutation selection according to (\ref{equ:SP:FPM:Rate-1})\\for each decoding path};
	
	\path[->] (16.75*\xs,-3.5*\ys) edge [bend right] (19*\xs, -6*\ys);
	\node[text width=6cm, font=\footnotesize\linespread{0.8}\selectfont, align=center] at (22*\xs,-6*\ys) {Select $L$ random\\permutations};
	
%	\node[text width=1.6cm, align=center, draw=black] at (6.15*\xs,-8*\ys) {\footnotesize{$\mathcal{RM}(r,m)$}};
%	\node[text width=6 cm, align=left, font=\footnotesize] at (18.5*\xs,-8*\ys) {Apply SPC-List decoding to $\mathcal{RM}(m-1,m)$};
%	
%	\draw (6.15*\xs,-11*\ys) circle [x radius=1.4, y radius=0.5, rotate=0];
%	\node[text width=1.35cm, align=center] at (6.15*\xs,-11*\ys) {\footnotesize{$\mathcal{RM}(1,m)$}};
%	\node[text width=5 cm, align=left, font=\footnotesize] at (17*\xs,-11*\ys) {Apply FHT-List decoding to $\mathcal{RM}(1,m)$};
%	
%	\draw[line width=1.5pt, color=green] (4.75*\xs,-13.5*\ys)--(7.5*\xs,-13.5*\ys);
%	\node[text width=5 cm, align=left, font=\footnotesize] at (17*\xs,-13.5*\ys) {$f$ functions};
%	
%	\draw[line width=1.5pt, color=gray] (4.75*\xs,-15*\ys)--(7.5*\xs,-15*\ys);
%	\node[text width=5 cm, align=left, font=\footnotesize] at (17*\xs,-15*\ys) {$g$ functions};
	
\end{tikzpicture}

%% file: example_FSC_simple.tikz.tex
\usetikzlibrary{patterns}
\usetikzlibrary{arrows, decorations, patterns}

% for double arrows a la chef
% adapt line thickness and line width, if needed
\tikzstyle{vecArrow} = [thick, decoration={markings,mark=at position
	1 with {\arrow[semithick]{open triangle 60}}},
double distance=1.4pt, shorten >= 5.5pt,
preaction = {decorate},
postaction = {draw,line width=1.4pt, white,shorten >= 4.5pt}]
\tikzstyle{innerWhite} = [semithick, white,line width=1.4pt, shorten >= 4.5pt]

\begin{tikzpicture}[scale=0.7]
\footnotesize	
\def\N{16}
\def\xM{7.5}
\def\xss{\xM/\N}
\def\xs{\xss/1}
\def\ys{0.4}
\def\Ygain{1.075}
\def\Xgain{0.8}
\def\markSize{3}
\def\PEmarkSize{3.8}

% Stage 1
\draw (6.125*\xs,2*\ys) circle [x radius=1.1, y radius=0.5, rotate=0];	
\node[text width=1.3cm, align=center] at (6.125*\xs,2*\ys) {\footnotesize{$\mathcal{RM}(1,3)$}};
\node[text width=1.3cm, draw=black, align=center] at (6.125*\xs,-2*\ys) {\footnotesize{$\mathcal{RM}(2,3)$}};

% Stage 2
\draw[line width=1.5pt, color=green] (8.5*\xs,2*\ys)--(9.75*\xs,0*\ys);
\draw[line width=1.5pt, color=gray] (8.5*\xs,-2*\ys)--(9.75*\xs,0*\ys);
%\draw[preaction={fill=red}, draw=blue] (9.75*\xs,-1.5*\ys) rectangle (10.25*\xs,1.5*\ys);
\node[text width=1.22cm] at (12.25*\xs,0*\ys) {\footnotesize{$\mathcal{RM}(2, 4)$}};
\node[text width=1.3cm, draw=black] at (12.45*\xs,-4*\ys) {\footnotesize{$\mathcal{RM}(3, 4)$}};

% Stage 3
\draw[line width=1.5pt, color=green] (14.75*\xs,0*\ys)--(16*\xs,-2*\ys);
\draw[line width=1.5pt, color=gray] (14.75*\xs,-4*\ys)--(16*\xs,-2*\ys);

\draw[preaction={fill=cyan}, draw=blue] (16*\xs,-3.5*\ys) rectangle (16.5*\xs,-0.5*\ys);
\node[text width=1.22cm] at (19.5*\xs,-2*\ys) {\footnotesize{$\mathcal{RM}(3, 5)$}};

%Stage 4
\draw[preaction={fill=orange}, pattern=north east lines, pattern color=blue, draw=blue] (16.5*\xs,-3.5*\ys) rectangle (17*\xs,-0.5*\ys);

%\path[->] (10*\xs,1.5*\ys) edge [bend right] (11.25*\xs,5*\ys);
\node[text width=3.8cm, font=\footnotesize\linespread{0.8}\selectfont, align=center, color=black] at (8*\xs,6*\ys) {};

\path[->] (16.25*\xs,-0.5*\ys) edge [bend left] (18*\xs,2*\ys);
\node[text width=6cm, font=\footnotesize\linespread{0.8}\selectfont, align=center] at (22*\xs,3*\ys) {Permutation selection according to (\ref{equ:SP:FPM:Rate-1})\\for each decoding path};

\path[->] (16.75*\xs,-3.5*\ys) edge [bend right] (19*\xs, -6*\ys);
\node[text width=6cm, font=\footnotesize\linespread{0.8}\selectfont, align=center] at (22*\xs,-6*\ys) {Select $L$ random\\permutations};

%	\node[text width=1.6cm, align=center, draw=black] at (6.15*\xs,-8*\ys) {\footnotesize{$\mathcal{RM}(r,m)$}};
%	\node[text width=6 cm, align=left, font=\footnotesize] at (18.5*\xs,-8*\ys) {Apply SPC-List decoding to $\mathcal{RM}(m-1,m)$};
%	
%	\draw (6.15*\xs,-11*\ys) circle [x radius=1.4, y radius=0.5, rotate=0];
%	\node[text width=1.35cm, align=center] at (6.15*\xs,-11*\ys) {\footnotesize{$\mathcal{RM}(1,m)$}};
%	\node[text width=5 cm, align=left, font=\footnotesize] at (17*\xs,-11*\ys) {Apply FHT-List decoding to $\mathcal{RM}(1,m)$};
%	
%	\draw[line width=1.5pt, color=green] (4.75*\xs,-13.5*\ys)--(7.5*\xs,-13.5*\ys);
%	\node[text width=5 cm, align=left, font=\footnotesize] at (17*\xs,-13.5*\ys) {$f$ functions};
%	
%	\draw[line width=1.5pt, color=gray] (4.75*\xs,-15*\ys)--(7.5*\xs,-15*\ys);
%	\node[text width=5 cm, align=left, font=\footnotesize] at (17*\xs,-15*\ys) {$g$ functions};

\end{tikzpicture}

%% file: example_FSC_legend.tikz.tex
\usetikzlibrary{patterns}
\usetikzlibrary{arrows, decorations, patterns}

% for double arrows a la chef
% adapt line thickness and line width, if needed
\tikzstyle{vecArrow} = [thick, decoration={markings,mark=at position
	1 with {\arrow[semithick]{open triangle 60}}},
double distance=1.4pt, shorten >= 5.5pt,
preaction = {decorate},
postaction = {draw,line width=1.4pt, white,shorten >= 4.5pt}]
\tikzstyle{innerWhite} = [semithick, white,line width=1.4pt, shorten >= 4.5pt]

\begin{tikzpicture}[scale=0.7]
	\footnotesize	
	\def\N{16}
	\def\xM{7.5}
	\def\xss{\xM/\N}
	\def\xs{\xss/1}
	\def\ys{0.4}
	\def\Ygain{1.075}
	\def\Xgain{0.8}
	\def\markSize{3}
	\def\PEmarkSize{3.8}

	\node[text width=1.5cm, align=center, draw=black] at (6.15*\xs,-8*\ys) {\footnotesize{$\mathcal{RM}(r,m)$}};
	\node[text width=6.25 cm, align=left, font=\footnotesize] at (18.9*\xs,-8*\ys) {Apply SPC-List decoding to $\mathcal{RM}(m-1,m)$};
	
	\draw (6.125*\xs,-11*\ys) circle [x radius=1.15, y radius=0.5, rotate=0];
	\node[text width=1.5cm, align=center] at (6.15*\xs,-11*\ys) {\footnotesize{$\mathcal{RM}(1,m)$}};
	\node[text width=6 cm, align=left, font=\footnotesize] at (18.6*\xs,-11*\ys) {Apply FHT-List decoding to $\mathcal{RM}(1,m)$};
	
	\draw[line width=1.5pt, color=green] (4.75*\xs,-13.5*\ys)--(7.5*\xs,-13.5*\ys);
	\node[text width=5 cm, align=left, font=\footnotesize] at (17*\xs,-13.5*\ys) {$f$ functions};
	
	\draw[line width=1.5pt, color=gray] (4.75*\xs,-15*\ys)--(7.5*\xs,-15*\ys);
	\node[text width=5 cm, align=left, font=\footnotesize] at (17*\xs,-15*\ys) {$g$ functions};
	
\end{tikzpicture}

%% file: ferRM512_46.tikz.tex
\begin{tikzpicture}[spy using outlines = {rectangle, magnification=2.0, connect spies}]

	\pgfplotsset{
		label style = {font=\fontsize{9pt}{7.2}\selectfont},
		tick label style = {font=\fontsize{7pt}{7.2}\selectfont}
	}
	
	\begin{axis}[
		scale = 1,
		ymode=log,
		xlabel={$E_b/N_0$ [\text{dB}]}, xlabel style={yshift=0.5em},
		ytick={1e-6, 1e-5,1e-4,1e-3,1e-2,1e-1,1e-0},
		xtick={1.5,2,2.5,3,3.5,4},
		ylabel={FER}, ylabel style={yshift=-0.75em},
		grid=both,
		ymajorgrids=true,
		xmajorgrids=true,
		grid style=dashdotted,
		width=0.35\linewidth, height=7cm,
		thin,
		mark size=2,
		% use the following key so the baseline of all ticklabel entries is the same
		% (compare this image to the one from marmot)
		typeset ticklabels with strut,
		% there is one default value for the `legend pos' that is outside the axis
		legend pos=outer north east,
		% (so the legend looks a bit better)
		legend cell align=left,
		% (moved this common key here)
		smooth,
	  legend style={
	  	  legend pos=outer north east,
	    anchor={center},
	    %cells={anchor=west},
	    column sep= 1mm,
	    font=\fontsize{7pt}{7.2}\selectfont,
	  },
	  legend to name=perf-legend-SP-RLD-2-9,
	  legend columns=6,
		]

		\addplot[
			color=blue,
			mark=o,
			thin,
			mark size=2,
			solid,
			mark options={solid},
		]
		table {
%			0	5.57E-01
%			0.5	4.02E-01
			%1	2.56E-01
			%1.5	1.35E-01
			2	6.21E-02
			2.5	2.31E-02
			3	6.12E-03
			3.5	1.49E-03
			4.0	2.33E-04
		};
		\addlegendentry{SP-RLD-$1$}
		
		\addplot[
		color=black,
		mark=square,
		thin,
		mark size=2,
		solid,
		mark options={solid},
		]
		table {
%			0	3.90E-01
%			0.5	2.42E-01
			%1	1.26E-01
			%1.5	5.15E-02
			2	1.66E-02
			2.5	4.09E-03
			3	6.62E-04
			3.5	7.39E-05
		};
		\addlegendentry{SP-RLD-$2$}
		
		\addplot[
			color=red,
			mark=triangle,
			thin,
			mark size=2,
			solid,
			mark options={solid},
		]
		table {
%			0	2.54E-01
%			0.5	1.28E-01
			%1	5.17E-02
			%1.5	1.57E-02
			2	3.80E-03
			2.5	7.11E-04
			3	6.01E-05
		};
		\addlegendentry{SP-RLD-$4$}
		
		\addplot[
		color=brown,
		mark=pentagon,
		thin,
		mark size=2,
		solid,
		mark options={solid},
		]
		table {
%			0	1.63E-01
%			0.5	7.02E-02
			%1	2.33E-02
			%1.5	5.32E-03
			2	1.04E-03
			2.5	9.56E-05
			2.75	2.09E-05
%			3	5.94E-06
		};
		\addlegendentry{SP-RLD-$8$}

		\addplot[
			color=magenta,
			mark=diamond,
			thin,
			mark size=2,
			mark options={solid},
		]
		table {
%			0	1.08E-01
%			0.5	3.95E-02
			%1	1.05E-02
			%1.5	2.49E-03
			2	3.60E-04
			2.5	2.35E-05
		};
		\addlegendentry{SP-RLD-$16$}
		
		\addplot[
		color=green,
		mark=star,
		thin,
		mark size=2,
		mark options={solid},
		]
		table {
%			0	8.24E-02
%			0.5	2.81E-02
			%1	7.25E-03
			%1.5	1.50E-03
			2	1.90E-04
			2.25	4.47094E-05
		};
		\addlegendentry{SP-RLD-$32$}
		
	\addplot[
		color=blue,
		mark=o,
		thin,
		mark size=2,
		dashdotted,
		mark options={solid},
	]
	table {
%		0	5.68E-01
%		0.5	4.09E-01
		%1	2.62E-01
		%1.5	1.43E-01
		2	6.46E-02
		2.5	2.34E-02
		3	6.57E-03
		3.5	1.56E-03
		4.0	2.36E-04
	};
	\addlegendentry{SSP-RLD-$S$-$1$}
	
	\addplot[
	color=black,
	mark=square,
	thin,
	mark size=2,
	dashdotted,
	mark options={solid},
	]
	table {
%		0	3.96E-01
%		0.5	2.44E-01
		%1	1.28E-01
		%1.5	5.22E-02
		2	1.70E-02
		2.5	4.21E-03
		3	6.42E-04
		3.5	7.55E-05
	};
	\addlegendentry{SSP-RLD-$S$-$2$}

	\addplot[
		color=red,
		mark=triangle,
		thin,
		mark size=2,
		dashdotted,
		mark options={solid},
	]
	table {
%		0	2.62E-01
%		0.5	1.36E-01
		%1	5.59E-02
		%1.5	1.71E-02
		2	4.16E-03
		2.5	6.96E-04
		3	6.39E-05
	};
	\addlegendentry{SSP-RLD-$S$-$4$}

		\addplot[
		color=brown,
		mark=pentagon,
		thin,
		mark size=2,
		dashdotted,
		mark options={solid},
		]
		table {
%			0	1.68E-01
%			0.5	7.31E-02
			%1	2.33E-02
			%1.5	6.34E-03
			2	1.11E-03
			2.5	1.11E-04
			2.75	2.32E-05
		};
	\addlegendentry{SSP-RLD-$S$-$8$}
		
		\addplot[
		color=magenta,
		mark=diamond,
		thin,
		dashdotted,
		mark size=2,
		mark options={solid},
		]
		table {
%			0	1.15E-01
%			0.5	4.09E-02
			%1	1.16E-02
			%1.5	2.53E-03
			2	3.76E-04
			2.5	2.52E-05
		};
		\addlegendentry{SSP-RLD-$S$-$16$}
		
		\addplot[
		color=green,
		mark=star,
		thin,
		mark size=2,
		dashdotted,
		mark options={solid},
		]
		table {
			%1.00E+00	0.0076125
			%1.50E+00	0.001575
			2.00E+00	0.0001995
			2.25E+00	4.694487E-05
			
		};
		\addlegendentry{SSP-RLD-$S$-$32$}
		
		\addplot[
			color=blue,
			mark=otimes,
			thin,
			mark size=2,
			solid,
			mark options={solid},
		]
		table {
			%1	2.41E-02
			%1.5	5.74E-03
			2	1.09E-03
			2.5	1.39E-04
			2.75	4.49E-05
		};
		\addlegendentry{SP-RLD-$32$ (cyclic)}
		
%		\addplot[
%			color=black,
%			mark=none,
%			thin,
%			mark size=2,
%			solid,
%			mark options={solid},
%		]
%		table {
%			%1	2.29E-02
%			%1.5	7.32E-03
%			2	1.85E-03
%			2.5	2.95E-04
%			3	3.18E-05
%		};
%		\addlegendentry{RLDP-$64$ \cite{Dumer06}}

		\addplot[
			color=cyan,
			mark=none,
			line width=0.75,
			mark size=2,
			solid,
			mark options={solid},
		]
		table {
			3.0	3.60E-03
			3.5	6.37E-04
			4.0	8.21E-05
		};
		\addlegendentry{FSCL-$32$ \cite{Ali_FSSCL}}
		
		\addplot[
			color=green,
			mark=+,
			thin,
			mark size=2,
			solid,
			mark options={solid},
		]
		table {
			%			2.5	3.16E-02
			3	6.29E-03
			3.5	7.11E-04
			4	3.68E-05
			%S=4096
		};
		\addlegendentry{SCS \cite{niu2012stack, TrifonovScore}}
		
		\addplot[
			color=red,
			mark=x,
			thin,
			mark size=2,
			solid,
			mark options={solid},
		]
		table {
			3	1.56E-3
			3.5	2.51E-4
			4	2.25E-5
		};
		\addlegendentry{SFP-SCL-32}
		
		\node[anchor=north east, fill=white] at (rel axis cs:1,1) {\footnotesize{\makecell{$\mathcal{RM}(2,9)$\\$D=4096$\\$M=32$}}};
	
	\coordinate (spypoint1) at (axis cs:2.5, 1.25e-4);
	\coordinate (magnifyglass1) at (axis cs:1.65, 5e-5);
		
	\end{axis}

%\spy [blue, width=0.75cm, height=1cm] on (spypoint1) in node[fill=white] at (magnifyglass1);

\end{tikzpicture}

%% file: ferRM512_130.tikz.tex
\begin{tikzpicture}[spy using outlines = {rectangle, magnification=2.0, connect spies}]

	\pgfplotsset{
		label style = {font=\fontsize{9pt}{7.2}\selectfont},
		tick label style = {font=\fontsize{7pt}{7.2}\selectfont}
	}
	
	\begin{axis}[
		scale = 1,
		ymode=log,
		xlabel={$E_b/N_0$ [\text{dB}]}, xlabel style={yshift=0.5em},
		ytick={1e-6, 1e-5,1e-4,1e-3,1e-2,1e-1,1e-0},
		xtick={1.5,2,2.5,3,3.5,4,4,4.5},
		ylabel={}, ylabel style={yshift=-0.75em},
		grid=both,
		ymajorgrids=true,
		xmajorgrids=true,
		grid style=dashdotted,
		width=0.35\linewidth, height=7cm,
		thin,
		mark size=2,
		ymax = 1,
		% use the following key so the baseline of all ticklabel entries is the same
		% (compare this image to the one from marmot)
		typeset ticklabels with strut,
		% there is one default value for the `legend pos' that is outside the axis
		legend pos=outer north east,
		% (so the legend looks a bit better)
		legend cell align=left,
		% (moved this common key here)
		smooth,
		legend style={
			legend pos=outer north east,
			anchor={center},
			%cells={anchor=west},
			column sep= 1.5mm,
			font=\fontsize{7pt}{7.2}\selectfont,
		},
		legend to name=perf-legend-SP-RLD-3-9,
		legend columns=6,
		]
		
		\addplot[
			color=blue,
			mark=o,
			thin,
			mark size=2,
			solid,
			mark options={solid},
		]
		table {
			%1	7.82E-01
			%1.5	5.84E-01
			2	3.60E-01
			2.5	1.73E-01
			3	6.02E-02
			3.5	1.61E-02
			4	1.96E-03
			4.5	1.32E-04
		};
		\addlegendentry{SP-RLD-$1$}
		
		\addplot[
			color=black,
			mark=square,
			thin,
			mark size=2,
			solid,
			mark options={solid},
		]
		table {
			%1	6.26E-01
			%1.5	3.81E-01
			2	1.78E-01
			2.5	5.50E-02
			3	1.07E-02
			3.5	9.46E-04
			4	5.13E-05
		};
		\addlegendentry{SP-RLD-$2$}
		
		\addplot[
		color=red,
		mark=triangle,
		thin,
		mark size=2,
		solid,
		mark options={solid},
		]
		table {
			%1	4.44E-01
			%1.5	2.09E-01
			2	6.58E-02
			2.5	1.13E-02
			3	1.14E-03
			3.5	4.63E-05
		};
		\addlegendentry{SP-RLD-$4$}
		
		\addplot[
		color=brown,
		mark=pentagon,
		thin,
		mark size=2,
		solid,
		mark options={solid},
		]
		table {
			%1	2.79E-01
			%1.5	9.60E-02
			2	1.77E-02
			2.5	1.85E-03
			3	9.72E-05
			3.25	1.49E-05
		};
		\addlegendentry{SP-RLD-$8$}
		
		\addplot[
		color=magenta,
		mark=diamond,
		thin,
		mark size=2,
		mark options={solid},
		]
		table {
			%1.0	1.58E-01
			%1.5	3.89E-02
			2.0	5.33E-03
			2.5	2.38E-04
			2.75	4.76E-05
		};
		\addlegendentry{SP-RLD-$16$}
		
		\addplot[
		color=green,
		mark=star,
		thin,
		mark size=2,
		mark options={solid},
		]
		table {
			%1.0	8.46E-02
			%1.5	1.38E-02
			2.0	1.29E-03
			2.5	3.37E-05
		};
		\addlegendentry{SP-RLD-$32$}
		
		\addplot[
			color=blue,
			mark=o,
			thin,
			mark size=2,
			dashdotted,
			mark options={solid},
		]
		table {
			%1	8.25E-01
			%1.5	6.34E-01
			2	3.90E-01
			2.5	1.79E-01
			3	6.02E-02
			3.5	1.64E-02
			4	2.22E-03
			4.5	1.58E-04
		};
		\addlegendentry{SSP-RLD-$4$-$1$}
		
		\addplot[
		color=black,
		mark=square,
		thin,
		mark size=2,
		dashdotted,
		mark options={solid},
		]
		table {
			2	1.93E-01
			2.5	5.80E-02
			3	1.07E-02
			3.5	1.20E-03
			4	6.04E-05
		};
		\addlegendentry{SSP-RLD-$4$-$2$}
		
		\addplot[
		color=red,
		mark=triangle,
		thin,
		mark size=2,
		dashdotted,
		mark options={solid},
		]
		table {
			2	7.51E-02
			2.5	1.35E-02
			3	1.29E-03
			3.5	5.62E-05
		};
		\addlegendentry{SSP-RLD-$4$-$4$}
		
		\addplot[
		color=brown,
		mark=pentagon,
		thin,
		mark size=2,
		dashdotted,
		mark options={solid},
		]
		table {
			2	2.41E-02
			2.5	2.31E-03
			3	1.28E-04
			3.25	1.56E-05
		};
		\addlegendentry{SSP-RLD-$4$-$8$}
		
		\addplot[
		color=magenta,
		mark=diamond,
		thin,
		dashdotted,
		mark size=2,
		mark options={solid},
		]
		table {
			%1.0	1.95E-01
			%1.5	5.04E-02
			2.0	6.39E-03
			2.5	3.53e-04
			2.75 5.52e-05
		};
		\addlegendentry{SSP-RLD-$4$-$16$}

		\addplot[
		color=green,
		mark=star,
		thin,
		mark size=2,
		dashdotted,
		mark options={solid},
		]
		table {
			%1.00E+00	0.09306
			%1.50E+00	0.01518
			2.00E+00	0.001419
			2.50E+00	3.707E-05
		};
		\addlegendentry{SSP-RLD-$4$-$32$}
		
		\addplot[
			color=blue,
			mark=otimes,
			thin,
			mark size=2,
			solid,
			mark options={solid},
		]
		table {
			%1	2.64E-01
			%1.5	9.13E-02
			2	1.98E-02
			2.5	2.05E-03
			3	1.35E-04
			3.25	2.7E-05
		};
		\addlegendentry{SP-RLD-$32$ (cyclic)}
	
		\addplot[
			color=green,
			mark=+,
			thin,
			mark size=2,
			solid,
			mark options={solid},
		]
		table {
			%			3	1.04E-01
			3.5	1.51E-02
			4	6.58E-04
			4.5	1.05E-05
			%S=16384
		};
		\addlegendentry{SCS-$D$ \cite{niu2012stack, TrifonovScore}}

		\addplot[
			color=red,
			mark=x,
			thin,
			solid,
			mark size=2,
			mark options={solid},
		]
		table {
%			3	1.35E-2
			3.5	2.27E-3
			4	1.89E-4
			4.5	7.74E-06
		};
		%		\addlegendentry{SP-SCL-$\{32,16,16\}$ \cite{Ali_SP}}
		\addlegendentry{SP-SCL-$M$ \cite{Ali_SP}}

%		\addplot[
%			color=black,
%			mark=none,
%			thin,
%			mark size=2,
%			solid,
%			mark options={solid},
%		]
%		table {
%			%%1.5		1.23E-01
%			2.0	3.05E-02
%			2.5	4.67E-03
%			3.0	3.33E-04
%			3.25	6.15E-5
%		};
%		\addlegendentry{RLD-$64$ \cite{Dumer06}}

		\addplot[
			color=cyan,
			mark=none,
			%thick,
			line width=0.75,
			mark size=2,
			solid,
			mark options={solid},
		]
		table {
			3.5	2.89E-03
			4.0	2.58E-04
			4.5	1.25E-05
		};
		\addlegendentry{FSCL-$32$ \cite{Ali_FSSCL}}
		
		\node[anchor=north east, fill=white] at (rel axis cs:1,1) {\footnotesize{\makecell{$\mathcal{RM}(3,9)$\\$D=16384$\\$M=16$}}};
		
		\coordinate (spypoint1) at (axis cs:3, 1.25e-4);
		\coordinate (magnifyglass1) at (axis cs:1.65, 7.5e-5);
	\end{axis}

%\spy [blue, width=0.75cm, height=1cm] on (spypoint1) in node[fill=white] at (magnifyglass1);

\end{tikzpicture}

%% file: ferRM512_256.tikz.tex
\begin{tikzpicture}[spy using outlines = {rectangle, magnification=3.0, connect spies}]

	\pgfplotsset{
		label style = {font=\fontsize{9pt}{7.2}\selectfont},
		tick label style = {font=\fontsize{7pt}{7.2}\selectfont}
	}
	
	\begin{axis}[
		scale = 1,
		ymode=log,
		xlabel={$E_b/N_0$ [\text{dB}]}, xlabel style={yshift=0.5em},
		ytick={1e-7, 1e-6, 1e-5,1e-4,1e-3,1e-2,1e-1,1e-0},
		xtick={2,2.5,3,3.5,4,4.5},
		ylabel={}, %ylabel style={yshift=-0.75em},
		grid=both,
		ymajorgrids=true,
		xmajorgrids=true,
		grid style=dashdotted,
		width=0.35\linewidth, height=7cm,
		thin,
		mark size=2,
		% use the following key so the baseline of all ticklabel entries is the same
		% (compare this image to the one from marmot)
		typeset ticklabels with strut,
		% there is one default value for the `legend pos' that is outside the axis
		legend pos=outer north east,
		% (so the legend looks a bit better)
		legend cell align=left,
		% (moved this common key here)
		smooth,
		%    legend style={
		%   	  legend pos=outer north east,
		%      anchor={center},
		%      %cells={anchor=west},
		%      column sep= 2mm,
		%      font=\fontsize{7pt}{7.2}\selectfont,
		%    },
		%    legend to name=perf-legend-SPRM128-fast,
		%    legend columns=5,
		]

		\addplot[
			color=blue,
			mark=o,
			thin,
			mark size=2,
			solid,
			mark options={solid},
		]
		table {
			%2.00E+00	7.21E-01
			2.50E+00	4.47E-01
			3	1.95E-01
			3.5	5.17E-02
			4	8.06E-03
			4.5	6.21E-04
%			5.0	2.47E-05
		};
		%{SP-RLD-$1$}
		
		\addplot[
			color=black,
			mark=square,
			thin,
			mark size=2,
			solid,
			mark options={solid},
		]
		table {
			%2	5.46E-01
			2.5	2.48E-01
			3	6.48E-02
			3.5	8.32E-03
			4	4.68E-04
			4.5	1.05E-05
		};
		%{SP-RLD-$2$}
		
		\addplot[
			color=red,
			mark=triangle,
			thin,
			mark size=2,
			solid,
			mark options={solid},
		]
		table {
			%2	3.58E-01
			2.5	1.08E-01
			3	1.57E-02
			3.5	8.48E-04
			4	1.39E-05
			
		};
		%{SP-RLD-$4$}
		
		\addplot[
			color=brown,
			mark=pentagon,
			thin,
			mark size=2,
			solid,
			mark options={solid},
		]
		table {
			%2	2.00E-01
			2.5	3.80E-02
			3	2.72E-03
			3.5	6.64E-05
		};
		%{SP-RLD-$8$}
		
		\addplot[
			color=magenta,
			mark=diamond,
			thin,
			mark size=2,
			mark options={solid},
		]
		table {
			%2.0	1.02E-01
			2.5	1.16E-02
			3.0	3.84E-04
			3.25	4.19E-05
		};
		%{SP-RLD-$16$}
		
		\addplot[
			color=green,
			mark=star,
			thin,
			mark size=2,
			mark options={solid},
		]
		table {
			%2.0	4.68E-02
			2.50	3.10E-03
			3.0	5.07E-05
		};
		%{SP-RLD-$32$}

		\addplot[
			color=blue,
			mark=o,
			thin,
			mark size=2,
			dashdotted,
			mark options={solid},
		]
		table {
			%2	7.82E-01
			2.5	5.04E-01
			3	2.23E-01
			3.5	6.10E-02
			4	8.84E-03
			4.5	6.29E-04
%			5	2.87E-05
		};
		%{SP-RLD*-$1$}
		
		\addplot[
			color=black,
			mark=square,
			thin,
			mark size=2,
			dashdotted,
			mark options={solid},
		]
		table {
			%2	6.04E-01
			2.5	2.85E-01
			3	7.61E-02
			3.5	1.03E-02
			4	5.50E-04
			4.5	1.26E-05
		};
		%{SP-RLD*-$2$}
		
		\addplot[
			color=red,
			mark=triangle,
			thin,
			mark size=2,
			dashdotted,
			mark options={solid},
		]
		table {
			%2	4.11E-01
			2.5	1.30E-01
			3	1.87E-02
			3.5	1.18E-03
			4.0	1.89E-05
		};
		%{SP-RLD*-$4$}
		
		\addplot[
		color=brown,
		mark=pentagon,
		thin,
		mark size=2,
		dashdotted,
		mark options={solid},
		]
		table {
			%2.0	2.47E-01
			2.5	4.95E-02
			3.0	4.33E-03
			3.5	1.02E-04
		};
		%{SP-RLD*-$8$}
		
		\addplot[
		color=magenta,
		mark=diamond,
		thin,
		dashdotted,
		mark size=2,
		mark options={solid},
		]
		table {
			%2.0	1.32E-01
			2.5	1.71E-02
			3.0	6.37E-04
			3.25 6.01E-05
		};
		%{SP-RLD*-$16$}
		
		\addplot[
			color=green,
			mark=star,
			thin,
			mark size=2,
			dashdotted,
			mark options={solid},
		]
		table {
			%2.0	6.36E-02
			2.5	3.89E-03
			3.0	7.00E-05
		};
		%{SP-RLD*-$32$}
				
		\addplot[
			color=blue,
			mark=otimes,
			thin,
			mark size=2,
			solid,
			mark options={solid},
		]
		table {
			%2	1.76E-01
			2.5	3.31E-02
			3	2.63E-03
			3.5	5.21E-05
		};
		%		\addlegendentry{SP-RLD-$32$ (cyclid)}
		
		\addplot[
		color=green,
		mark=+,
		thin,
		mark size=2,
		solid,
		mark options={solid},
		]
		table {
			3.5	2E-02
			4	6.51E-04
			4.25	4.64E-05
			%S=4096
		};
		%\addlegendentry{SCS \cite{niu2012stack, TrifonovScore}}
		
		\addplot[
			color=red,
			mark=x,
			thin,
			mark size=2,
			solid,
			mark options={solid},
		]
		table {
%			3	0.03580358
			3.5	0.0050005
			4	0.00030255
			4.25	6.00E-05
			4.5	1.108E-05
		};
		%\addlegendentry{SP-SCL-16 \cite{Ali_SP}}
		
%		\addplot[
%			color=black,
%			mark=none,
%			thin,
%			mark size=2,
%			solid,
%			mark options={solid},
%		]
%		table {
%			%2	1.98E-01
%			2.5	3.85E-02
%			3.0	3.39E-03
%			3.5	1.11E-04
%			3.625	4.20E-05
%		};
%		%		\addlegendentry{RLDP-$64$ \cite{Dumer06}}	

		\addplot[
		color=cyan,
		mark=none,
		%thick,
		line width=0.75,
		mark size=2,
		solid,
		mark options={solid},
		]
		table {
			3.5	7.30E-03
			4.0	4.66E-04
			4.5	1.15E-05
		};
		%\addlegendentry{FSCL-$32$ \cite{Ali_FSSCL}}
		
		\node[anchor=north east, fill=white] at (rel axis cs:1,1) {\footnotesize{\makecell{$\mathcal{RM}(4,9)$\\$D=4096$\\$M=16$}}};
		
		\coordinate (spypoint1) at (axis cs:4.25, 6.0e-5);
		\coordinate (magnifyglass1) at (axis cs:2.6, 3e-5);
\end{axis}

\spy [blue, width=0.75cm, height=1.25cm] on (spypoint1) in node[fill=white] at (magnifyglass1);
\end{tikzpicture}

%% file: comp_RM512_46.tikz.tex
\begin{tikzpicture}
	\pgfplotsset{
		label style = {font=\fontsize{9pt}{7.2}\selectfont},
		tick label style = {font=\fontsize{7pt}{7.2}\selectfont}
	}
	
	\begin{axis}[
		scale = 1,
		ymode=log,
		ylabel={$\Gamma$}, ylabel style={yshift=-1em},
%		xlabel={$L$},
%		scaled y ticks=base 10:-5,
		xlabel style={yshift=0.5em},
		ytick={1E4,1E5,1E6},
		xtick={1,2,3,4,5,6},
		xticklabels={1,2,4,8,16,32},
		%		ymin = 3E-6,
		%		ymax = 1,
		grid=both,
		ymajorgrids=true,
		xmajorgrids=true,
		grid style=dashdotted,
		width=0.34\linewidth, height=4cm,
		thin,
		mark size=2,
		% use the following key so the baseline of all ticklabel entries is the same
		% (compare this image to the one from marmot)
		%		typeset ticklabels with strut,
		%		% there is one default value for the `legend pos' that is outside the axis
		%		legend pos=outer north east,
		%		% (so the legend looks a bit better)
		legend cell align=left,
		%		% (moved this common key here)
		%		smooth,
		legend style={
			%		legend pos=outer north east,
			%	      anchor={center},
			%cells={anchor=west},
			column sep= 2mm,
			font=\fontsize{7pt}{8}\selectfont,
		},
%				legend to name=perf-legend-comp,
%				legend columns=2,
		]
		\addplot[
			color=red,
			mark=o,
			thin,
			mark size=2,
			solid,
			mark options={solid},
		]
		table {
			1	43273
			2	86612
			3	173571
			4	349016
			5	704792
			6	1435359
		};
		%\addlegendentry{SP-RLD-$L$ (sequential)}
		
		\addplot[
			color=blue,
			mark=triangle,
			thin,
			mark size=2,
			solid,
			mark options={solid},
		]
		table {
			1	43273
			2	86612
			3	173571
			4	349016
			5	704792
			6	1435359
		};
		%\addlegendentry{SP-RLD-$L$ (parallel)}
		
		\addplot[
			color=red,
			mark=o,
			thin,
			mark size=2,
			dashdotted,
			mark options={solid},
		]
		table {
			1	42721
			2	85492
			3	171338
			4	344580
			5	695897
			6	1417635
		};
		%\addlegendentry{SSP-RLD-$S$-$L$ (sequential)}
		
		\addplot[
			color=blue,
			mark=triangle,
			thin,
			mark size=2,
			dashdotted,
			mark options={solid},
		]
		table {
			1	42721
			2	85492
			3	171338
			4	344580
			5	695897
			6	1417635
		};
		%\addlegendentry{SSP-RLD-$S$-$L$ (parallel)}
		
		\node[anchor=south east, fill=white] at (rel axis cs:1,0) {\footnotesize{{{$\mathcal{RM}(2,9)$}}}};
		
	\end{axis}
\end{tikzpicture}

%% file: comp_RM512_130.tikz.tex
\begin{tikzpicture}
	\pgfplotsset{
		label style = {font=\fontsize{9pt}{7.2}\selectfont},
		tick label style = {font=\fontsize{7pt}{7.2}\selectfont}
	}
	
	\begin{axis}[
		scale = 1,
		ymode=log,
		ylabel={$\Gamma$},	ylabel style={yshift=-1em},
%		scaled y ticks=base 10:-5,
%		xlabel={$L$},
		xlabel style={yshift=0.5em},
		xtick={1,2,3,4,5,6},
		xticklabels={1,2,4,8,16,32},
		%		ymin = 3E-6,
		%		ymax = 1,
		grid=both,
		ymajorgrids=true,
		xmajorgrids=true,
		grid style=dashdotted,
		width=0.34\linewidth, height=4cm,
		thin,
		mark size=2,
		% use the following key so the baseline of all ticklabel entries is the same
		% (compare this image to the one from marmot)
		%		typeset ticklabels with strut,
		%		% there is one default value for the `legend pos' that is outside the axis
		%		legend pos=outer north east,
		%		% (so the legend looks a bit better)
		legend cell align=left,
		%		% (moved this common key here)
		%		smooth,
		legend style={
			%		legend pos=outer north east,
			%	      anchor={center},
			%cells={anchor=west},
			column sep= 2mm,
			font=\fontsize{7pt}{8}\selectfont,
		},
		legend to name=perf-legend-comp-3-9,
		legend columns=2,
		]
		
		\addplot[
			color=red,
			mark=o,
			thin,
			mark size=2,
			solid,
			mark options={solid},
		]
		table {
			1	37238
			2	74663
			3	150429
			4	306748
			5	629677
			6	1302525
		};
		\addlegendentry{SP-RLD-$L$ (sequential SP)}

		\addplot[
		color=red,
		mark=o,
		thin,
		mark size=2,
		dashdotted,
		mark options={solid},
		]
		table {
			1	25377
			2	50939
			3	102962
			4	211810
			5	439815
			6	923082
		};
		\addlegendentry{SSP-RLD-\textcolor{red}{$4$}-$L$ (sequential SP)}		
		
		\addplot[
			color=blue,
			mark=triangle,
			thin,
			mark size=2,
			solid,
			mark options={solid},
		]
		table {
			1	37238
			2	74663
			3	150429
			4	306748
			5	629677
			6	1302525
		};
		\addlegendentry{SP-RLD-$L$ (parallel SP)}
				
		\addplot[
			color=blue,
			mark=triangle,
			thin,
			mark size=2,
			dashdotted,
			mark options={solid},
		]
		table {
			1	25377
			2	50939
			3	102962
			4	211810
			5	439815
			6	923082
		};
		\addlegendentry{SSP-RLD-\textcolor{red}{$4$}-$L$ (parallel SP)}

		\node[anchor=south east, fill=white] at (rel axis cs:1,0) {\footnotesize{{{$\mathcal{RM}(3,9)$}}}};
		
	\end{axis}
\end{tikzpicture}

%% file: comp_RM512_256.tikz.tex
\begin{tikzpicture}
	\pgfplotsset{
		label style = {font=\fontsize{9pt}{7.2}\selectfont},
		tick label style = {font=\fontsize{7pt}{7.2}\selectfont}
	}
	
	\begin{axis}[
		scale = 1,
		ymode=log,
		ylabel={$\Gamma$}, ylabel style={yshift=-1em},
		scaled y ticks=base 10:-5,
		xlabel style={yshift=0.5em},
		xtick={1,2,3,4,5,6},
		xticklabels={1,2,4,8,16,32},
		grid=both,
		ymajorgrids=true,
		xmajorgrids=true,
		grid style=dashdotted,
		width=0.34\linewidth, height=4cm,
		thin,
		mark size=2,
		legend cell align=left,
		legend style={
			column sep= 2mm,
			font=\fontsize{7pt}{8}\selectfont,
		},
		]
		
		\addplot[
			color=red,
			mark=o,
			thin,
			mark size=2,
			solid,
			mark options={solid},
		]
		table {
			1	31692
			2	63705
			3	129322
			4	268811
			5	562247
			6	1170342
		};
		%\addlegendentry{SP-RLD-$L$}
		
		\addplot[
			color=blue,
			mark=triangle,
			thin,
			mark size=2,
			solid,
			mark options={solid},
		]
		table {
			1	31692
			2	63705
			3	129322
			4	268811
			5	562247
			6	1170342
		};
		%\addlegendentry{SP-RLD-$L$}
		
		\addplot[
			color=red,
			mark=o,
			thin,
			mark size=2,
			dashdotted,
			mark options={solid},
		]
		table {
			1	17736
			2	35766
			3	73458
			4	157069
			5	338548
			6	723625
		};
		%\addlegendentry{SSP-RLD-$L$}
		
		\addplot[
			color=blue,
			mark=triangle,
			thin,
			mark size=2,
			dashdotted,
			mark options={solid},
		]
		table {
			1	17736
			2	35766
			3	73458
			4	157069
			5	338548
			6	723625
		};
		%\addlegendentry{SSP-RLD-$L$}
		
		\node[anchor=south east, fill=white] at (rel axis cs:1,0) {\footnotesize{$\mathcal{RM}(4,9)$}};
		
	\end{axis}
\end{tikzpicture}

%% file: lat_RM512_46.tikz.tex
\begin{tikzpicture}
	\pgfplotsset{
		label style = {font=\fontsize{9pt}{7.2}\selectfont},
		tick label style = {font=\fontsize{7pt}{7.2}\selectfont},
		scaled y ticks=base 10:3,
	}
	
	\begin{axis}[
		scale = 1,
		ymode=log,
%		scaled y ticks=base 10:-2,
		ylabel={$\Upsilon$}, ylabel style={yshift=-1em},
		xlabel={$L$},
		xlabel style={yshift=0.5em},
		ytick={1E1,1E2,1E3},
		xtick={1,2,3,4,5,6},
		xticklabels={1,2,4,8,16,32},
		grid=both,
		ymajorgrids=true,
		xmajorgrids=true,
		grid style=dashdotted,
		width=0.34\linewidth, height=4cm,
		thin,
		mark size=2,
		% use the following key so the baseline of all ticklabel entries is the same
		% (compare this image to the one from marmot)
		%		typeset ticklabels with strut,
		%		% there is one default value for the `legend pos' that is outside the axis
		%		legend pos=outer north east,
		%		% (so the legend looks a bit better)
		legend cell align=left,
		%		% (moved this common key here)
		%		smooth,
		legend style={
			%		legend pos=outer north east,
			%	      anchor={center},
			%cells={anchor=west},
			column sep= 2mm,
			font=\fontsize{7pt}{8}\selectfont,
		},
%		legend to name=perf-legend-comp,
%		legend columns=4,
		]
		\addplot[
			color=red,
			mark=o,
			thin,
			mark size=2,
			solid,
			mark options={solid},
		]
		table {
			1	366
			2	382
			3	404
			4	440
			5	459
			6	477
		};
		%\addlegendentry{SP-RLD-$L$}
		
		\addplot[
			color=blue,
			mark=triangle,
			thin,
			mark size=2,
			solid,
			mark options={solid},
		]
		table {
			1	101
			2	117
			3	139
			4	175
			5	194
			6	212
		};
		%\addlegendentry{SP-RLD-$L$}
		
		\addplot[
		color=red,
		mark=o,
		thin,
		mark size=2,
		dashdotted,
		mark options={solid},
		]
		table {
			1	321
			2	337
			3	359
			4	395
			5	414
			6	432
		};
		%\addlegendentry{SSP-RLD-$L$}

		\addplot[
			color=blue,
			mark=triangle,
			thin,
			mark size=2,
			dashdotted,
			mark options={solid},
		]
		table {
			1	95
			2	111
			3	133
			4	169
			5	188
			6	206
		};
		%\addlegendentry{SSP-RLD-$L$}
		
		\node[anchor=south east, fill=white] at (rel axis cs:1,0) {\footnotesize{{{$\mathcal{RM}(2,9)$}}}};
		
	\end{axis}
\end{tikzpicture}

%% file: lat_RM512_130.tikz.tex
\begin{tikzpicture}
	\pgfplotsset{
		label style = {font=\fontsize{9pt}{7.2}\selectfont},
		tick label style = {font=\fontsize{7pt}{7.2}\selectfont}
	%	scaled y ticks=base 10:3,
	}
	
	\begin{axis}[
		scale = 1,
		ymode=log,
		ylabel={$\Upsilon$},	ylabel style={yshift=-1em},
		xlabel={$L$},
		xlabel style={yshift=0.5em},
		xtick={1,2,3,4,5,6},
		xticklabels={1,2,4,8,16,32},
		ytick={1e2,1e3},
		grid=both,
		ymajorgrids=true,
		xmajorgrids=true,
		grid style=dashdotted,
		width=0.34\linewidth, height=4cm,
		thin,
		mark size=2,
		% use the following key so the baseline of all ticklabel entries is the same
		% (compare this image to the one from marmot)
		%		typeset ticklabels with strut,
		%		% there is one default value for the `legend pos' that is outside the axis
		%		legend pos=outer north east,
		%		% (so the legend looks a bit better)
		legend cell align=left,
		%		% (moved this common key here)
		%		smooth,
		legend style={
			%		legend pos=outer north east,
			%	      anchor={center},
			%cells={anchor=west},
			column sep= 2mm,
			font=\fontsize{7pt}{8}\selectfont,
		},
		%		legend to name=perf-legend-comp,
		%		legend columns=4,
		]
		
		\addplot[
		color=red,
		mark=o,
		thin,
		mark size=2,
		solid,
		mark options={solid},
		]
		table {
			1	747
			2	796
			3	881
			4	1059
			5	1166
			6	1238
			
		};
		%\addlegendentry{SP-RLD-$L$}
		
		\addplot[
			color=blue,
			mark=triangle,
			thin,
			mark size=2,
			solid,
			mark options={solid},
		]
		table {
			1	252
			2	301
			3	386
			4	555
			5	671
			6	743
			
		};
		%\addlegendentry{SP-RLD-$L$}
		
		\addplot[
		color=red,
		mark=o,
		thin,
		mark size=2,
		dashdotted,
		mark options={solid},
		]
		table {
			1	365
			2	414
			3	499
			4	668
			5	784
			6	856
			
		};
		%\addlegendentry{SSP-RLD-$L$}
		
		\addplot[
		color=blue,
		mark=triangle,
		thin,
		mark size=2,
		dashdotted,
		mark options={solid},
		]
		table {
			1	203
			2	252
			3	337
			4	506
			5	622
			6	694
			
		};
		%\addlegendentry{SSP-RLD-$L$}
		
		\node[anchor=south east, fill=white] at (rel axis cs:1,0) {\footnotesize{{{$\mathcal{RM}(3,9)$}}}};
		
	\end{axis}
\end{tikzpicture}

%% file: lat_RM512_256.tikz.tex
\begin{tikzpicture}
	\pgfplotsset{
		label style = {font=\fontsize{9pt}{7.2}\selectfont},
		tick label style = {font=\fontsize{7pt}{7.2}\selectfont}
	}
	
	\begin{axis}[
		scale = 1,
		ymode=log,
		scaled y ticks=base 10:-2,
		ylabel={$\Upsilon$},	ylabel style={yshift=-1em},
		xlabel={$L$},
		xlabel style={yshift=0.5em},
		ytick={1E2,1E3,1E4,1E5},
		xtick={1,2,3,4,5,6},
		xticklabels={1,2,4,8,16,32},
		%		ymin = 3E-6,
		%		ymax = 1,
		grid=both,
		ymajorgrids=true,
		xmajorgrids=true,
		grid style=dashdotted,
		width=0.34\linewidth, height=4cm,
		thin,
		mark size=2,
		% use the following key so the baseline of all ticklabel entries is the same
		% (compare this image to the one from marmot)
		%		typeset ticklabels with strut,
		%		% there is one default value for the `legend pos' that is outside the axis
		%		legend pos=outer north east,
		%		% (so the legend looks a bit better)
		legend cell align=left,
		%		% (moved this common key here)
		%		smooth,
		legend style={
			%		legend pos=outer north east,
			%	      anchor={center},
			%cells={anchor=west},
			column sep= 2mm,
			font=\fontsize{7pt}{8}\selectfont,
		},
		%		legend to name=perf-legend-comp,
		%		legend columns=4,
		]
		
		\addplot[
		color=red,
		mark=o,
		thin,
		mark size=2,
		solid,
		mark options={solid},
		]
		table {
			1	973
			2	1059
			3	1235
			4	1621
			5	1997
			6	2279
		};
		%\addlegendentry{SP-RLD-$L$}
		
				\addplot[
		color=red,
		mark=o,
		thin,
		mark size=2,
		dashdotted,
		mark options={solid},
		]
		table {
			1	396
			2	482
			3	658
			4	1044
			5	1420
			6	1702
		};
		%\addlegendentry{SSP-RLD-$L$}
		
		\addplot[
			color=blue,
			mark=triangle,
			thin,
			mark size=2,
			solid,
			mark options={solid},
		]
		table {
			1	374
			2	460
			3	636
			4	1022
			5	1398
			6	1680
		};
		%\addlegendentry{SP-RLD-$L$}
		
		\addplot[
			color=blue,
			mark=triangle,
			thin,
			mark size=2,
			dashdotted,
			mark options={solid},
		]
		table {
			1	283
			2	360
			3	545
			4	931
			5	1307
			6	1589
		};
		%\addlegendentry{SSP-RLD-$L$}
				
		\node[anchor=south east, fill=white] at (rel axis cs:1,0) {\footnotesize{$\mathcal{RM}(4,9)$}};
		
	\end{axis}
\end{tikzpicture}

%% file: mem_RM512_46.tikz.tex
\begin{tikzpicture}
	\pgfplotsset{
		label style = {font=\fontsize{9pt}{7.2}\selectfont},
		tick label style = {font=\fontsize{7pt}{7.2}\selectfont},
		scaled y ticks=base 10:3,
	}
	
	\begin{axis}[
		scale = 1,
		ymode=log,
		scaled y ticks=base 10:-1,
		ylabel={$\Phi$}, ylabel style={yshift=-1em},
		xlabel={$L$},
		xlabel style={yshift=0.5em},
		xtick={1,2,3,4,5,6},
		xticklabels={1,2,4,8,16,32},
		grid=both,
		ymajorgrids=true,
		xmajorgrids=true,
		grid style=dashdotted,
		width=0.7\linewidth, height=4cm,
		thin,
		mark size=2,
		% use the following key so the baseline of all ticklabel entries is the same
		% (compare this image to the one from marmot)
		%		typeset ticklabels with strut,
		%		% there is one default value for the `legend pos' that is outside the axis
		%		legend pos=outer north east,
		%		% (so the legend looks a bit better)
		legend cell align=left,
		%		% (moved this common key here)
		%		smooth,
		legend style={
			%		legend pos=outer north east,
			%	      anchor={center},
			%cells={anchor=west},
			column sep= 2mm,
			font=\fontsize{7pt}{8}\selectfont,
		},
		legend to name=perf-legend-mem,
		legend columns=2,
		]
		\addplot[
			color=red,
			mark=o,
			thin,
			mark size=2,
			solid,
			mark options={solid},
		]
		table {
			1	4.1
			2	6.29
			3	10.54
			4	19.04
			5	36.04
			6	70.04
		};
		\addlegendentry{SP-RLD-$L$ (sequential SP)}
		
		\addplot[
		color=red,
		mark=o,
		thin,
		mark size=2,
		dashdotted,
		mark options={solid},
		]
		table {
			1	4.1
			2	6.29
			3	10.54
			4	19.04
			5	36.04
			6	70.04
		};
		\addlegendentry{SSP-RLD-$4$-$L$ (sequential SP)}
		
		\addplot[
			color=blue,
			mark=triangle,
			thin,
			mark size=2,
			solid,
			mark options={solid},
		]
		table {
			1	20.1
			2	38.3
			3	74.5
			4	147
			5	292
			6	582.04
		};
		\addlegendentry{SP-RLD-$L$ (parallel SP)}
		
		\addplot[
			color=blue,
			mark=triangle,
			thin,
			mark size=2,
			dashdotted,
			mark options={solid},
		]
		table {
			1	20.1
			2	38.3
			3	74.5
			4	147
			5	292
			6	582.04
		};
		\addlegendentry{SSP-RLD-$4$-$L$ (parallel SP)}
		
		\node[anchor=north west, fill=white] at (rel axis cs:0,1) {\footnotesize{{{$\mathcal{RM}(r,9)$ - $r\in \{2,3,4\}$}}}};
		
	\end{axis}
\end{tikzpicture}

%% file: ferRM256_37_FSC_per_rpa.tikz.tex
\begin{tikzpicture}[spy using outlines = {rectangle, magnification=2.5, connect spies}]

	\pgfplotsset{
		label style = {font=\fontsize{9pt}{7.2}\selectfont},
		tick label style = {font=\fontsize{7pt}{7.2}\selectfont}
	}
	
\begin{axis}[
scale = 1,
ymode=log,
xlabel={$E_b/N_0$ [\text{dB}]}, xlabel style={yshift=0.5em},
ytick={1e-6, 1e-5,1e-4,1e-3,1e-2,1e-1,1e-0},
xtick={1,1.5,2,2.5,3},
ylabel={FER}, ylabel style={yshift=-0.5em},
grid=both,
ymajorgrids=true,
xmajorgrids=true,
grid style=dashdotted,
width=0.35\linewidth, height=7cm,
thin,
mark size=2.25,
% use the following key so the baseline of all ticklabel entries is the same
% (compare this image to the one from marmot)
%		typeset ticklabels with strut,
%		% there is one default value for the `legend pos' that is outside the axis
%		legend pos=outer north east,
%		% (so the legend looks a bit better)
legend cell align=left,
%		% (moved this common key here)
%		smooth,
legend style={
	%		legend pos=outer north east,
	%	      anchor={center},
	%cells={anchor=west},
	column sep= 2mm,
	font=\fontsize{7pt}{7.2}\selectfont,
},
		legend to name=perf-legend-RPA-fer-2-8,
		legend columns=4,
]

\node[anchor=north east, fill=white] at (rel axis cs:1,1) {\scriptsize{\makecell{$M=32$, $P=48$\\$(S,L)=(3,8)$\\$(L', T)= (1, 8)$}}};
\node[anchor=south west, fill=white] at (rel axis cs:0,0) {\footnotesize{{{$\mathcal{RM}(2,8)$}}}};

		\addplot[
		color=blue,
		mark=o,
		thin,
		mark size=2,
		dashdotted,
		mark options={solid},
		]
		table {
			1	3.61E-02
			1.5	1.19E-02
			2	3.37E-03
			2.5	5.99E-04
		};
		\addlegendentry{RLDP-$M$ \cite{Dumer06}}
		
		\addplot[
		color=blue,
		mark=o,
		thin,
		mark size=2,
		solid,
		mark options={solid},
		]
		table {
			1	2.53E-02
			1.5	7.58E-03
			2	1.78E-03
			2.5	2.83E-04
		};
		\addlegendentry{RLDA-$M$ \cite{Geiselhart, Dumer06}}

		\addplot[
		color=red,
		mark=square,
		thin,
		mark size=2,
		dashdotted,
		mark options={solid},
		]
		table {
			1	4.16E-02
			1.5	1.41E-02
			2	3.72E-03
			2.5	8.87E-04			
		};
		\addlegendentry{Per-SSC-FHT-$P$ \cite{dumerFHT, Kamenev19}}
		
		\addplot[
		color=red,
		mark=square,
		thin,
		mark size=2,
		solid,
		mark options={solid},
		]
		table {			
			1	2.11E-02
			1.5	5.37E-03
			2	1.19E-03
			2.5	1.82E-04			
		};
		\addlegendentry{Aut-SSC-FHT-$P$ \cite{dumerFHT, Geiselhart}}
		
		\addplot[
			color=gray,
			mark=none,
			thin,
			mark size=2,
			solid,
			mark options={solid},
		]
		table {			
			1	1.912e-02
			1.5 4.9e-03
			2	0.00111636
			2.5	0.00017401
%			3.0	2.00E-05
		};
		\addlegendentry{ML (lower bound) \cite{Dumer06}}
		
		\addplot[
			color=blue,
			mark=triangle,
			thin,
			mark size=2,
			solid,
			mark options={solid},
		]
		table {
			1	3.60E-02
			1.5	1.13E-02
			2	2.62E-03		
			2.5	4.46E-04
		};
		\addlegendentry{SRPA \cite{fathollahi2020sparse}}

		\addplot[
		color=cyan,
		mark=diamond,
		thin,
		mark size=2,
		solid,
		mark options={solid},
		]
		table {
			1	2.62E-02
			1.5	6.60E-03
			2	1.38E-03
			2.5	2.23E-04
		};
		\addlegendentry{SSP-RLD-$S$-$L$}
		
		\addplot[
			color=cyan,
			mark=pentagon,
			thin,
			mark size=2,
			solid,
			mark options={solid},
		]
		table {
			1	2.91E-02
			1.5	7.14E-03
			2	1.73E-03
			2.5	2.37E-04
		};
		\addlegendentry{Ens-SSP-RLD-$S$-$L'$-$T$}
				
\coordinate (spypoint1) at (axis cs:2.5, 2.25e-4);
\coordinate (magnifyglass1) at (axis cs:1.25, 1.25e-3);

\end{axis}

\spy [blue, width=1cm, height=1.6cm] on (spypoint1) in node[fill=white] at (magnifyglass1);
\end{tikzpicture}

%% file: ferRM256_93_FSC_per_rpa.tikz.tex
\begin{tikzpicture}[spy using outlines = {rectangle, magnification=2.5, connect spies}]
	
	\pgfplotsset{
		label style = {font=\fontsize{9pt}{7.2}\selectfont},
		tick label style = {font=\fontsize{7pt}{7.2}\selectfont}
	}
	
	\begin{axis}[
		scale = 1,
		ymode=log,
		xlabel={$E_b/N_0$ [\text{dB}]}, xlabel style={yshift=0.5em},
		ytick={1e-6, 1e-5,1e-4,1e-3,1e-2,1e-1,1e-0},
		xtick={1,1.5,2,2.5,3},
		ylabel style={yshift=-0.5em},
		grid=both,
		ymajorgrids=true,
		xmajorgrids=true,
		grid style=dashdotted,
		width=0.35\linewidth, height=7cm,
		thin,
		mark size=2.25,
		% use the following key so the baseline of all ticklabel entries is the same
		% (compare this image to the one from marmot)
		%		typeset ticklabels with strut,
		%		% there is one default value for the `legend pos' that is outside the axis
		%		legend pos=outer north east,
		%		% (so the legend looks a bit better)
		legend cell align=left,
		%		% (moved this common key here)
		%		smooth,
		legend style={
			%		legend pos=outer north east,
			%	      anchor={center},
			%cells={anchor=west},
			column sep= 2mm,
			font=\fontsize{7pt}{7.2}\selectfont,
		},
		legend to name=perf-legend-rpa-fer-3-8,
		legend columns=3,
		]		
		
\node[anchor=north east, fill=white] at (rel axis cs:1,1) {\scriptsize{\makecell{$M=64$, $P=113$\\$(S,L)=(2,32)$\\$(L',T)=(4,8)$}}};
\node[anchor=south west, fill=white] at (rel axis cs:0,0) {\footnotesize{{{$\mathcal{RM}(3,8)$}}}};

		\addplot[
			color=gray,
			mark=none,
			thin,
			mark size=2,
			solid,
			mark options={solid},
		]
		table {
			1	2.42000E-02
			1.5	4.04000E-03
			2	5.07614E-04
			2.5	4.80538E-05
			%			3	3.22126E-06
		};
		%\addlegendentry{ML (lower bound) \cite{Dumer06}}
							
		\addplot[
			color=red,
			mark=square,
			thin,
			mark size=2,
			solid,
			mark options={solid},
		]
		table {
			1.00E+00	6.68E-02
			1.50E+00	1.33E-02
			2.00E+00	2.20E-03
			2.50E+00	1.20E-04
		};
		%\addlegendentry{Aut-SSC-$P$ \cite{Geiselhart}}
		
		\addplot[
		color=red,
		mark=square,
		thin,
		mark size=2,
		dashdotted,
		mark options={solid},
		]
		table {
			1	1.13E-01
			1.5	3.17E-02
			2.0	5.82E-03
			2.5	6.92E-04
		};
		%\addlegendentry{Per-SSC-$P$ \cite{Kamenev19}}
		
		\addplot[
		color=blue,
		mark=triangle,
		thin,
		mark size=2,
		solid,
		mark options={solid},
		]
		table {
			1	7.82E-02
			1.5	2.30E-02
			2	3.57E-03
			2.5	3.88E-04
		};
		%\addlegendentry{SRPA \cite{fathollahi2020sparse}}
		
		\addplot[
			color=blue,
			mark=o,
			thin,
			mark size=2,
			solid,
			mark options={solid},
		]
		table {
			1	6.55E-02
			1.5	1.43E-02
			2	1.98E-03
			2.5	1.50E-04
		};
		%\addlegendentry{RLDA-$M$ \cite{Geiselhart, Dumer06}}
		
		\addplot[
		color=blue,
		mark=o,
		thin,
		mark size=2,
		dashdotted,
		mark options={solid},
		]
		table {
			1	1.12E-01
			1.5	3.19E-02
			2	6.20E-03
			2.5	7.86E-04
		};
		%\addlegendentry{RLDP-$M$ \cite{Dumer06}}

		\addplot[
			color=cyan,
			mark=diamond,
			thin,
			mark size=2,
			solid,
			mark options={solid},
		]
		table {
			1.0	6.43E-02
			1.5	1.36E-02
			2.0	1.68E-03
			2.5 9.14E-05
		};
		%\addlegendentry{SSP-RLD-$S$-$L$}
		
		\addplot[
			color=cyan,
			mark=pentagon,
			thin,
			mark size=2,
			solid,
			mark options={solid},
		]
		table {
			1	7.22E-02
			1.5	1.60E-02
			2	2.54E-03
			2.5	1.64E-04
		};
		%\addlegendentry{Aut-SSP-RLD-$S'$-$L'$-$T$}
		
\coordinate (spypoint1) at (axis cs:2.5, 1.25e-4);
\coordinate (magnifyglass1) at (axis cs:1.25, 8e-4);

\end{axis}

\spy [blue, width=1cm, height=1.6cm] on (spypoint1) in node[fill=white] at (magnifyglass1);
\end{tikzpicture}

%% file: ferRM256_163_FSC_per_rpa.tikz.tex
\begin{tikzpicture}[spy using outlines = {rectangle, magnification=2.5, connect spies}]
	
	\pgfplotsset{
		label style = {font=\fontsize{9pt}{7.2}\selectfont},
		tick label style = {font=\fontsize{7pt}{7.2}\selectfont}
	}
	
	\begin{axis}[
		scale = 1,
		ymode=log,
		xlabel={$E_b/N_0$ [\text{dB}]}, xlabel style={yshift=0.5em},
		ytick={1e-6, 1e-5,1e-4,1e-3,1e-2,1e-1,1e-0},
		xtick={2,2.5,3,3.5},
		ylabel style={yshift=-0.5em},
		grid=both,
		ymajorgrids=true,
		xmajorgrids=true,
		grid style=dashdotted,
		width=0.35\linewidth, height=7cm,
		thin,
		mark size=2.25,
		% use the following key so the baseline of all ticklabel entries is the same
		% (compare this image to the one from marmot)
		%		typeset ticklabels with strut,
		%		% there is one default value for the `legend pos' that is outside the axis
		%		legend pos=outer north east,
		%		% (so the legend looks a bit better)
		legend cell align=left,
		%		% (moved this common key here)
		%		smooth,
		legend style={
			%		legend pos=outer north east,
			%	      anchor={center},
			%cells={anchor=west},
			column sep= 2mm,
			font=\fontsize{7pt}{7.2}\selectfont,
		},
				legend to name=perf-legend-rpa-fer-4-8,
				legend columns=3,
		]
		
		\addplot[
			color=gray,
			mark=none,
			thin,
			mark size=2,
			solid,
			mark options={solid},
		]
		table {
			2	2.91000E-02
			2.5	2.71053E-03
			3	2.81690E-04
			3.5	2.21926E-05
		};
		%\addlegendentry{ML (lower bound) \cite{Dumer06}}
		
		\addplot[
			color=blue,
			mark=triangle,
			thin,
			mark size=2,
			solid,
			mark options={solid},
		]
		table {
			2	1.36E-1
			2.5	2.80E-2
			3	3.34E-3
			3.5	2.26E-04
		};
		%\addlegendentry{SRPA \cite{fathollahi2020sparse}}
				
		\addplot[
			color=blue,
			mark=o,
			thin,
			mark size=2,
			solid,
			mark options={solid},
		]
		table {
			2	6.02E-02
			2.5	9.22E-03
			3.0	7.31E-04
			3.5	4.75E-05
		};
		%\addlegendentry{RLDA-$64$ \cite{Dumer06, Geiselhart}}
		
		\addplot[
			color=blue,
			mark=o,
			thin,
			mark size=2,
			dashdotted,
			mark options={solid},
		]
		table {
			2.0	1.05E-01
			2.5	2.28E-02
			3.0	2.22E-03
			3.5	1.32E-04
		};
		%\addlegendentry{RLDP-$M$ \cite{Dumer06}}
		
		\addplot[
			color=red,
			mark=square,
			thin,
			mark size=2,
			solid,
			mark options={solid},
		]
		table {
			2	7.87E-02
			2.5	1.31E-02
			3	1.08E-03
			3.5	4.65E-05
		};
		%\addlegendentry{Aut-SSC-$96$ \cite{Geiselhart}}
		
		\addplot[
		color=red,
		mark=square,
		thin,
		mark size=2,
		dashdotted,
		mark options={solid},
		]
		table {
			2	1.09E-01
			2.5	2.21E-02
			3.0	2.36E-03
			3.5	1.17E-04
		};
		%\addlegendentry{Per-SSC-$P$ \cite{Kamenev19}}
		
		\addplot[
			color=cyan,
			mark=diamond,
			thin,
			mark size=2,
			solid,
			mark options={solid},
		]
		table {
			2	6.06E-02
			2.5	8.88E-03
			3.0	6.84E-04
			3.5	3.45E-05
		};
		%\addlegendentry{SSP-RLD-$3$-$32$}
		
		\addplot[
			color=cyan,
			mark=pentagon,
			thin,
			mark size=2,
			solid,
			mark options={solid},
		]
		table {
			2	7.46E-02
			2.5	1.20E-02
			3.0	8.56E-04
			3.5	4.37E-05
		};
		%\addlegendentry{Aut-SSP-RLD-$2$-$4$-$8$}
				
		\node[anchor=north east, fill=white] at (rel axis cs:1,1) {\scriptsize{\makecell{$M=64$, $P=96$\\$(S,L)=(2, 32)$\\$(L', T)= (4, 8)$}}};
		\node[anchor=south west, fill=white] at (rel axis cs:0,0) {\footnotesize{{{$\mathcal{RM}(4,8)$}}}};
		
\coordinate (spypoint1) at (axis cs:3.5, 4.0e-5);
\coordinate (magnifyglass1) at (axis cs:2.25, 5.0e-4);

\end{axis}

\spy [blue, width=1cm, height=1.6cm] on (spypoint1) in node[fill=white] at (magnifyglass1);
\end{tikzpicture}

%% file: ferRM512_46_FSC_per_rpa.tikz.tex
\begin{tikzpicture}[spy using outlines = {rectangle, magnification=2.5, connect spies}]
	
	\pgfplotsset{
		label style = {font=\fontsize{9pt}{7.2}\selectfont},
		tick label style = {font=\fontsize{7pt}{7.2}\selectfont}
	}
	
	\begin{axis}[
		scale = 1,
		ymode=log,
		xlabel={$E_b/N_0$ [\text{dB}]}, xlabel style={yshift=0.5em},
		ytick={1e-6, 1e-5,1e-4,1e-3,1e-2,1e-1,1e-0},
		xtick={0,0.5,1,1.5,2,2.5},
		ylabel={FER}, ylabel style={yshift=-0.5em},
		grid=both,
		ymajorgrids=true,
		xmajorgrids=true,
		grid style=dashdotted,
		width=0.35\linewidth, height=7cm,
		thin,
		mark size=2.25,
		% use the following key so the baseline of all ticklabel entries is the same
		% (compare this image to the one from marmot)
		%		typeset ticklabels with strut,
		%		% there is one default value for the `legend pos' that is outside the axis
		%		legend pos=outer north east,
		%		% (so the legend looks a bit better)
		legend cell align=left,
		%		% (moved this common key here)
		%		smooth,
		legend style={
			%		legend pos=outer north east,
			%	      anchor={center},
			%cells={anchor=west},
			column sep= 2mm,
			font=\fontsize{7pt}{7.2}\selectfont,
		},
				legend to name=perf-legend-rpa-fer-2-9,
				legend columns=3,
		]
		
\node[anchor=north east, fill=white] at (rel axis cs:1,1) {\scriptsize{\makecell{$M=128$, $P=116$\\$(S,L)=(3,16)$\\$(L',T)= (2,8)$}}};
\node[anchor=south west, fill=white] at (rel axis cs:0,0) {\footnotesize{{{$\mathcal{RM}(2,9)$}}}};
		
		\addplot[
		color=gray,
		mark=none,
		thin,
		mark size=2,
		solid,
		mark options={solid},
		]
		table {
%			0	6E-02
			0.5	2.01E-02
			1	5.5E-03
			1.5	1.0E-03
			2.0	1.7E-04
%			2.5	2.2E-05
		};
		%\addlegendentry{ML (lower bound) \cite{Dumer06}}
				
		\addplot[
			color=blue,
			mark=o,
			thin,
			mark size=2,
			solid,
			mark options={solid},
		]
		table {
%			0.0	1.02E-01
			0.5	3.81E-02
			1.0	1.03E-02
			1.5	2.28E-03
			2.0	2.83E-04
		};
		%\addlegendentry{RLDA-$64$ \cite{Dumer06, Geiselhart}}
		
		\addplot[
			color=blue,
			mark=o,
			thin,
			mark size=2,
			dashdotted,
			mark options={solid},
		]
		table {
			0.5	8.05E-02
			1.0	2.86E-02
			1.5	7.76E-03
			2.0	1.39E-03
		};
		%\addlegendentry{RLDP-$128$ \cite{Dumer06}}
		
		\addplot[
			color=red,
			mark=square,
			thin,
			mark size=2,
			solid,
			mark options={solid},
		]
		table {
			0.5	3.71E-02
			1.0	1.13E-02
			1.5	1.84E-03
			2.0	2.65E-04
		};
		%\addlegendentry{Aut-SSC-$128$ \cite{Geiselhart}}
		
		\addplot[
			color=cyan,
			mark=diamond,
			thin,
			mark size=2,
			solid,
			mark options={solid},
		]
		table {
			0.5	4.10E-02
			1.0	1.16E-02
			1.5	2.43E-03
			2.0	3.26E-04
		};
		%\addlegendentry{SSP-RLD-$3$-$16$}
		
		\addplot[
			color=blue,
			mark=triangle,
			thin,
			mark size=2,
			solid,
			mark options={solid},
		]
		table {
			0.5	5.73E-02
			1.0	1.73E-02
			1.5	4.06E-03
			2.0	5.00E-04
		};
		%\addlegendentry{SRPA \cite{fathollahi2020sparse}}
		
		\addplot[
		color=red,
		mark=square,
		thin,
		mark size=2,
		dashdotted,
		mark options={solid},
		]
		table {
			0.5	1.01E-01
			1.0	4.30E-02
			1.5	1.52E-02
			2.0	4.21E-03
		};
		%\addlegendentry{Per-SSC-$P$ \cite{Kamenev19}}
		
		\addplot[
			color=cyan,
			mark=pentagon,
			thin,
			mark size=2,
			solid,
			mark options={solid},
		]
		table {
			0.5	4.60E-02
			1.0	1.45E-02
			1.5	2.88E-03
			2.0	3.70E-04
		};
		%\addlegendentry{Aut-SSP-RLD-$S'$-$L'$-$T$}
		
\coordinate (spypoint1) at (axis cs:2, 3e-4);
\coordinate (magnifyglass1) at (axis cs:0.75, 1.5e-3);

\end{axis}

\spy [blue, width=1cm, height=1.6cm] on (spypoint1) in node[fill=white] at (magnifyglass1);
\end{tikzpicture}

%% file: ferRM512_130_FSC_per_rpa.tikz.tex
\begin{tikzpicture}[spy using outlines = {rectangle, magnification=2.5, connect spies}]
	
	\pgfplotsset{
		label style = {font=\fontsize{9pt}{7.2}\selectfont},
		tick label style = {font=\fontsize{7pt}{7.2}\selectfont}
	}
	
	\begin{axis}[
		scale = 1,
		ymode=log,
		xlabel={$E_b/N_0$ [\text{dB}]}, xlabel style={yshift=0.5em},
		ytick={1e-6, 1e-5,1e-4,1e-3,1e-2,1e-1,1e-0},
		xtick={0.5, 1,1.5,2,2.5,3},
		ylabel style={yshift=-0.5em},
		grid=both,
		ymajorgrids=true,
		xmajorgrids=true,
		grid style=dashdotted,
		width=0.35\linewidth, height=7cm,
		thin,
		mark size=2.25,
		% use the following key so the baseline of all ticklabel entries is the same
		% (compare this image to the one from marmot)
		%		typeset ticklabels with strut,
		%		% there is one default value for the `legend pos' that is outside the axis
		%		legend pos=outer north east,
		%		% (so the legend looks a bit better)
		legend cell align=left,
		%		% (moved this common key here)
		%		smooth,
		legend style={
			%		legend pos=outer north east,
			%	      anchor={center},
			%cells={anchor=west},
			column sep= 2mm,
			font=\fontsize{7pt}{7.2}\selectfont,
		},
		legend to name=perf-legend-rpa-fer-3-8,
		legend columns=3,
		]		
		
		\node[anchor=north east, fill=white] at (rel axis cs:1,1) {\scriptsize{\makecell{$M=128$, $P=119$\\$(S,L)=(2,32)$\\$(L',T)=(4,8)$}}};
		\node[anchor=south west, fill=white] at (rel axis cs:0,0) {\footnotesize{{{$\mathcal{RM}(3,9)$}}}};
		
		\addplot[
		color=gray,
		mark=none,
		thin,
		mark size=2,
		solid,
		mark options={solid},
		]
		table {
			0.5	2.39E-03
			1.0	3.37E-04
%			1.5	2.20E-05
		};
		%\addlegendentry{ML (lower bound) \cite{Dumer06}}
		
		\addplot[
		color=red,
		mark=square,
		thin,
		mark size=2,
		solid,
		mark options={solid},
		]
		table {
			0.5	4.63E-01
			1.0	2.10E-01 
			1.5	4.77E-02
			2.0	5.78E-03
			2.5	2.38E-04
		};
		%\addlegendentry{Aut-SSC-$P$ \cite{Geiselhart}}
		
		\addplot[
			color=blue,
			mark=o,
			thin,
			mark size=2,
			solid,
			mark options={solid},
		]
		table {
			0.5	3.48E-01
			1.0	1.25E-01
			1.5 2.75E-02
			2.0	3.25E-03
			2.5	1.58E-04
		};
		%\addlegendentry{RLDA-$M$ \cite{Geiselhart, Dumer06}}
		
		\addplot[
			color=red,
			mark=square,
			thin,
			mark size=2,
			dashdotted,
			mark options={solid},
		]
		table {
			0.5	5.47E-01
			1.0	2.79E-01
			1.5	9.14E-02
			2.0	1.73E-02
			2.5	1.84E-03
		};
		%\addlegendentry{Per-SSC-$P$ \cite{Kamenev19}}
		
		\addplot[
		color=blue,
		mark=o,
		thin,
		mark size=2,
		dashdotted,
		mark options={solid},
		]
		table {
			0.5	4.95E-01
			1.0	2.21E-01
			1.5	7.28E-02
			2.0	1.40E-02
			2.5	1.50E-03
		};
		%\addlegendentry{RLDP-$M$ \cite{Dumer06}}
		
		\addplot[
		color=cyan,
		mark=diamond,
		thin,
		mark size=2,
		solid,
		mark options={solid},
		]
		table {
			0.5	3.71E-01
			1	1.45E-01
			1.5	3.45E-02
			2.0	3.72E-03
			2.5	1.82E-04
		};
		%\addlegendentry{SSP-RLD-$S$-$L$}
		
		\addplot[
		color=cyan,
		mark=pentagon,
		thin,
		mark size=2,
		solid,
		mark options={solid},
		]
		table {
			0.5	4.54E-01
			1.0	1.94E-01
			1.5	4.82E-02
			2.0	5.80E-03
			2.5	2.83E-04
		};
		%\addlegendentry{Aut-SSP-RLD-$S'$-$L'$-$T$}
				
		\coordinate (spypoint1) at (axis cs:2.5, 2.2e-4);
		\coordinate (magnifyglass1) at (axis cs:1.25, 2e-3);
		
	\end{axis}
	
	\spy [blue, width=1cm, height=1.6cm] on (spypoint1) in node[fill=white] at (magnifyglass1);
\end{tikzpicture}

%% file: ferRM512_256_FSC_per_rpa.tikz.tex
\begin{tikzpicture}[spy using outlines = {rectangle, magnification=2.5, connect spies}]
	
	\pgfplotsset{
		label style = {font=\fontsize{9pt}{7.2}\selectfont},
		tick label style = {font=\fontsize{7pt}{7.2}\selectfont}
	}
	
	\begin{axis}[
		scale = 1,
		ymode=log,
		xlabel={$E_b/N_0$ [\text{dB}]}, xlabel style={yshift=0.5em},
		ytick={1e-6, 1e-5,1e-4,1e-3,1e-2,1e-1,1e-0},
		xtick={1,1.5,2,2.5,3},
		ylabel style={yshift=-0.5em},
		grid=both,
		ymajorgrids=true,
		xmajorgrids=true,
		grid style=dashdotted,
		width=0.35\linewidth, height=7cm,
		thin,
		mark size=2.25,
		% use the following key so the baseline of all ticklabel entries is the same
		% (compare this image to the one from marmot)
		%		typeset ticklabels with strut,
		%		% there is one default value for the `legend pos' that is outside the axis
		%		legend pos=outer north east,
		%		% (so the legend looks a bit better)
		legend cell align=left,
		%		% (moved this common key here)
		%		smooth,
		legend style={
			%		legend pos=outer north east,
			%	      anchor={center},
			%cells={anchor=west},
			column sep= 2mm,
			font=\fontsize{7pt}{7.2}\selectfont,
		},
		legend to name=perf-legend-rpa-fer-3-8,
		legend columns=3,
		]	
		
		\addplot[
		color=gray,
		mark=none,
		thin,
		mark size=2,
		solid,
		mark options={solid},
		]
		table {
			1.0	1.40E-03
			1.5	8.86E-05
		};
		%\addlegendentry{ML (lower bound) \cite{Dumer06}}
		
		\addplot[
		color=red,
		mark=square,
		thin,
		mark size=2,
		solid,
		mark options={solid},
		]
		table {
			1.0	8.30E-01
			1.5	5.10E-01
			2.0	1.60E-01
			2.5	1.62E-02
			3.0	6.20E-04
		};
		%\addlegendentry{Aut-SSC-$P$ \cite{Geiselhart}}
		
		\addplot[
		color=red,
		mark=square,
		thin,
		mark size=2,
		dashdotted,
		mark options={solid},
		]
		table {
			1.0	8.35E-01
			1.5	5.16E-01
			2.0	1.79E-01
			2.5	3.00E-02
			3.0	1.96E-03
		};
		%\addlegendentry{Per-SSC-$P$ \cite{Kamenev19}}
		
		\addplot[
		color=blue,
		mark=o,
		thin,
		mark size=2,
		solid,
		mark options={solid},
		]
		table {
			1.0	7.63E-01
			1.5	4.23E-01
			2.0	1.25E-01
			2.5	1.51E-02
			3.0	5.85E-04
		};
		%\addlegendentry{RLDA-$M$ \cite{Geiselhart, Dumer06}}
		
		\addplot[
		color=blue,
		mark=o,
		thin,
		mark size=2,
		dashdotted,
		mark options={solid},
		]
		table {
			1.0	8.63E-01
			1.5	5.16E-01
			2.0	1.99E-01
			2.5	3.88E-02
			3.0	3.13E-03
		};
		%\addlegendentry{RLDP-$M$ \cite{Dumer06}}
		
		\addplot[
		color=cyan,
		mark=diamond,
		thin,
		mark size=2,
		solid,
		mark options={solid},
		]
		table {
			1.0 7.14E-01
			1.5	3.63E-01
			2.0	9.16E-02
			2.5	9.94E-03
			3.0	3.81E-04
		};
		%\addlegendentry{SSP-RLD-$S$-$L$}
		
		\addplot[
		color=cyan,
		mark=pentagon,
		thin,
		mark size=2,
		solid,
		mark options={solid},
		]
		table {
			1.0	7.64E-01
			1.5	4.29E-01
			2.0	1.20E-01
			2.5	1.32E-02
			3.0	5.00E-04
		};
		%\addlegendentry{Aut-SSP-RLD-$S'$-$L'$-$T$}
				
		\coordinate (spypoint1) at (axis cs:3, 5e-4);
		\coordinate (magnifyglass1) at (axis cs:1.75, 1.5e-3);
		
		\node[anchor=north east, fill=white] at (rel axis cs:1,1) {\scriptsize{\makecell{$M=64$, $P=85$\\$(S, L)=(2,32)$\\$(L',T)=(4,8)$}}};
		\node[anchor=south west, fill=white] at (rel axis cs:0,0) {\footnotesize{{{$\mathcal{RM}(4,9)$}}}};	
		
	\end{axis}
	
	\spy [blue, width=1cm, height=1.6cm] on (spypoint1) in node[fill=white] at (magnifyglass1);
\end{tikzpicture}